\documentclass{aa}  

\usepackage{url}
\usepackage{hyperref}
\usepackage{subcaption}
\usepackage{natbib}
\usepackage{multirow}
\usepackage{graphicx}
\usepackage{txfonts}
\usepackage{enumitem}
\usepackage{placeins}
\usepackage{enumitem,kantlipsum}
\usepackage{booktabs}
\usepackage{pifont}
\usepackage{mathtools}
\usepackage[thinc]{esdiff}
\usepackage{mathabx}
\usepackage{wasysym}
\usepackage{xspace}

\hypersetup{
    colorlinks = true,
    citecolor = blue,
}

\bibpunct{(}{)}{;}{a}{}{,}
\setcitestyle{authoryear,open={(},close={)}} 

\newcommand{\fOtwo}{\textit{f}O$_2$\xspace}
\newcommand{\dIW}[1]{$\Delta$IW#1\xspace}
\newcommand{\logvmf}{$\Psi_\text{VMF}$\xspace}
\newcommand{\metal}{$\xi$\xspace}
\newcommand{\micrometer}{$\mu$m\xspace}
\newcommand{\micron}{$\mu$m\xspace}
\newcommand{\tirr}{$T_\text{irr}$\xspace}
\newcommand{\tint}{$T_\text{int}$\xspace}
\newcommand{\atmodeller}{\texttt{Atmodeller}\xspace}

\begin{document} 

    \title {Volatile-bearing mineral atmospheres of hot rocky exoplanets as probes of interior state and composition }
    \titlerunning{Volatile-bearing mineral atmospheres of hot rocky exoplanets as probes...}
        
    \author{Fabian L. Seidler \inst{1} \and Paolo A. Sossi \inst{1} \and Dan J. Bower \inst{1} \and Brice-Olivier Demory \inst{2, 3}}
    
    \institute{
        ETH Zürich, Department of Earth and Planetary Sciences, Institute for Geochemistry and Petrology, Zürich, Switzerland \\
        \email{fabian.seidler@eaps.ethz.ch}
        \and
        Universität Bern, Center for Space and Habitability, Bern, Switzerland
        \and
        Universität Bern, Physikalisches Institut, Weltraumforschung und Planetologie, Gesellschaftsstrasse 6, 3012 Bern, Switzerland
    }
    
    \date{Received --; accepted --}

  \abstract
   {Spectra collected by the James Webb Space Telescope (JWST) hint at a volatile-rich and perhaps CO/CO$_2$-bearing atmosphere on the hot rocky exoplanet (HRE) 55 Cancri e. Such atmospheres, should they persist, are thought to be products of mass exchange with underlying magma oceans and thereby carry an imprint of the geochemical state of HREs. }
   {Here, we aim to identify diagnostic features in emission and transmission that can be used to infer the composition and geochemical state of HREs.  }
   {We constructed a coupled atmosphere-interior model that computes the equilibrium gas speciation in the system Si-Mg-Fe-O-C-H-S-N-He. The model accounts for both the equilibrium vaporisation of mineral gases and the partitioning of volatile species between the magma ocean and atmosphere. Using a fiducial planet with the properties of 55 Cancri e, we explored a parameter space that spans volatile mass fractions from 0.1 to ten times that of the Earth, solar- to Earth-like metallicities, and oxygen fugacities (\fOtwo) -6 to +6 log$_{10}$ units relative to the iron-wüstite (IW) buffer. } 
   {
    We find the \fOtwo of the mantle to be the major control on emission and transmission spectra. 
    At low \fOtwo (\dIW{} $<$ -3), SiO and CO appear in both emission and transmission. Between $-3 < \Delta$IW $<$ +3, CO and CO$_2$ dominate infrared spectra. At high \fOtwo (\dIW{} $\ge$ +3), CO$_2$ in high-metallicity atmospheres is joined by SO$_2$, producing strong absorption at 9$\mu$m. H$_2$O features appear in all \dIW{} > -3 atmospheres, but are sensitive to H$_2$ content. Condensation of silicates and iron oxides is common in the upper atmospheres of oxidised planets and is driven by cooling from triatomic molecules.
    Observed mass-radius trends suggest that a substantial fraction of HREs, including 55 Cancri e, have modest atmospheres of mixed heritage that are degenerate in \fOtwo, volatile mass, and metallicity. Combined with JWST Mid-Infrared Instrument observations, high metallicity (Earth-like) atmospheres as well as highly reduced solar-like atmospheres are precluded on 55 Cancri e, while the Near Infrared Camera data remain inconclusive.
    Future observations at wavelengths beyond 8 $\mu$m are key to discerning between potential scenarios. 
   }
   {}


   \maketitle

\section{Introduction}
\label{intro}
Hot rocky exoplanets (HREs) denote a class of ultra-short period exoplanets that are similar in density to those of the terrestrial planets of the Solar System, implying an interior structure dominated by a rocky mantle and a core of iron-nickel alloy. Unlike the terrestrial planets, they have dayside temperatures ($T_\text{day}$) in excess of the melting point of silicates ($\gtrsim$1000~K).
They usually exceed Earth in mass and radius \citep{luque2022density}, thus forming a subset of the Super-Earth population, and revolve on tight orbits around their host stars usually at a distance of a few stellar radii, $d_\text{orb}\sim2-10$  $R_\star$). Consequently, also owing to the age of the stellar system in which they occur, these planets are likely to be tidally locked \citep{leger2011extreme}, resulting in the day side likely showing extreme hot spots even in the presence of atmospheres \citep{hammond2017linking, zhang2017effects}.
Hot rocky exoplanets are also numerous. Today, roughly 1000 candidates with substellar temperatures above 1000 K have been identified \citep{zilinskas2022observability}.

Though their densities are broadly consistent with an interior structure similar to that of Earth \citep{zeng2016mass}, the range permits planets with large metallic cores akin to Mercury \citep{adibekyan2021compositional}, while others have uncompressed densities lower than Earth's and thus could harbour a smaller core \citep{elkins2008coreless}, a gaseous or supercritical water-rich layer \citep[e.g.][]{aguichine2021mass}, or a (comparatively) thick atmosphere \citep{peng2024puffy}; for some, only the last two options are consistent with their low density. 
 
Recent observations with the James Webb Space Telescope (JWST) of the old \citep[8.6$\pm$1 Gyr,][]{bourrier2018_55cnce} HRE 55 Cancri e indicate the possible existence of an atmosphere \citep{hu2024secondary, patel2024jwst} that appears compatible with a CO/CO$_2$ composition \citep{hu2024secondary}, supporting earlier claims based on data from the Spitzer Space Telescope \citep{demory2016map, demory2016variability}. Other potential HRE candidates are TOI-431 b \citep{monaghan2025low} and TOI-561 b \citep{teske2025thick}, showing lower than expected day-side emission consistent with heat-redistribution by an atmosphere.
Therefore, despite the likelihood of atmospheric escape owing to the high degree of stellar irradiation experienced by HREs (e.g. \citealt{salz2016energy, cherubim2025oxidation}), volatile-rich atmospheres appear to persist.

If present, the atmosphere may act as a thermal blanket for the interior, sustaining a deep molten magma ocean over its lifetime \citep{nicholls2024magma, nicholls2025convective}.
The high temperatures further facilitate short chemical reaction timescales \citep{arrhenius1889dissociationswarme}, implying that the chemistry of the atmosphere should adjust to the thermochemical conditions of the mantle, provided the mass of the atmosphere is small compared to the reservoir of chemically active melt (\citealt{seidler2024impact}; otherwise, the mantle and atmosphere tend to a common equilibrium; see e.g. \citealt{schlichting2022chemical}).
Physically, this equilibrium manifests as the mutual solubility of volatile and silicate liquid components at the magma ocean--atmosphere interface (MAI). Volatile species that show particularly high solubilities in silicate liquids are water \citep{sossi2023solubility}, S-bearing gases \citep{BW22, BW23}, and N$_2$, provided N$^{3-}$ is stable in the liquid \citep{libourel2003nitrogen, dasgupta2022fate}. As volatile-bearing gases dissolve in the magma, the atmospheric composition must change as a result \citep{sossi2020redox, kite2020atmosphere, seo2024role}.
In addition, evaporation of the underlying silicate melt releases gases such as SiO(g), Mg(g), and Fe(g), colloquially known as mineral gases. On dry (volatile-free) HREs, the mineral vapour may form an atmosphere on its own \citep[e.g.][]{zilinskas2022observability, seidler2024impact}, but if volatiles are present, a mixed atmosphere emerges \citep{zilinskas2023observability, charnoz2023effect, piette2023rocky, van2025lavatmos}.
Therefore, interactions at the MAI, which is likely not directly observable by telescopes, imposes the speciation and mass of the atmosphere at the interface, thereby permitting the nature of the underlying magma ocean to be inferred from observations of atmospheric spectra.

A handful of earlier studies have attempted to model mixed mineral-volatile atmospheres of HREs. Typically, they assumed injection of pre-formed silicate vapour into a volatile background gas, closing the chemical network a posteriori \citep{zilinskas2023observability, piette2023rocky}.
This approach, though instructive, is not thermodynamically self-consistent because the \fOtwo of the system after the addition of volatile elements (such as H and C) decreases due to consumption of O$_2$ in the initial mineral atmosphere by H$_2$O and CO/CO$_2$  \citep[cf.][]{falco2024hydrogenated}.
Therefore, the final \fOtwo will no longer correspond to the initial guess, which was used to determine the mass and composition of the mineral atmosphere.
This is important because the equilibrium partial pressures of mineral gas species, such as SiO(g), depend on \fOtwo \citep{wolf2023vaporock, seidler2024impact}.

Later studies by \citet{charnoz2023effect} and \citet{falco2024hydrogenated} combined a stoichiometric vaporisation approach with titration of H$_2$-gas into the system.
The resulting speciation in the atmospheres is chemically and thermodynamcally self-consistent but imposes the condition that \fOtwo depends upon the amount of H$_2$ gas titrated into the system, precluding a treatment of \fOtwo as an independent variable.
This requirement arises because under the small-atmosphere approximation, the thermochemistry (including \fOtwo) of the vapour should be set by the mantle, and the conserved quantity for gas species in exchange with the melt is therefore not their amount but their chemical potential, $\mu = \mu^0 + RT \ln f$, characterised by their fugacity, $f$, at a given temperature ($\mu^0$ is the chemical potential at the arbitrary reference state, typically 1 bar and the $T$ of interest, where $T$ is the temperature and $R$ the is universal gas constant).
As shown in \citet{seidler2024impact} (their Eq. 8), the fugacity of a mineral gas species is a function of the thermodynamic activity, $a$, of the parent species\footnote{For example, SiO$_2$(l) for SiO(g).
However, the melt does not contain SiO$_2$ as a species per se. Instead, it is composed of weakly covalently bound Si$^{4+}$O$^{2-}_4$ tetrahedrons that may or many not polymerise to form a network, with the metal ions (Mg$^{2+}$, Fe$^{2+}$, ...) occupying interstitial sites.
However, via mass-balance, the mixture can be reduced to a set of functional endmembers (pseudospecies) that simplify the system, such as the oxides SiO$_2$ and MgO.} and the oxygen fugacity:
\begin{equation}
    \label{eq:seidler_2024}
    f ~\propto~ a^\alpha(P, T, \Vec{x}_{melt}) \cdot f\text{O}_2^\beta,
\end{equation}

where $\alpha$ and $\beta$ are constants depending solely on the stoichiometry of the vaporisation reaction.

The importance of \fOtwo is further demonstrated in pure volatile atmospheres, wherein the speciation shifts from reduced components such as CO and H$_2$ to oxidised CO$_2$ and H$_2$O \citep{sossi2020redox}, and the solubility of N$_2$ and S$_2$ increase significantly as \fOtwo decreases \citep{libourel2003nitrogen, bernadou2021nitrogen, BW22, BW23}. Correspondingly, the atmospheric signature should change; in particular, the strong increase in SO$_2$ partial pressure and its high opacity make it an excellent probe in emission spectra for the redox state of magma ocean atmospheres in planets with an Earth-like volatile budget \citep{hu2024secondary, bello2025evidence, nicholls2025convective}. However, the degree to which its intensity varies as a function of the partial pressures of other gases has yet to be explored. Moreover, depending on the extent to which mineral gases evaporate, the production of new species occurs, such as silane (SiH$_4$), in atmospheres of solar-like composition under reducing conditions \citep{charnoz2023effect}. 

The mantle redox state, as expressed by \fOtwo, is expected to exert crucial controls on the atmospheric composition.  Inversion thereof can be used to constrain information about the hidden interior. 
Yet a systematic and self-consistent treatment of how the presence of volatile elements interact with mineral species in the presence of a magma ocean have not been performed with self-consistent thermodynamics. Such models are crucial for identifying degeneracies in atmospheric retrievals of \fOtwo, volatile mass fractions, and metallicities of HREs as well as diagnostic signals thereof. 

Here, we present a coupled interior-atmosphere model that computes the key observable properties of volatile-bearing HREs of a given mass: their radius, $R_p$; the emission and transmission spectra as function of the \fOtwo at the magma ocean-atmosphere interface; the volatile mass fraction (VMF); and the metallicity of volatile elements (\metal).
In Section \ref{methods}, we introduce this model, and we investigate the main results in Section \ref{results}, where we present a systematic investigation of the atmospheric mass and speciation as function of \fOtwo, VMF, and \metal as well as the structure and emission and transmission spectra of these atmospheres. We relate our models to the observed properties, notably mass-radius and spectral observations, where available, of HREs in Section \ref{discussion}. Finally we conclude and summarise our work in Section \ref{summary}, where we also highlight our assumptions and caveats.

\section{Methods}
\label{methods}

\subsection{Model pipeline}
\label{methods:pipeline}

The coupled atmosphere-interior model is composed of two main components, \textit{a.)} the (independent) model for computing the speciation and mass of gaseous species and dissolved components at the MAI,
\atmodeller\footnote{\url{https://github.com/ExPlanetology/atmodeller}} \citep{bower2025diversity}, and \textit{b.)} the (dependent) radiative transfer pipeline \texttt{phaethon}\footnote{\url{https://github.com/ExPlanetology/phaethon}} \citep{seidler2024impact} that computes the speciation and structure of the overlying atmosphere layer-by-layer based on the input boundary conditions at the MAI (step a.) and the top of the atmosphere (TOA).
In order to ensure consistency, steps a.) and b.) have to be brought into equilibrium, that is, the temperature at the MAI ($T_\text{MAI}$) must match the temperature at the bottom of the atmosphere ($T_\text{BOA}$) computed by the radiative transfer calculation for a given \tirr, for which an iterative computation is required. 
The components of the pipeline are explained in more detail below.

\subsubsection{Atmosphere-interior equilibrium model}
\label{methods:pipeline:outgassing}

The fundamental assumption of the atmosphere-interior model is that the melt and gas are at chemical equilibrium.
A precise statement of this condition is that the chemical potential ($\mu$) of any species ($i$) in the melt must equal that in the gas,
\begin{equation}
    \label{eq:chem_eq_methods}
    \mu_{i, \text{melt}} = \mu_{i, \text{gas}},
\end{equation}
as it enables the partitioning of species between mantle and atmosphere to be derived by linking their fugacities (effective pressures) in the gas phase with their activities (effective concentrations) in the melt, which is performed by \atmodeller via the extended law of mass action.\footnote{This law simultaneously enforces the law of mass action, which relates the free energies of compounds to their activities and fugacities and subsequently their mole fractions and partial pressures, and mass balance, which ensures mass is conserved.}

We imposed two sets of constraints on the system of equations.
First, we ensured mass conservation for each volatile element H, He, C, N, and S and thus their combined total mass.
The individual masses were determined as described in Sec. \ref{methods:volatile_budget}.

The second set of constraints are the fugacities of species supplied by the magma ocean. The mantle, composed of molten silicates, is the largest reservoir of O in the system and can effectively act as an infinite reserve thereof.
Hence, the masses of O$_2$, the mineral gases, or any gas species containing oxygen as well as the total atmospheric mass are not conserved. Instead, we imposed the oxygen fugacity (\fOtwo) as an independent variable, which is reported here relative to the iron-wüstite buffer IW, \dIW$:= f\text{O}_2(P,T) - IW(P,T)$, for which we used the formulation of \citet{hirschmann2021iron}.
The fugacities of the mineral species, here \textit{f}SiO, \textit{f}Mg and \textit{f}Fe \citep[one per included element; see][]{sossi2019evaporation}, are set by the melt composition, \fOtwo and $T_\text{MAI}$, and were computed using a modified version of the \texttt{MAGMA} code \citep{fegley1987vaporization, schaefer2004thermodynamic, seidler2024impact}. The initially volatile-free melt is composed of SiO$_2$, MgO and FeO in identical ratios to the bulk silicate Earth for all simulations \citep{palme2013cosmochemical}.

To evaluate partitioning of a (volatile) species $i$ (e.g. CO$_2$, H$_2$O) between mantle and atmosphere via a solubility law, its fugacity $f_i$ at the MAI has to be found. Fugacity relates to the partial pressure $p_i$ via an equation of state, taken here as the ideal gas law (for which $f_i = p_i$), that in turn hinges on the radius of the planet at the interface (i.e. $R_{p,\text{MAI}}$), the surface gravity ($g_\text{MAI}$), the species atmospheric mass ($M_i$), its molecular weight ($m_i$), and the average mean molecular weight (MMW) of the atmosphere ($\bar{m}$; \citet{bower2019linking}):
\begin{equation}
    \label{eq:mmw_effect}
    f_i = p_i = \frac{M_i\cdot g_\text{MAI}}{4 \pi R_{p,\text{MAI}}^2} \cdot \frac{\bar{m}}{m_i}.
\end{equation}
Internally, some of these quantities are further related by the temperature at the interface, $T_\text{MAI}$.
$R_{p,\text{MAI}}$ and $g_\text{MAI}$ are computed from the mass of the planet, $M_p$, and its core mass fraction (CMF) by interior structure models for rocky planets \citep{zeng2016mass}.
The effect of dissolved volatiles on the density of the planet is neglected.
As shown a posteriori (Sec. \ref{results:atmo_comp}), the concentration of water - the most soluble species - reaches, at most, 3 wt\%, for which we estimate a density decrease of $\sim 0.1$ g cm$^{-3}$ \citep{matsukage2005density, dorn2021hidden}. Relative to differences in mass-radius induced by atmospheric extent, the effect of H$_2$O is likely negligible; indeed, \citet{boley2023fizzy} found no major change in planet density even when incorporating 5.2 wt\% of water.
Additionally, we ignore the density difference introduced by melting of the mantle.
Again, the density difference between a fully molten and fully solid mantle becomes small for planets with $M_p \gtrsim 2.7 M_\oplus$ ($\sim 2$\% in density $\Leftrightarrow \sim 0.7$ \% in radius, \citealt{boley2023fizzy}), but might be relevant for Earth-sized planets ($\sim 11.1$\% in density $\Leftrightarrow \sim 3.7$ \% in radius; \citealt{boley2023fizzy, bower2019linking}), though this remains small compared to variations induced by atmospheric properties.
Unless mentioned otherwise, the mass is kept constant at $M_p=8 M_\oplus$ to mimic that of 55 Cancri e \citep{bourrier2018_55cnce}.
To study Earth-like interiors only, we keep CMF=0.325 throughout. With these parameters we obtain $R_{p,\text{MAI}}=1.74 R_\oplus$.
The melt mass is internally computed by \atmodeller from the mantle melt fraction $x_\text{melt}$ and the total mantle mass, $M_\text{melt} = x_\text{melt} \cdot M_p(1-\text{CMF})$. Because our model ignores any volatile species that can be incorporated into solids, $M_\text{melt}$ defines the effective mass into which volatile elements may dissolve.
Throughout this study, assume a fully molten mantle, $x_\text{melt} = 1$.
Lower melt fractions are not expected to substantially modify the atmospheric composition unless $x_\text{melt} \lesssim 0.3$  \citep{bower2022,maurice2024volatile}.
The melt fraction of volatile rich HREs are expected to be $\gtrsim 0.75$, as the planets struggle to cool under the intense irradiation \citep{nicholls2024magma}.

The set of desired target gas species (including species of both mineral and volatile gases) is added to \atmodeller together with their respective solubility laws, if available \citep{bower2025diversity}. The full list of gas species, including their solubility laws, can be found in Table \ref{table:target_species}.
\atmodeller then solves for the mass distribution and partial pressures of all species given the aforementioned constraints. 
Once all the partial pressures (and thus the total pressure), together with the dissolved quantities in the melt is known, its output is passed on to the radiative model (see next Sec. \ref{methods:pipeline:radtrans}). If \atmodeller detects the formation of condensates, the condensate mass is removed and the atmospheric composition is calculated from the remaining gas mass. Here, only graphite is considered due to the high temperature at the MAI.

\subsubsection{Radiative transfer model}
\label{methods:pipeline:radtrans}

To match $T_\text{MAI}$ with $T_\text{BOA}$ and find the equilibrium atmospheric pressure, structure and chemistry, we integrate the outgassing model into the radiative transfer pipeline \texttt{phaethon}. It operates as follows:

\begin{enumerate}

    \item  The energy budget of the planet is determined from the imposed irradiation temperature \tirr. The atmospheric P-T-structure for a given bulk composition and intrinsic temperature (see point 5, below) depends only on the incoming spectral energy distribution (SED), given in W m$^{-2}$ \micron$^{-1}$. In integrated form, this is reasonably well approximated by the Stefan-Boltzmann law, $\sigma T_\text{irr}^4$, where \tirr is the irradiation temperature at the TOA. Thus, planets with identical \tirr (and identical atmospheric composition) have identical atmospheric structures for equal stellar SEDs, regardless of other characteristics (e.g. stellar radius $R_\star$ or orbital separation $d$ between star and planet).

    \item Initially, we set $T_\text{MAI}$ to \tirr. Alternatively, an informed initial guess can be given to accelerate convergence.

    \item The atmospheric elemental abundances as well as the pressure at the MAI ($P_\text{MAI}$) are computed by the atmosphere-interior model (Sec. \ref{methods:pipeline:outgassing}) using the current $T_\text{MAI}$, and passed to the radiative transfer routine as constraints.
    
    \item \texttt{FastChem COND} \citep{kitzmann2024fastchem} is used to calculate look-up tables for the gas speciation as function of pressure and temperature (i.e. altitude), assuming that the gas retains the elemental composition set at the MAI throughout the atmospheric column, and that thermochemical equilibrium is established at every layer. This additionally assumes that the reaction timescale is negligibly short compared to the timescale of vertical diffusion of molecular species, for example by quenching, an assumption that is supported by the (expected) high temperatures of the atmosphere. We allowed for the formation of condensates but did not separate them from the gas (equilibrium condensation mode, as opposed to the rainout mode which would permanently remove condensing elements). The spectral properties of condensates, as either clouds or hazes, are ignored during the radiative transfer computation. All included gas and condensate species are listed in Table \ref{tab:fastchem_reactions}.
    
    \item \texttt{HELIOS} \citep{Malik:2017, Malik:2019} derives the P-T-structure of the atmosphere, including a new value for $T_\text{BOA}$, by solving the equations of radiative transfer using the opacities of the individual gas species and the chemistry determined in step 4. The atmosphere is plane-parallel and confined between the top-of-the-atmosphere pressure ($P_\text{TOA}$, fixed at $10^{-8}$ bar) and bottom-of-the-atmosphere pressure ($P_\text{BOA} = P_\text{MAI}$, computed by the outgassing model, step 3). The energy budget is characterised by \tirr (step 1); however, \texttt{HELIOS} internally operates on the physical parameters of the star-planet system. Therefore, we fix the properties of the star to solar values, $R_\star = R_\odot$, and take the spectrum from \citet{Gueymard:2003}. The orbital distance is reconstructed from \tirr via
    \begin{equation}
    \label{eq:orb_sep}
        d = \sqrt{\mathfrak{f} \cdot (1-A_B)} R_\star \left(\frac{T_\star}{T_{\text{irr}}} \right)^2,
    \end{equation}
    where $A_B$ is the bond albedo of the planet and $\mathfrak{f}$ is the dilution factor. We assume $A_B=0$ (black-body) and $\mathfrak{f} = 2/3$ (ineffective heat redistribution; see \citealt{hansen2008absorption}); as elucidated in point 1, neither the characteristics of the atmosphere nor the resulting emission and transmission spectra depend on these parameters.
    For the intrinsic temperature, we take a conservative estimate of $T_\text{int}=0$ K. However, deviations therefrom have been shown to severely modify atmospheric structures of cooler lava planets, particularly by inducing convection \citep[e.g.][]{nicholls2025self}. Consequently, we independently examine the effect of $T_\text{int} \in \{200, 250, 300\}$ K in Appendix \ref{apx:intrinsic_temp} and justify the choice of $T_\text{int}=0$ a posteriori.
    Convective adjustment is enabled with an adiabatic exponent for diatomic gases, $\kappa=2/7$:
    \begin{equation}
         \diff{\ln T}{\ln P} = \kappa := -\frac{1-\gamma}{\gamma},
     \end{equation} where $\gamma$ is the adiabatic index which is 7/5 for a diatomic, ideal gas. 
    The opacity is computed on-the-fly via the random-overlap method \citep{amundsen2017treatment}, using the mixing ratios of each layer inferred from the \texttt{FastChem} look-up tables (point 4). The selection of species opacities are outlined in Sec. \ref{methods:opacities}.
    
    \item  If the new $T_\text{BOA}$ is within $\Delta T_\text{abstol}$ of the prior guess of $T_\text{MAI}$, the model is taken to have converged. Otherwise, a new $T_\text{MAI}$ is estimated via root-finding (see Appendix \ref{apx:rootfinding}), and the model repeats from step 3. We choose $\Delta T_\text{abstol}=35$ K to minimise the number of calls to the computationally expensive radiative transfer solver, but find that most models converge with $\Delta T := T_\text{BOA} - T_\text{MAI} < 10$ K or even < 2 K. We expect negligible differences in the results if the constraints were to be tightened.
    
    \item Once converged, \texttt{phaethon} yields the atmospheric structure, surface pressure, gas speciation in each layer, and emission spectra. The latter cannot be used directly, as HELIOS does not take into account the wavelength dependent radius of the planet (based on the varying atmospheric opacity). \texttt{phaethon} therefore reconstructs the emission spectrum using the vertical structure and contribution function of the atmosphere (see Appendix \ref{apx:radius_wavelength_dependent}). 
    \item Transmission spectra were computed from the P-T-structure and chemistry with \texttt{petitRADTRANS} \citep{molliere2019petitradtrans}. The opacity sources are kept the same.
\end{enumerate}

\subsection{Opacities}
\label{methods:opacities}

The introduction of volatiles to a silicate mineral atmosphere vastly expands the list of species for which opacities are required \citep{zilinskas2023observability, zilinskas2025characterising}.
In most of our simulations, we use a selection that was optimised for accuracy and computational time and is based on the results from our speciation and radiative transfer simulations presented in Sec. \ref{results}.
We obtained the raw opacity sources for the atomic and molecular species from the DACE\footnote{\url{https://dace.unige.ch/opacityDatabase/}} or the former Exoclimes simulation platform, which hosted the data prior to their upload to DACE (no longer available at the time of publication).
The opacities hosted on either were computed with \texttt{HELIOS-K} \citep{grimm2015helios, grimm2021helios} assuming Voigt profiles, and with a resolution of $\Delta\nu=0.01~\text{cm}^{-1}$.
Molecular species have a line wing cut-off at 100 cm$^{-1}$ from the core, whereas atomic species have none to preserve the shape of wide resonant lines (e.g. Mg at 285 nm).
Collision-induced absorption (CIA) opacities were generated with \texttt{HELIOS-K} with the same specifications as the atomic and molecular opacities, using the data from HITRAN.\footnote{\url{https://hitran.org/cia/}}

The tables of correlated-k coefficients used by \texttt{HELIOS} were constructed from the raw opacities via the \texttt{k-table} program (part of \texttt{HELIOS}) using a resolution $R:=\lambda/\Delta\lambda=200$ in the wavelength range 0.18-200 \micron.
The detailed list of species is shown in Table \ref{tab:opacities} and comprises the "active" absorbers SiO, MgO, Mg, Fe, CO$_2$, CO, H$_2$O, CH$_4$, NO, SO, SO$_2$, HS, H$_2$S, OH,  O$_2$, and HCN, in addition to the "inactive" H, He, N$_2$ and H$_2$.
The latter do not show strong opacities (Fig. \ref{fig:opacities}), but may occur in large abundances and are therefore required by \texttt{HELIOS} to reproduce the proper MMW\footnote{While \texttt{FastChem} does compute the MMW correctly, it is not adopted by \texttt{HELIOS}, which instead computes it internally based on the species for which opacities are given.}.
C$_2$H$_2$ was not used as absorber due to limitations outlined in Appendix \ref{apx:acetylene_worlds}.
As Rayleigh scatterers, we allowed for O$_2$, H$_2$O, He, e-, H, N$_2$, CO, CO$_2$, and H$_2$.
For CIAs, only H$_2$-H$_2$, H$_2$-He, H$_2$-H and He-H were available at the temperatures of interest; since we found negligible influence on HRE spectra, we omitted them unless otherwise mentioned (Appendix \ref{apx:cia}).

\subsection{Volatile budget}
\label{methods:volatile_budget}

\subsubsection{Total volatile mass}
\label{methods:volatile_budget:total_mass}

We defined the volatile mass fraction (VMF) as the mass of H-He-C-N-S ($M_V$) relative to the mass of the entire planet ($M_P$): 

\begin{equation}
    \text{VMF} := \frac{M_V}{M_P}.
\end{equation}

For the Earth, $VMF_\oplus=0.00028485$ \citep{palme2013cosmochemical}. We further defined the volatile mass factor of any model planet by

\begin{equation}
    \Psi_{\text{VMF}} := \log_{10} \left( \frac{\text{VMF}}{\text{VMF}_\oplus} \right),
\end{equation}

which represents the $\text{VMF}$ of a planet of any mass, $M_P$, compared to that of Earth scaled up to $M_P$. That is, \logvmf=0 if the planet has the same mass fraction of volatile elements as Earth (though these are not necessarily present in the same relative abundances; see Sec. \ref{methods:volatile_budget:linear_mixing}, below). We caution that \logvmf is not identical to the atmospheric mass fraction since oxygen is not included in its calculation. Instead, oxygen is a special case as it is set by a combination of \logvmf, metallicity (see Sec. \ref{methods:volatile_budget:linear_mixing} below) and the \fOtwo of the mantle (for the same reason, we cannot use the atmospheric mass fraction as free parameter).
We investigated \logvmf $\in \{-1, 0, 1\}$, corresponding to 0.1, 1 and 10 times the Earth's volatile budget in mass, respectively.

\subsubsection{Mixing between solar and bulk-silicate Earth-like metallicities}
\label{methods:volatile_budget:linear_mixing}
\begin{table}
    \centering
    \caption{Mass ratios of volatiles for the two endmember cases.}
    \begin{tabular}{ccccc}
        \hline
         & He/H & C/H & N/H & S/H \\
         \hline
        SOLAR & 0.392518 & 0.004315 & 0.001205 & 0.000552\\
        VIBSE & 0 & 0.8300  & 0.0167 & 1.6667 \\
        \hline
    \end{tabular}
    \label{tab:volatile_ratios}
\end{table}

\begin{figure}[!t]
    \centering
    \includegraphics[width=\linewidth]{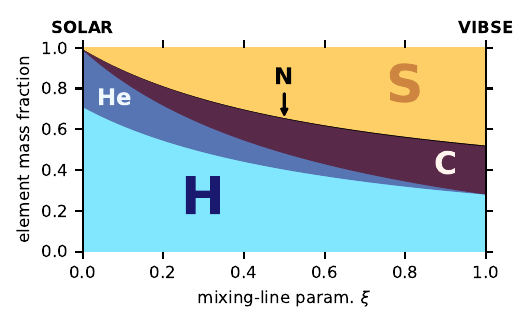}
    \caption{Bulk volatile (mass) ratios of volatiles as a function of \metal.}
    \label{fig:volatile_mass_ratios_z}
\end{figure}

Due to the computational demand of our forward model, and our inability to constrain the relative abundances of volatile elements in planetary bodies from stellar spectra in a similar vein to the major rock-forming elements \citep[e.g.][]{wang2019enhanced, wang2022detailed, spaargaren2023plausible}, we restrict ourselves to a more compact and efficient framework.
To incorporate variations in He/H, C/H, N/H and S/H ratios into a single parameter, we define a mixing line between a composition that mimics a solar gas (SOLAR, \citealt{lodders2021relative}) and one that corresponds to the volatile budget in bulk silicate Earth (VIBSE, \citealt{palme2013cosmochemical}, their Table 4)\footnote{We distinguish between BSE and VIBSE. 
The former is used only when describing the relative amounts of the rock-forming elements, Si-Mg-Fe-O, in the exact proportions stated by \citet{palme2013cosmochemical}, their Table 4, whereas VIBSE refers exclusively to the relative abundances of C-H-N-S-He, but at a potentially different $\text{VMF}$ from that of the BSE.
The distinction is important because, even in the case of SOLAR gas, the underlying interior melt composition remains that of the BSE; see Sec. \ref{methods:pipeline:outgassing}, point 3.}, for which we introduce the parameter \metal.
It formally corresponds to the fraction of VIBSE-like material in the total amount of volatiles, where \metal=0 is SOLAR and \metal=1 denotes VIBSE.
The mass ratios of volatiles relative to hydrogen, $\vec{\Phi} :=$ [He/H, C/H, N/H, S/H], are defined as

\begin{equation}
    \vec{\Phi}_\text{planet} = \vec{\Phi}_\text{SOLAR} + \xi \cdot \left( \vec{\Phi}_\text{VIBSE} - \vec{\Phi}_\text{SOLAR} \right),
\end{equation}

where the reference ratios can be found in Table \ref{tab:volatile_ratios}. 
The mass fraction of the elements as function of \metal is shown in Fig. \ref{fig:volatile_mass_ratios_z}. Phosphorus, while often listed among the crucial volatiles for exoplanet atmospheres \citep{zilinskas2025characterising}, was not considered in this work due to \textit{a.)} its low abundance in VIBSE ($\sim 87$ ppm, \citealt{palme2013cosmochemical}) and \textit{b.)} its comparatively low volatility compared to C-H-N-S \citep{gillmann2024interior}.

\section{Results}
\label{results}

\subsection{Atmospheric and interior chemistry}
\label{results:atmo_comp}

The atmospheric and interior chemistry is set by exchange at the MAI, and thus determined by the atmosphere-interior equilibrium model (Sec. \ref{methods:pipeline:outgassing}).
We describe element speciation in the interior and atmosphere of a fully molten planet with $M_p=8 M_\oplus$, $R_{p,\text{MAI}}=1.74 R_\oplus$ and $T_\text{MAI}=3000$ K, which mimics the broad characteristics of 55 Cancri e ($M_p=7.99\pm0.33~M_\oplus$ and $R_p=1.875\pm0.029~R_\oplus$, \citealt{bourrier2018_55cnce}, and $T_\text{MAI}\sim$ 3000 K in accordance with later findings; see Sec. \ref{results:atmo_struc:opac_and_thermal}).
We examine variations in the surface partial pressures at the MAI (Fig. \ref{fig:outgassing}) and in the elemental mass fractions in the atmosphere relative to that dissolved in the mantle (Fig. \ref{fig:ingassing}) as a function of the three independent variables investigated in this work: \logvmf, \metal and \fOtwo.

\subsubsection{Helium}
\begin{figure}[!t]
    \centering
    \includegraphics[width=\linewidth]{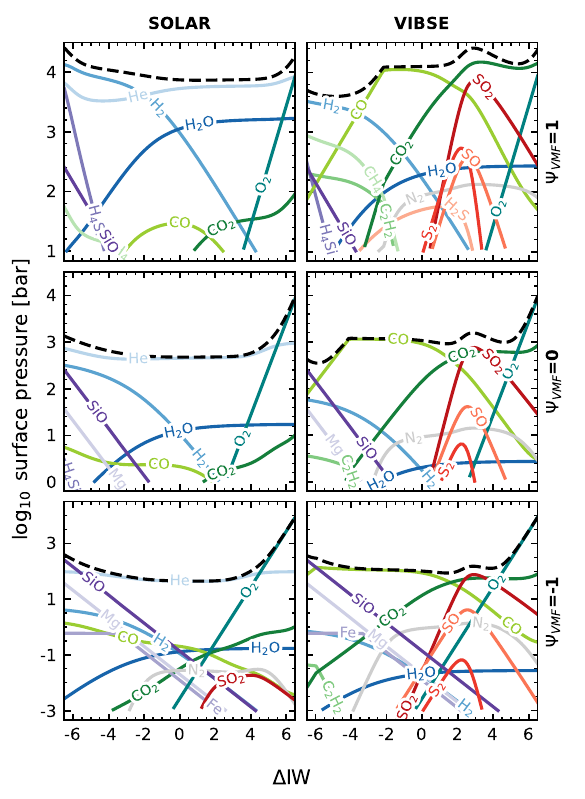}
    \caption{Partial pressures of gas species at the magma ocean-atmosphere interface (MAI) as function of \fOtwo, \logvmf, and \metal. All atmospheres were simulated at $T_\text{MAI}=3000$ K and $M_p=8 M_\oplus$.}
    \label{fig:outgassing}
\end{figure}

Helium is abundant in low-\metal (SOLAR) atmospheres, with the vast majority of its inventory remaining in the atmosphere (>90\%; see Fig. \ref{fig:ingassing}; melt concentrations are $< 0.07$ wt\% over the tested range).
This reflects its low solubility, based on mechanical incorporation into interstitional sites in the melt structure \citep{carroll1994noble}.
Its solubility law is therefore independent of \fOtwo, yet, there is an apparent dependence of $p$He on \dIW{} in Fig. \ref{fig:outgassing}, which arises due to evolving $\bar{m}$ of the atmosphere with \fOtwo, particularly affecting light gases such as He (cf. Eq. \ref{eq:mmw_effect}).
The changes in $\bar{m}$ are driven by Si-outgassing in reducing and O$_2$-formation in oxidising atmospheres (see Sec. \ref{results:atmo_comp:Si} and \ref{results:atmo_comp:O}).
Consequently, increased $p$He leads to greater dissolution of He under these conditions (Fig. \ref{fig:ingassing}).

\subsubsection{Hydrogen}
We recovered the $f$O$_2$-dependent behaviour of the partial pressures of the major hydrogen gases H$_2$ and H$_2$O found in earlier studies \citep{sossi2020redox, gaillard2022redox}. 
While $p$H$_2$O/$p$H$_2$ is proportional to \fOtwo$^{0.5}$, the high solubility of water \citep[as OH$^-$][]{sossi2023solubility,thompson2025water} relative to H$_2$ \citep[as H$_2$,][]{hirschmann2012solubility, foustoukos2025molecular, chaudhari2025hydrogen} means that $p$H$_2$ reaches higher limits than does $p$H$_2$O, by $\sim$ 1-2 orders of magnitude.
Correspondingly, the atmospheric H content becomes a strong function of \fOtwo, with expected H contents an order of magnitude greater in reducing (\dIW{$\lesssim$ 0}) than in oxidising atmospheres.

\begin{figure}[!t]
    \centering
    \includegraphics[width=\linewidth]{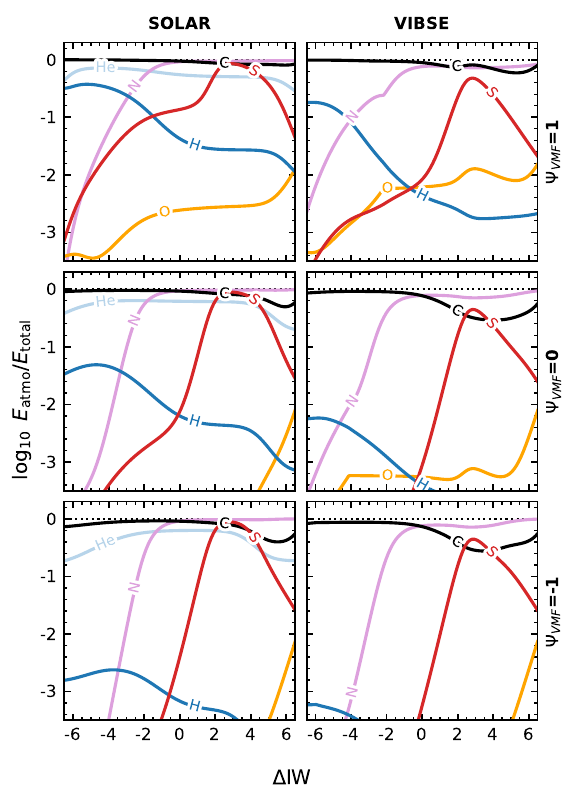}
    \caption{Faction of the budget of a given element $E \in \{H, He, C, N, O, S\}$ that resides in the atmosphere ($E_\text{atmo}$/$E_\text{total}$), where "total" refers to the mass-weighted sum of atmosphere plus mantle. The dotted black line, $\log_{10}$-zero, denotes the level at which the entirety of an element would reside in the atmosphere.}
    \label{fig:ingassing}
\end{figure}

Relative to its atmospheric abundance, H is almost always more concentrated in the mantle, especially under oxidised conditions (Fig. \ref{fig:ingassing}).
This leads to comparatively high concentrations, up to $\sim$ 0.5 wt\% of H$_2$O in the melt for \logvmf$\lesssim 1$; in volatile-rich and intermediate-to-oxidised systems (\logvmf=1, \dIW{}$\geq 0$), concentrations of up to 3 wt\% are reached.
Although Fig. \ref{fig:ingassing} shows lower H fractions in mantles in equilibrium with SOLAR-like metallicity (\metal) atmospheres, they actually contain higher absolute concentrations of H$_2$O than for \metal=1 (VIBSE) cases, roughly by a factor of $\sim 2$.
This is because the total hydrogen budgets are higher in SOLAR than in VIBSE cases (Fig. \ref{fig:volatile_mass_ratios_z}).

Water and hydrogen are the two dominant H-bearing gas species for most of the explored parameter space. Another important hydrogen carrier, CH$_4$, only achieves a low partial pressure at the MAI temperatures of HRE atmospheres. However, it becomes the most abundant H-bearing gas for cases with VIBSE-like \metal and high \logvmf ($>$1; Fig. \ref{fig:vapser_batch_T3000}), owing to the homogeneous equilibrium:

\begin{equation}
   \mathrm{ 3H_2(g) + CO(g) = CH_4(g) + H_2O(g)},
    \label{eq:methane_prod}
\end{equation}

which proceeds to the right at high pressures. It should be noted, however, that the total pressure in such atmospheres exceeds 1 GPa, leading to non-ideal behaviour of gases \citep[e.g.][]{bower2025diversity, hakim2026silane}. 

\subsubsection{Carbon}
Neither member of the CO-CO$_2$ redox pair has a high solubility in silicate melts \citep{dixon1995, yoshioka2019}, nor does CH$_4$ \citep{ardia2013}; out of those, CO$_2$ is the most soluble under intermediate to oxidising conditions (\dIW{} > 0) and may reach concentrations of up to $\sim 0.15$ wt\% in \logvmf=1 atmospheres (CO $<0.01$ wt\%, CH$_4$ $<0.0014$ wt\%). As a result, the majority of the carbon budget of a given planet resides in the atmosphere (Fig. \ref{fig:ingassing}).
Therefore, a nearly continuous transition with similar $p$CO to $p$CO$_2$ occurs with increasing \fOtwo$^{0.5}$ at constant \logvmf. For constant \fOtwo, carbon may also precipitate as graphite, provided the partial pressure of CO is sufficiently high:

\begin{equation}
  \mathrm{  CO(g) = C(cr) + \frac{1}{2}O_2(g). }
    \label{eq:graphite_prod}
\end{equation}

This shows as a sharp decrease in the partial pressures of carbon-bearing gases, mainly CO, under reducing conditions (Fig. \ref{fig:outgassing}).
As illustrated in Eq. \ref{eq:graphite_prod}, graphite precipitation occurs when the system is reducing (i.e. low \fOtwo) and/or rich in carbon (i.e. high \metal and \logvmf). Because the condensation of C(cr) is arrested by the formation of methane (Eq. \ref{eq:methane_prod} shifts to the right, meaning Eq. \ref{eq:graphite_prod} shifts to the right), C precipitation does not occur in low \metal atmospheres (Fig. \ref{fig:outgassing}, also Fig. \ref{fig:vapser_batch_T3000}). 

In highly reducing (\dIW{}$\lesssim -3$) and heavy atmospheres (\logvmf$\gtrsim 1$), methane might replace CO as the dominant C-bearing gas species. Owing to the entropy change of Eq. \ref{eq:methane_prod}, this exchanges is more pronounced in cooler temperatures ($\lesssim$ 2000 K) due to the enhanced stability of methane; cold and highly reducing HREs could therefore also harbour methane-rich atmospheres even for \logvmf < 1 \citep[cf.][]{liggins2022growth}. However, the production of CH$_4$ (and C-bearing gases more generally) is, to some degree, stifled in high H/C atmospheres (SOLAR) owing to a lack of carbon availability.

\subsubsection{Sulfur}
Aside from gases in the C-O-H(-He) system, which otherwise dominate the speciation of the atmospheres of HREs, S-bearing species may become abundant under certain, oxidising conditions \citep{gaillard2022redox, gillmann2024interior}.
This is also borne out in our study, in which SO$_2$ is the dominant species in the VIBSE case near \dIW{+3} (Fig. \ref{fig:outgassing}).
Briefly, this arises because the solubility of S$_2$ depends on \fOtwo \citep{o2002sulfide,BW22,BW23} which manifests as two competing limits: under reducing conditions (\dIW{}$\lesssim 0$), nearly all sulfur is dissolved as S$^{2-}$ in the silicate melt, substituting for oxygen \citep{BW23}, whereas under highly oxidising conditions, it dissolves predominantly as S$^{6+}$ in the SO$_4^{2-}$ moiety \citep{BW22}.
Compared to water, its dissolved concentration remains low: $\lesssim 0.025$ wt\% of S$_2$-equivalent resides in the melt for \logvmf $\lesssim 1$, while under \logvmf=1, the maximum concentration reached is $\sim 0.2$ wt\%.
Roughly around \dIW{+3}, there exists a maximum where sulfur accumulates in the atmosphere ($\sim$ 60-100\% of all S in the gas phase, cf. Fig. \ref{fig:ingassing}) and S-rich gases, particularly SO$_2$, become the dominant species \citep[see][for a detailed thermodynamic explanation]{hughes2023sulfur}.
Other sulfur-bearing gases are not as important, but SO is found to accompany SO$_2$ when the latter is abundant, and H$_2$S is a minor species under very high \logvmf ($\geq$1), high \metal and reducing conditions, reaching mixing ratios of up to 5~\% (with the caveat again that these atmospheres are likely to be non-ideal).
All else being equal, high-\metal (VIBSE) cases host greater concentrations of S in their mantles, typically $\gtrsim 2$ times higher than in low-\metal (SOLAR), and this maximum increases with \logvmf.

\subsubsection{Nitrogen}
The partial pressure of N$_2$ is sensitive to \fOtwo because it dissolves preferentially in the silicate melt under highly reducing conditions (\dIW{}$\lesssim$ -3), at which N$^{3-}$ is stable \citep{libourel2003nitrogen,bernadou2021nitrogen}.
Consequently, highly reduced atmospheres are likely to be nitrogen-depleted relative to oxidised atmospheres \citep{shorttle2024distinguishing} (Fig. \ref{fig:ingassing}).
Therefore, even at high H/C ratios, the formation of gaseous ammonia, a reducing species, is inhibited, a result compounded by its thermal instability at the temperatures of HRE atmospheres.
Gaseous nitrogen therefore mostly occurs as N$_2$(g), with oxidising atmospheres (\dIW{} $\gtrsim +3$) potentially forming nitrous oxide, NO(g), in similar or greater quantities than N$_2$(g).

\subsubsection{Mineral gases and hydrides - SiO, Mg, Fe, SiH$_4$}
\label{results:atmo_comp:Si}
The partial pressures of the mineral gases are set by the melt activities and \fOtwo (Eq. \ref{eq:seidler_2024}) and thus behave like their dry lava planet counterparts \citep[cf.][]{seidler2024impact}.
In atmospheres with bulk silicate Earth-like amounts of volatiles (\logvmf=0), their partial pressures remain low and only approach those of the major volatiles ($\sim$100 bar) under highly reducing conditions (\dIW{}$\lesssim -5$).
This remains true even when the planetary mass is scaled down, as the total pressure remains of the order of 100 bar for planets with $M_p \geq 1~M_\oplus$, to which the mineral gases contribute less than 1-10 bar \citep{seidler2024impact} for \dIW{}>-4.
For mineral gases to dominate, the planet must be depleted in volatiles (\logvmf $\ll$ 0), highly reducing, or extremely hot such that evaporation of mineral gases is promoted \citep{wolf2023vaporock, seidler2024impact}.
The transition from mineral-dominated to volatile-dominated atmospheres is therefore highly case dependent.

Metal hydrides such as SiH$_4$ (silane), MgH, FeH or hydroxides (MgOH) are only minor constituents of most atmospheres; only under very reducing (\dIW{}$\lesssim -4$ ) and massive (\logvmf$\gtrsim$ 1) atmospheres does silane emerge (Fig. \ref{fig:vapser_batch_T3000}. Its stability is, however, only weakly dependent on \metal). It overtakes H$_2$ as the dominant species below $\Delta$IW = -5 for \logvmf$\sim$ 2. However, at these conditions, the total atmospheric pressure reaches 10s of GPa, well beyond the applicability of the ideal gas law and is more typical of the Sub-Neptune regime \citep{charnoz2023effect, misener2023atmospheres, seo2024role, hakim2026silane}.

\subsubsection{Oxygen}
\label{results:atmo_comp:O}
Highly oxidising atmospheres (\dIW{}$\gtrsim +5$) are dominated by free O$_2$. The \fOtwo (relative to IW) at which O$_2$ becomes the most abundant gas species increases with decreasing temperature since the IW buffer (Fe + 1/2O$_2$ = FeO) fixes higher $p$O$_2$ (or $f$O$_2$) at higher temperature.
While we do not explicitly track the total amount of oxygen as it is instead set by the \fOtwo, we can reconstruct its share between the atmosphere and mantle by assuming the mantle abundance of oxygen is given by the amount of dissolution-sequestered O by oxygen-bearing species (H$_2$O, CO$_2$ and CO) together with the 44.33 wt\% oxygen residing in mantle silicates \citep{palme2013cosmochemical}.
We find that the atmospheres studied here have $\ll$ 1 wt\% of their entire oxygen budget in the gas (Fig. \ref{fig:ingassing}), a limit that is only reached under highly oxidising conditions; planets with \logvmf$\leq 0$ generally host $\ll$ 1 \permil~ of their oxygen budget in their atmosphere, validating the notion that magma oceans buffer the atmospheric $f$O$_2$.

\begin{figure*}[p]
    \centering
    \includegraphics[width=0.98\textwidth]{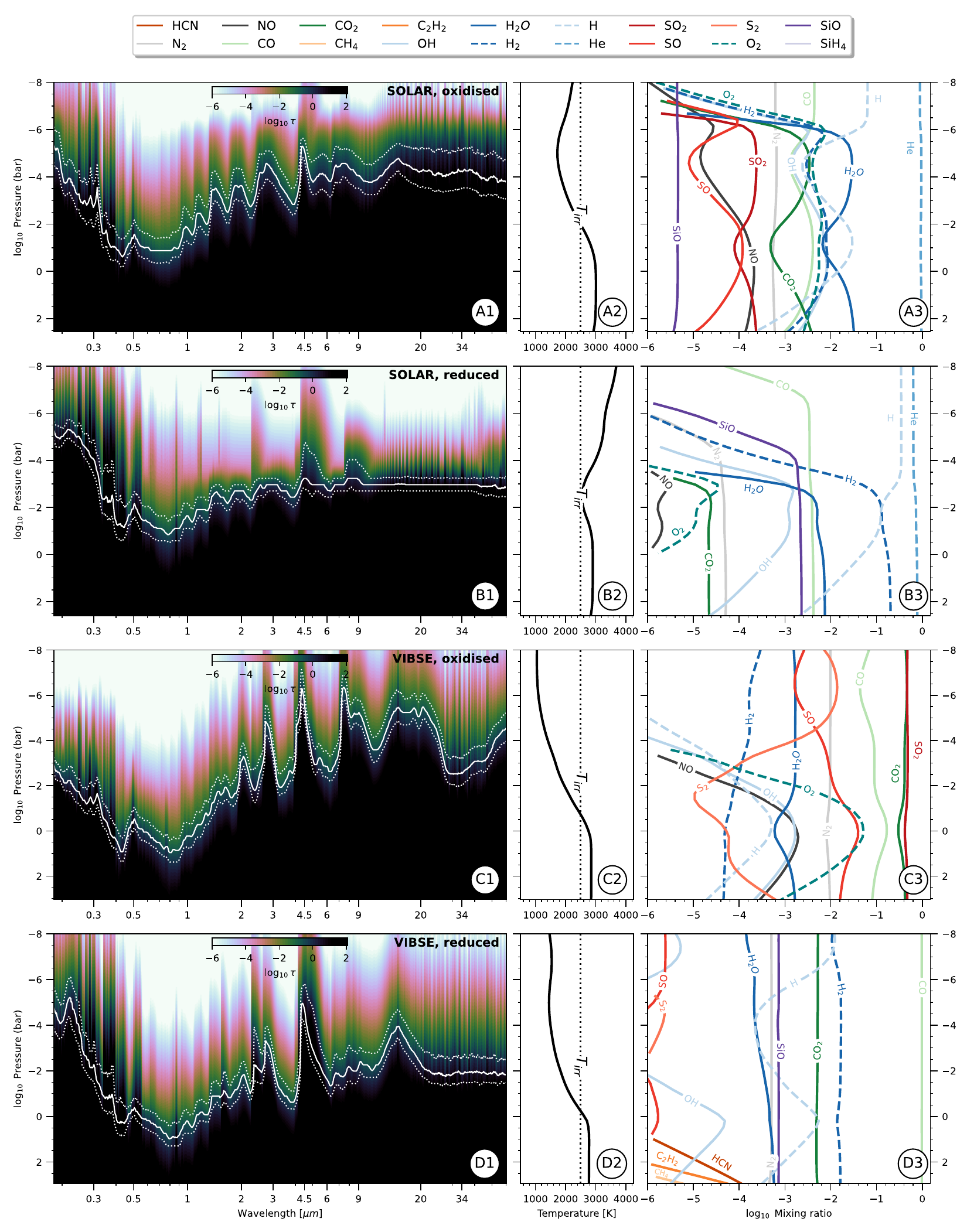}
    \caption{Detailed atmospheric structures for a selected set of synthetic hot rocky exoplanets (\logvmf=0, $M_p=8 M_\oplus$, \tirr=2500 K). \textbf{Column 1:} Optical depth $\tau$ as function of wavelength and atmospheric pressure altitude. The photosphere, defined where the atmosphere has absorbed 50\% of all incoming light, is indicated as solid white line; the 25\% and 75\% levels are shown in dotted white lines. \textbf{Column 2:} Pressure-temperature structure. \textbf{Column 3:} Speciation as function of pressure altitude.}
    \label{fig:atmospheric_structure}
\end{figure*}

\begin{figure}[!t]
    \centering
    \includegraphics[width=\linewidth]{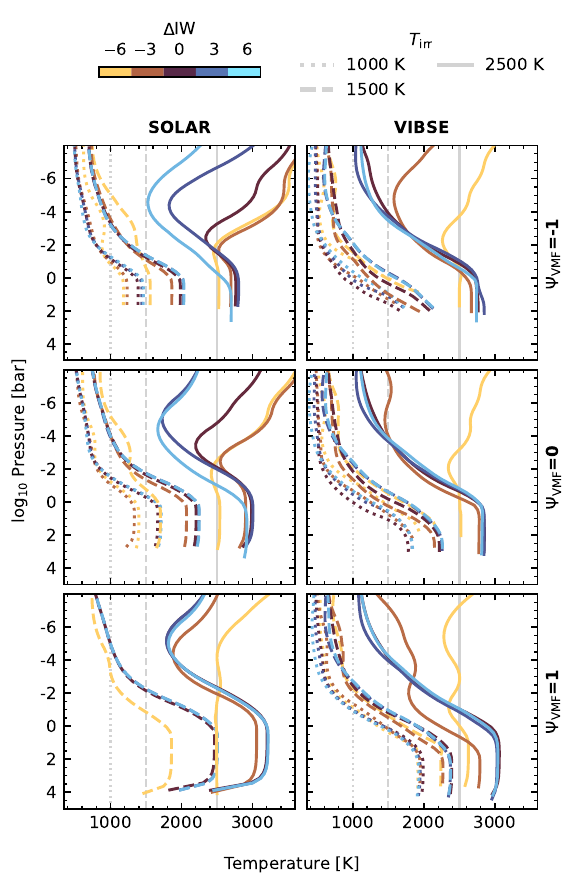}
    \caption{Atmospheric thermal structure for representative scenarios. The pressure-temperature profiles for each simulated model is shown in colour, according to its oxygen fugacity, and the line style shows the \tirr of the model. Vertical grey lines indicate the irradiation temperature of the respective simulation.
    }
    \label{fig:pt_structure}
\end{figure}

\subsection{Thermal structure and opacity}
\label{results:atmo_struc:opac_and_thermal}

\subsubsection{Phenomenology}

In Fig. \ref{fig:atmospheric_structure}, we show the detailed atmospheric wavelength-dependent optical-depth (subpanel 1), temperature profile (subpanel 2), and mixing ratios of abundant gas species as a function of pressure (altitude, subpanel 3), organised in four archetypal planetary states; A - SOLAR oxidised (\dIW{+3}), B - SOLAR reduced (\dIW{-3}), C - VIBSE oxidised (\dIW{+3}), D - VIBSE reduced (\dIW{-3}). Each was calculated assuming a synthetic, fully molten archetype similar to 55 Cancri e, as outlined in Sec. \ref{methods:pipeline:outgassing} ($8 M_\oplus$, average dayside temperature $T_{irr}= $2500 K). Here, only cases for an Earth-like amount of volatiles (\logvmf= 0) are shown.

Many of the dominant gas species, namely CO$_2$ and H$_2$O, are classical greenhouse gases that cause heating in the lower atmosphere due to entrapment of outgoing infrared emission \citep[e.g.][]{ledley1999climate}, but cooling in the upper layers due to efficient emission in the same wavelength bands \citep[e.g.][]{bougher1999comparative}.
Thus, most of the simulated atmospheres show a notable greenhouse effect up to 500 K in conjunction with an upper atmosphere ($P \lesssim 0.1$ bar) that can be 500-1000 K cooler than the irradiation temperature, particularly for the VIBSE cases (Fig. \ref{fig:atmospheric_structure}, C2, D2).
The major gas species maintain a roughly constant mixing ratio throughout the atmospheric column, in spite of changing temperatures ($\Delta T \sim 1000$ K) as discussed above.
Only when the atmosphere shows a thermal inversion (i.e. when the upper layers are hotter than lower layers), is the speciation of the major gases tangibly impacted; this occurs mainly by dissociation into atoms or ions, which predominantly affects H$_2$O (Fig. \ref{fig:atmospheric_structure}, A3, B3). 
Therefore, an inversion can weaken the radiative activity of the atmosphere, Fig. \ref{fig:atmospheric_structure}, B1; consequently, a flat photosphere develops beyond $\sim$1.5~$\mu$m at 10$^{-3}$ bar, and only the thermally more resilient gases SiO(g) and CO(g) survive as molecules above this layer. 

Mainly responsible for the formation of an inversion is SiO(g), a strong ultra-violet (UV) absorber (see Fig. \ref{fig:opacities}); other (atomic) metal species that have abundant lines in the UV and the visible (UVIS), namely Mg, Fe and Si, contribute as well \citep[e.g.][also see Fig. \ref{fig:opacities}]{zilinskas2022observability, zilinskas2023observability, piette2023rocky, seidler2024impact}.
These gases are prevalent in atmospheres of low \fOtwo and high temperatures due to enhanced silicate vaporisation \citep{seidler2024impact}, particularly when \logvmf is (comparatively) low to prevent the major volatile species (CO, H$_2$, etc.) from overwhelming them.
Indeed, the UV range ($\lesssim$ 0.3 \micrometer) is more opaque in reduced atmospheres (Fig. \ref{fig:atmospheric_structure} B1, D1) than in their oxidised counterparts (Fig. \ref{fig:atmospheric_structure} A1, C1), a consequence of stronger SiO(g) and Fe(g) absorption.

The VIBSE cases exhibit clear peaks in optical depth, owing to the presence of CO(g) in the reducing case (4.5~$\mu$m and 13~$\mu$m) and both CO$_2$ (2~$\mu$m, 3~$\mu$m and 4.5~$\mu$m) and SO$_2$ ($\sim$9~$\mu$m and $\sim$20~$\mu$m) in the oxidised case. Their high IR opacities relative to the UV permit substantial cooling in the upper atmospheres of these planets, particularly in the BSE oxidised case, where the photospheric pressure can be as low as $\sim$10$^{-6}$ bar in the IR compared to $\sim$10$^{-2}$ in the UV.
The cooler upper atmospheres of the VIBSE cases leads to a positive feedback loop, wherein the mineral gases condense (see Sec. \ref{results:atmo_struc:condensation}), thereby lowering their mixing ratios even further, and thus also the UVIS opacity.

Below the photosphere (indicated by the white bands in Fig. \ref{fig:atmospheric_structure}, A1-D1), radiation can only be transferred diffusively, and the pressure-temperature profile will therefore become isothermal in the absence of internal heating \citep[cf.][]{zilinskas2023observability}.
The atmosphere remains stable against convection in all cases, as the intense irradiation sets a gradient less steep than the adiabat \citep{nicholls2025convective}.

\subsubsection{Variation with oxygen fugacity, volatile mass fraction, metallicity, and temperature}
\label{results:atmo_struc:varaition_with_params}

To further test the effect of the compositional parameters, we expanded our analysis to include \dIW{}$\in \{-6, -3, 0, +3, +6\}$, \logvmf$\in \{-1, 0, 1\}$, and varying \tirr $\in \{1000, 1500, 2500\}$ K as shown in Fig. \ref{fig:pt_structure}. We restrict the discussion to \metal $\in \{0, 1\}$ (SOLAR vs. VIBSE), but we find that atmospheres with \metal $\ge 0.1$ exhibit similar characteristics to the \metal= 1 (VIBSE) cases.

In general, we find the volume-mixing ratios (VMRs) of major volatile species (e.g. CO, CO$_2$, H$_2$O, H$_2$, He) to behave broadly similarly as in Fig. \ref{fig:atmospheric_structure} for all cases investigated, meaning stable VMRs with altitude unless a thermal inversion induces dissociation of H$_2$ and H$_2$O.
The tendency for atmospheres to undergo thermal inversions increases as \tirr increases, a result of the enhanced evaporation (and thus higher mixing ratios) of mineral species \citep[cf.][]{zilinskas2023observability, piette2023rocky}. Consequently, atmospheres formed around planets with \tirr $< 2000$~K show uniformly cooler upper atmospheres compared to their lower atmospheres that are subject to the greenhouse effect. As above, the proclivity for thermal inversion is greater at lower \metal (i.e. in SOLAR atmospheres) and lower \logvmf, all else being equal, again owing to the higher mixing ratio of mineral gases that arises.

The pronounced greenhouse effect ensures that the vast majority of planets modelled show surface temperatures that exceed \tirr and in many cases cross the solidus of dry peridotite, $\sim 1600$ K \citep{katz2003new}. The extent of the greenhouse effect is relatively insensitive to $\Delta$IW for VIBSE-like atmospheres, and translates into $\Delta T := T_\text{MAI} - T_\text{irr}$ of $\sim$ 800~K for \tirr = 1000~K,  $\sim$600~K for \tirr = 1500~K and roughly 300~K for \tirr = 2500~K.
By contrast, the magnitude of $\Delta T $ for SOLAR cases is \fOtwo dependent and appears similar to VIBSE for oxidised cases, but decreases markedly for reduced cases, which can be attributed to the lack of effective greenhouse gases like CO$_2$ (due to the high H/C ratio) or H$_2$O (due to the low \fOtwo) and the cooling effect of the inversion driven by the introduction of the anti-greenhouse gas SiO (see Sec. \ref{results:atmo_struc:opac_and_thermal}); similar results were found by \citet{falco2024hydrogenated} for mixed H$_2$-silicate atmospheres once the implicit shift in \fOtwo is accounted for. 
In VIBSE-like atmospheres (CO and CO$_2$ rich) even an irradiation temperature of 1000 K might be sufficient to reach the required surface temperature for melting, while a reduced SOLAR-like atmosphere with a similar energy budget might be too cold to host a molten mantle (Fig. \ref{fig:pt_structure}). 

In systems with \logvmf=1, the temperature at the bottom of the atmosphere suddenly declines below that of the isothermal radiative zone.
This effect is most pronounced in SOLAR-like cases (Fig. \ref{fig:pt_structure}) and results from a relative decrease in MIR opacity compared to the UVIS with increasing pressure.
We note that this occurs at the upper limit of the tabulated values for the opacity ($\sim$1000~bar) and requires further investigation.

\subsubsection{Condensation}
\label{results:atmo_struc:condensation}

\begin{figure*}[!t]
    \centering
    \includegraphics[width=\linewidth]{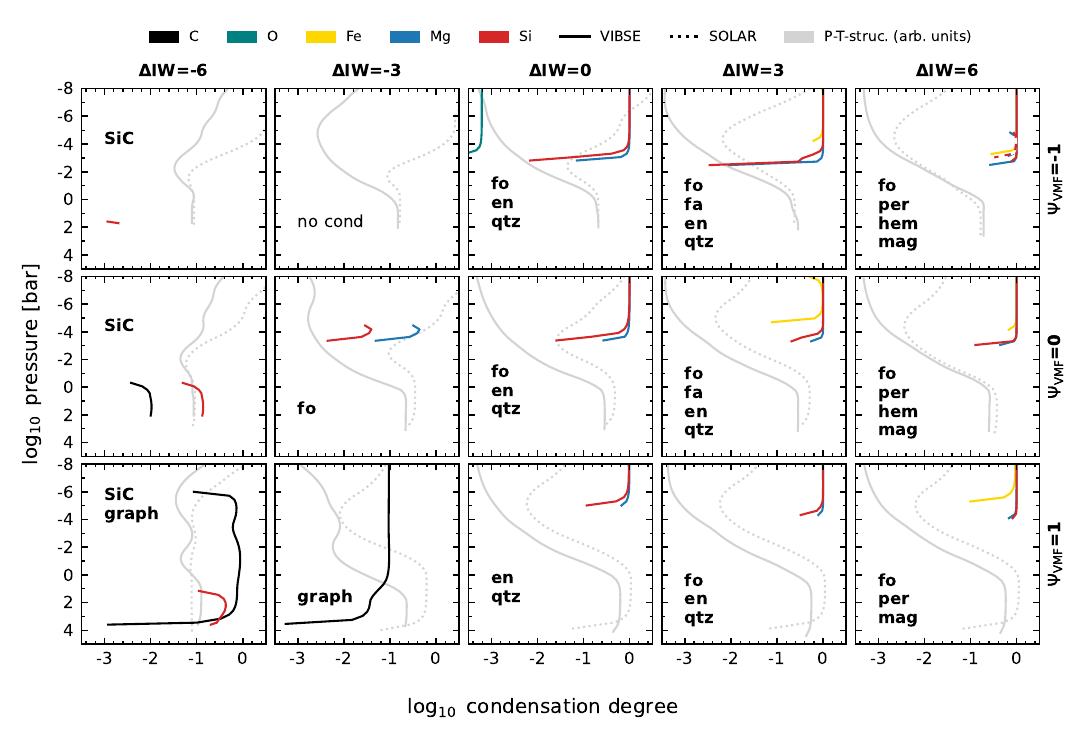}
    \caption{Condensation degrees of elements (i.e. number fraction of an element that has condensed relative to its total atmospheric abundance) as function of atmospheric altitude for the \tirr=2500 K cases in the grid in Fig. \ref{fig:pt_structure}. The colour denotes the element (red - Si, blue - Mg, teal - O, black - C, yellow - Fe), and the line style indicates the composition (solid - VIBSE; dashed - SOLAR). The pressure-temperature structure (at arbitrary units) is shown in light grey. The identity of condensates that occur in the respective atmospheres are SiC (silicon carbide), graph (graphite), fo (forsterite; Mg$_2$SiO$_4$), fa (fayalite; Fe$_2$SiO$_4$), qtz (quartz; SiO$_2$), en (enstatite; MgSiO$_3$), per (periclase; MgO), hem (hematite; Fe$_2$O$_3$), and mag (magnetite; Fe$_3$O$_4$).}
    \label{fig:cond_at_2500K}
\end{figure*}

The generally cooler upper atmospheric temperatures can, in many cases, induce partial condensation of the atmosphere. In Fig. \ref{fig:cond_at_2500K}, we show the condensation degrees of elements as function of altitude in the atmospheres from Fig. \ref{fig:pt_structure} for the \tirr=2500 K case. The first striking observation is that, at this temperature, the SOLAR cases are nearly devoid of condensates (only for \dIW{+6}, \logvmf=-1, does forsterite condense), a consequence of the thermal structures of these atmospheres, which are typically inverted in the uppermost atmosphere ($<$10$^{-4}$ bar) where condensation would otherwise occur.

Secondly, the nature of the condensing phases depend strongly on the \fOtwo.
Under highly reducing conditions (\dIW{-6}), silicon carbide precipitates, sometimes alongside graphite when \logvmf is high.
However, the fraction condensed remains modest; only $\sim$10\% of the total element budgets of Si and C are removed from the gas.
Further, the condensation of SiC occurs under high pressure (as high as several thousand bar), corresponding to the lower atmospheric layers; these regions generally lie beneath the photosphere (Fig. \ref{fig:atmospheric_structure}) and are thus hidden from observation. Graphite, on the other hand, may condense throughout the atmospheric column.

In neutral to oxidising atmospheres (\dIW{0} - \dIW{+3}), the minerals forsterite (Mg$_2$SiO$_4$), enstatite (MgSiO$_3$) and quartz (SiO$_2$) condense.
Iron condenses as fayalite (Fe$_2$SiO$_4$) in atmospheres initially equilibrated at \dIW{+3}, while in highly oxidising systems (\dIW{}$\geq +3$), it can condense to form more oxidised phases, namely, hematite (Fe$_2$O$_3$) or magnetite (Fe$_3$O$_4$). Excess Mg not incorporated into forsterite can also condense as periclase (MgO) in the most oxidising atmospheres, while quartz is typically absent.
In atmospheres with \dIW{}$\gtrsim 0$, the condensation degree is high; nearly 100\% of the Mg, Si and Fe is removed from the atmosphere.
Unlike in the high \logvmf, reduced atmospheres, condensation in oxidising atmospheres (\dIW{}$\gtrsim 0$) occurs at lower pressures, where $P\lesssim 10^{-4}$ bar. Thus, the minerals have the potential to form high-altitude hazes, provided they do not settle.

For \tirr < 2500 K, the condensation front is pushed to lower altitudes in the atmosphere, and SOLAR-like atmospheres start condensing silicates in a similar manner to the VIBSE cases. Further, the condensation of carbon is expressed for all \logvmf at \dIW{}<0, either as SiC or graphite. However, the reduced vapour pressures of silicates at lower temperatures also decrease the total mass of condensates in the atmosphere, such that cooler atmospheres, despite greater condensation fractions, contain a lower mass of condensed phases.

\subsection{Spectra}
\label{results:spectra}

\subsubsection{Emission}
\label{results:spectra:emission}

\begin{figure*}[!t]
    \centering
    \includegraphics[width=\linewidth]{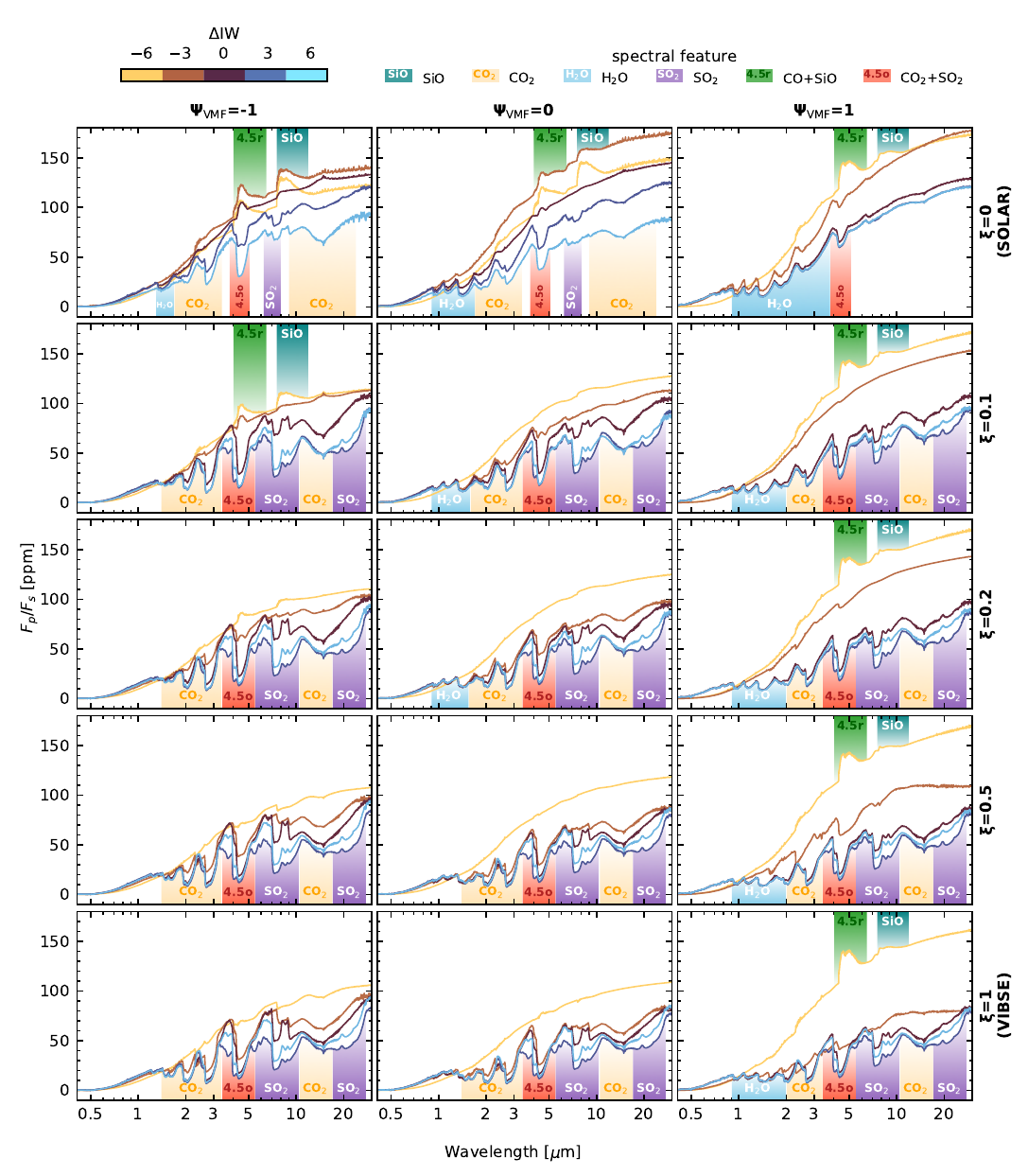}
    \caption{Synthetic emission spectra over a grid in \dIW{} $\in \{-6, -3, 0, +3, +6\}$ (colour), \metal $\in \{0, 0.1, 0.2, 0.5, 1\}$ (row) and \logvmf $\in \{-1, 0, 1\}$ (column). Other parameters were kept constant: $M_p=8M_\oplus$, $x_\text{melt}=1$ (fully molten mantle), \tirr=2500 K. Important broad-band spectral features are highlighted by coloured patches. The 4.5 \micron feature, where multiple bands coalesce, is labelled 4.5r when reducing species (CO + SiO) dominate and 4.5o when oxidising species (CO$_2$ + SO$_2$) dominate.}
    \label{fig:spectra_emission}
\end{figure*}

To investigate the effect of interior parameters on the emission spectrum, we follow the procedure outlined in Sec. \ref{results:atmo_struc:opac_and_thermal} over a wider range of \metal$\in\{0, 0.1, 0.2, 0.5, 1\}$.
The emission spectra are reported as the secondary eclipse depth (also called occultation depth or planet-to-star flux ratio):

\begin{equation}
    \frac{F_p}{F_s}(\lambda) = \frac{\varepsilon_p(\lambda)}{\varepsilon_s(\lambda)} \cdot \frac{R_p^2(\lambda)}{R_s^2},
\end{equation}
where $F$ is the totally emitted flux (erg s$^{-2}$ cm$^{-1}$), $\varepsilon$ denotes the spectral exitance (erg s$^{-2}$ cm$^{-3}$), and $R$ the radii of the respective bodies. 
The radius of the planet depends on the wavelength at which it is observed, as seen in the photospheres in Fig. \ref{fig:atmospheric_structure}.
Assuming a constant radius does not produce inconsistencies when the atmosphere is highly compressed compared to the total radius (i.e. the VIBSE case, or a dry mineral atmosphere), but if it is extended, then $R_p \neq R_{p,\text{MAI}}$, and wavelength dependence has to be incorporated; the method is highlighted in Appendix \ref{apx:radius_wavelength_dependent}.
The resulting emission spectra are shown in Fig. \ref{fig:spectra_emission} for \tirr= 2500 K, the average dayside temperature of 55 Cancri e that is consistent with phase curve observations \citep{demory2016map, angelo2017case}. Additional figures for lower \tirr are shown in Appendix \ref{apx:spectra_emission_cooler}. 

CO$_2$ features are common among the studied archetypes. Distinct bands can be found in the mid-infrared at 2-3, 4.5 and 15 \micron that similarly appear in almost all atmospheres where \dIW{} $\geq$ 0.
For \dIW{} $\le$ -3, these features either weaken or vanish, or, in the case of the 4.5 \micron feature, blend with a CO absorption band of similar shape.
Since the ratio $p$CO$_2$/$p$CO depends on $f$O$_2^{0.5}$, and both gases have a similar opacity in this region, the 4.5 \micron feature at a resolution accessible to the JWST Near Infrared Camera (NIRCam) has a similar depth for atmospheres of nearly all tested redox states and compositions except for the most reducing or metal-poor cases (\dIW{} $\le$ -3, \metal=0).
Further, the 4.5 \micron CO$_2$ band also shows contribution from SO$_2$ and CO, which deform the absorption trough at its blue and red ends, respectively ($\sim $ 4 \micron and $\sim$ 5 \micron).
In highly reduced, low-\metal or heavy \logvmf=1 atmospheres (\dIW{}=-6, \metal $\lesssim 0.1$, \logvmf=1), the 4.5 \micron CO feature is converted from absorption into emission as the thermal structure of the atmosphere undergoes inversion (see Fig. \ref{fig:pt_structure}).
However, in turn, CO(g) emission merges with a SiO(g) emission feature at the same wavelength.
The near-to-mid infrared CO$_2$ wavebands at 2-3 \micron also overlap with broad H$_2$O features at 1-3 \micron. Depending on the atmospheric composition, the dominant species will partially obscure the other; this effect is composition dependent.
Similar to CO$_2$, these water bands saturate at \dIW{} $\gtrsim$ -3, thus neither CO$_2$ nor H$_2$O features appear to be universal measurement devices for redox state. However, they can become indicative of highly reducing atmospheres (\dIW{} < -3) if absent. The features that are sensitive to higher \fOtwo are the SO$_2$ features at 8 and 21 \micron \citep{hu2024secondary, nicholls2025convective}. They confidently indicate \dIW{}$\gtrsim$0 due to the dissolution of sulfur under reducing conditions (cf. Fig. \ref{fig:outgassing}). Both features manifest as absorption troughs whose intensity mirrors $p$SO$_2$, and thus \fOtwo, which shows a pronounced maximum at \dIW{+3} whereas any \fOtwo higher or lower produces a weaker signal. 
Additionally, \fOtwo exerts a significant influence on the baseline secondary eclipse depth, $F_p/F_s$, with variation of up to $\sim$ 100 ppm for the planetary parameters over the 12 order-of-magnitude \fOtwo range explored here, caused by the conversion of (high-MMW) H$_2$O into low-MMW H$_2$, particularly affecting H-rich \metal=0 and \logvmf=1 atmospheres, which leads to extended atmospheres and thus higher baselines (cf. Appendix \ref{apx:radius_wavelength_dependent}).

The influence of \metal is milder than that of \fOtwo. Its principal effect is to lower the MMW as \metal decreases (cf. Fig. \ref{fig:mmw_2500}), and thus extending the atmosphere, leading to slightly higher baseline secondary eclipse depths, $F_p/F_s$. Spectroscopically however, only the features of planets with \metal=0 are visibly distinct from their higher-\metal peers, manifest as the weakening (at low \logvmf) or disappearance (at high \logvmf) of CO, CO$_2$ and SO$_2$ absorption in the 4.5, 8 and 15 \micron bands, and the emergence of emission features of CO(g) and SiO(g) for \dIW{-6}. In addition, the H$_2$O bands in the near-to-mid infrared ($\sim$ 1-3 µm) trace \metal by means of their width and depth, however, remain generally more sensitive to \logvmf (see below).

\logvmf technically implies an accumulation of atmospheric mass, and thus, in principle, radial expansion. However, only the emission spectra of reduced or low \metal atmospheres seem to bear out this expectation (i.e., \dIW{}$\le -3$ or \metal = 0), whereas the $F_p/F_s$ of oxidised or high-\metal atmospheres maintain a similar shape and intensity as the total mass of atmosphere increases (Fig. \ref{fig:spectra_emission}). This occurs because their atmospheres are dominated by greenhouse gases whose opacities are so high that they become saturated at identical pressures, even at low \logvmf. Specifically, even \logvmf=-1 atmospheres have their photospheres at altitudes significantly above the surface. Any additional mass from larger \logvmf accumulates below the photosphere, where they have effectively no spectroscopic expression. Due to the high compressibility of triatomic, high-MMW ideal gases (i.e. H$_2$O, CO$_2$ and SO$_2$), the contribution to planetary radii is also negligible, and thus $F_p/F_s$ remains near-constant despite increasing \logvmf.
The only exceptions are H$_2$-rich atmospheres due to their low-MMW (cf. Sec. \ref{results:spectra:transmission}), which arises when \fOtwo and \metal are low (\dIW{}$\le -3$ or \metal = 0); correspondingly, such atmospheres show a mild increase of the baseline $F_p/F_s$ with increasing \logvmf.
In terms of spectroscopic changes, only the near-to-mid infrared ($\sim$ 1-3 µm) water bands continuously strengthen in depth and width as \logvmf grows, from near absent at low \logvmf to pronounced troughs at high \logvmf, a result of greater H$_2$O-abundance in high-\logvmf atmospheres (Fig. \ref{fig:outgassing}).

Volatile rich atmospheres may still exhibit a SiO(g) feature in emission at 9 \micron, akin to dry mineral atmospheres \citep{zilinskas2022observability, piette2023rocky, falco2024hydrogenated, seidler2024impact}. The requirements for their emergence are low \logvmf(i.e. low volatile masses, cf. \citealt{piette2023rocky}), reducing conditions (\dIW{}$\lesssim -3$, \citealt{seidler2024impact}) and SOLAR-like gas mixtures with \metal$\sim 0$. Earlier results by \citet{zilinskas2023observability}, using an outgassing model in which mineral gas species and major volatile species are not in equilibrium \citep{falco2024hydrogenated}, suggested that SiO features should disappear in the presence of hydrogen. Our results do not support such a general statement; some low-\metal atmospheres still show distinct SiO(g) features in emission, particularly at low \logvmf (Fig. \ref{fig:spectra_emission}). Such emission features are only absent in carbon-rich atmospheres, as the opacities for CO(g) and CO$_2$(g) are sufficiently high in the infrared (see Fig. \ref{fig:opacities}) so as to lead to upper atmospheric cooling, leading to the removal of mineral gases via condensation (see Sec. \ref{results:atmo_struc:condensation}). 

The generalities described above  hold true for when \tirr < 2500 K, except that variation between redox regimes becomes increasingly non-linear with decreasing temperature (Fig. \ref{fig:spectra_emission_1500K}, \ref{fig:spectra_emission_2000K}), typified by sudden appearances of CO$_2$ features at \dIW{-3} and SO$_2$ at \dIW{+3}, and the overall decline in $F_p/F_s$ that makes detection more challenging (sensitivity of $\Delta F_p/F_s\ll25$ ppm required). Absorption features are also more readily observed, even in the most reducing cases, as thermal inversions, which would ordinarily be sustained by SiO(g), are not as marked, given the lower partial pressures of mineral gases and their tendency to condense; together, these processes lead first to a transition from SiO(g) emission to absorption features at \tirr=2000 K, and a complete disappearance of mineral gas features for \tirr=1500 K (Fig. \ref{fig:spectra_emission_2000K}).
Instead, methane absorption features unseen at higher temperatures appear in cold (\tirr $\leq 2000$ K) and highly reduced atmospheres (\dIW{}$\leq -3$) (cf. Appendix \ref{apx:acetylene_worlds}).

\subsubsection{Transmission}
\label{results:spectra:transmission}

\begin{figure*}[!t]
    \centering
    \includegraphics[width=\linewidth]{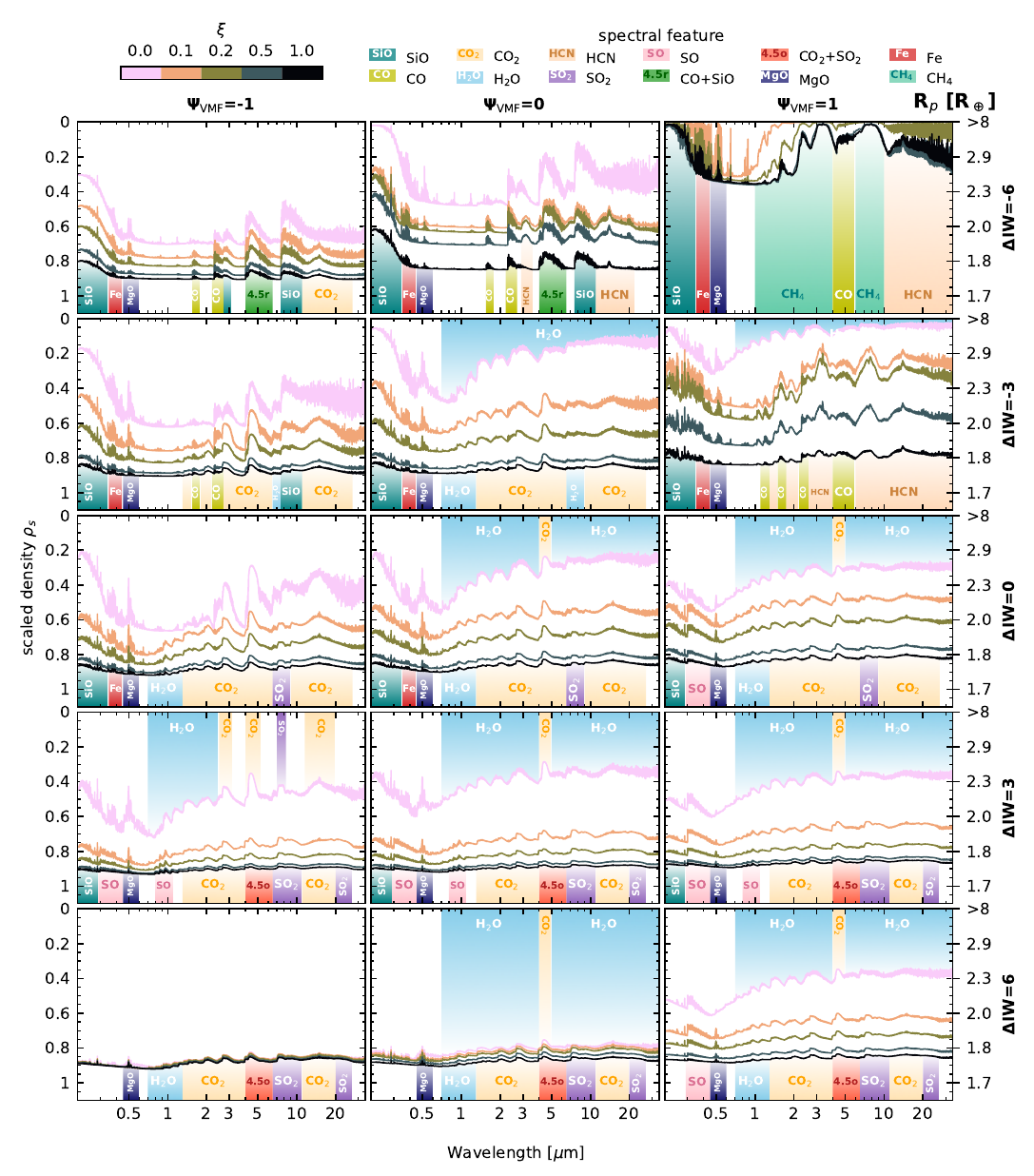}
    \caption{Transmission spectra of the synthetic atmospheres from Fig. \ref{fig:spectra_emission} rearranged to highlight the effect of \metal, which was (nearly) absent in emission.
    The spectra are reported in terms of the scaled density, Eq. \ref{eq:scaled_density}, which at constant mass (here, $M_p=8 M_\oplus$) can be directly related to the radius (right axis). Prominent spectral features are highlighted by coloured bands.
    }
    \label{fig:spectra_transmission}
\end{figure*}

In Fig. \ref{fig:spectra_transmission}, we show the transmission spectra associated with the models presented in Sec. \ref{results:spectra:emission}. 
Compared to the restricted range of secondary eclipse depths ($\sim$ 0-150 ppm), the transit depth varies by $\Delta R_p \gtrsim 1~ R_\oplus$ between the models (with the bare-rock baseline being $1.7 ~R_\oplus$), and thus we converted the usual transit depth 
\begin{equation}
    \label{eq:transit_depth}
    \delta = \left( \frac{R_p[\lambda]}{R_s} \right)^2
\end{equation}
into the (dimensionless) scaled density $\rho_s$ for convenience:
\begin{equation}
    \label{eq:scaled_density}
    \rho_s [\lambda]:= \frac{\rho_p[\lambda]}{\rho_\oplus(M_p)} = \frac{3 M_p}{4 \pi \delta^{3/2} [\lambda]\cdot R_s^3 \cdot \rho_\oplus(M_p)},
\end{equation}
where $\rho_p$ is the planets (mean) density, $\rho_\oplus(M_p)$ is the density of an upscaled Earth-like planet of mass $M_p$, and $R_s$ is the radius of the star.
Planets denser than Earth have $\rho_s > 1$ (likely corresponding to interiors dominated by a large iron-rich core), less dense objects have $0 < \rho_s < 1$ (e.g. when the planet has an atmosphere).
Since we assume here a constant $M_p=8~M_\oplus$, $\rho_s$ can be easily converted back to $R_p$.

The transit depth and thus $\rho_s$ is coupled to the physical extent of the atmosphere, which can be approximated to the first order by its scale height, $H$:
\begin{equation}
    \label{eq:scale_height}
    H = \frac{RT}{\bar{m} \cdot g},
\end{equation}
where $R$ is the gas constant, $T$ is the temperature, $\bar{m}$ is the MMW of the gas, and $g$ is the local gravitational acceleration (here, $g=g_\text{MAI}$; note that this implicitly assumes the thin atmosphere approximation in which $g$ is independent of altitude).
The temperature profiles up to the photosphere are similar among models of equal \tirr, \logvmf, and \metal (cf. Sec. \ref{results:atmo_struc:opac_and_thermal}), and therefore, the variation in $\rho_s$ in Fig. \ref{fig:spectra_transmission} is mostly explained by variation in the MMW which ranges from 2 to 52 g/mol. This variation is driven primarily by \metal, with secondary sensitivity to \fOtwo, and with a weaker dependence on \logvmf (Fig. \ref{fig:mmw_2500}).
At constant \logvmf{} and \fOtwo, increasing \metal leads to high MMWs ($\bar{m}\sim 28-44$ g/mol) and thus more compact atmospheres as composition switches from H-dominated to C($\pm$S) dominated. Low-\metal atmospheres are H$_2$-He dominated and thus lighter ($\bar{m} \sim 2$–5 g/mol), and thus significantly more extended. Consequently, they show lower scaled densities $\rho_s$, corresponding to greater transit radii.
This effect is enhanced under reducing conditions and \logvmf{}=1, both of which promote high $p$H$_2$. 
However, despite the changes in bulk elemental chemistry with varying \metal at constant \logvmf and \fOtwo (Fig. \ref{fig:volatile_mass_ratios_z}), we do not find major changes in the identity of C- and S-bearing absorbing species, nor in the relative strength of their features, a consequence of the relatively stable C/S ratio with \metal $> 0.1$. Only H$_2$O spectral bands are sensitive to \metal, as they increase in width as \metal becomes 0. Otherwise, \metal has only a minor influence on the shape of the transmission spectrum and instead chiefly governs the atmospheric extent. This implies a degeneracy with the radius of the condensed interior ($R_{p,\text{MAI}}$).

As in emission spectra, H$_2$O and CO$_2$ features are weak stand-alone indicators of \fOtwo due to their widespread appearance and similar strength across models with \dIW{} $\gtrsim$ -3, including the 1-3 \micron H$_2$O bands and the 2-3, 4.5 and 15 µm CO$_2$ bands.
However, again similar to emission, the relative strength of the SO$_2$ features becomes diagnostic of \fOtwo.
Alongside SO$_2$, SO is also produced (see Sec. \ref{results:atmo_comp}), exhibiting visible features between 0.3–0.4 µm and 0.8–1 µm in sulfur-bearing atmospheres. 
They also overlap with narrow but sharp, \fOtwo-sensitive OH lines located between 300-400 nm (not resolved in Fig. \ref{fig:spectra_transmission}). While $p$SO peaks near \dIW{+3} (similar to SO$_2$), $p$OH continues to increase with \fOtwo; distinguishing these features requires high resolution but can provide more stringent constraints on \fOtwo. However, features of sulfur-bearing gases are preferentially associated with atmospheres of low transit depth and thus may be challenging to observe. High-resolution ground-based spectroscopy could access both SO and OH features in the UVIS in addition to the nearby water bands at $\sim$ 1 \micron, but feasibility remains to be tested.

Mineral gases also appear in transmission spectra, and most prominently in the UV ($\lambda \le $300 nm) in the form of an intense SiO feature. Lines of Mg, MgO and Fe also occur in the UVIS, respectively at $\sim$285 nm, 500 nm and 300-600 nm. These spectral lines overlap with those of the aforementioned SO, OH and H$_2$O-bearing waveband, providing complementary \fOtwo indicators that are sensitive to changes below \dIW{}$<0$ (unlike the aforementioned SO$_2$, H$_2$O and CO$_2$).  
At longer wavelengths (9 \micron), SiO has additional features that are clearly visible in very reduced (\dIW{}$\leq$-3) atmospheres of any \metal and \logvmf $\leq 0$, and define viable targets for space-based transmission spectroscopy, for example with JWSTs Mid Infrared Instrument (MIRI), or ground based high-resolution spectroscopy \citep{dash2025detectability}.
If the partial pressures of mineral gases are high (i.e. under highly reducing conditions and high temperatures), they increase $\bar{m}$ of the H-He-dominated background gas, leading to lower transit depths. This effect is evident in low-\metal, low-\logvmf atmospheres, where the most reduced models are not the most extended ones. 

As in emission spectroscopy, \logvmf exerts a limited effect on the extent of the atmosphere for the same reasons (cf. \ref{results:spectra:emission}). Thus, atmospheres with varying \logvmf$\in \{-1,0\}$ appear similar in transmission spectra when \metal and \fOtwo are held constant, especially in oxidised, high-\metal cases, and only scale with \logvmf if the atmosphere is highly enriched in H$_2$ and He (i.e. \metal=0 or \logvmf=1 at \dIW{}$\leq-3$).
High-\logvmf atmospheres have more pronounced water bands that scale with \logvmf (as per emission spectra) and could, especially in comparison with the CO$_2$ band at 4.5 \micron, place constraints on the total atmospheric pressure.
Additionally, enhancement of the aforementioned SO(g)+OH(g) absorption bands between 300–400 nm with increasing \logvmf is observed.
Combined observation of these features might reveal the \logvmf. However, similar to \metal, the change in base-line induced by this parameter is expected to be degenerate with $R_{p, \text{MAI}}$.

Finally, the prominent CH$_4$ and HCN features that appear in the NIR and MIR for reduced, high-\logvmf atmospheres may not be realistic, as the abundance of acetylene is so high that it may start dominating the opacity. However, its spectral line list remains incomplete, and the models presented herein have to be interpreted with caution (see Sec. \ref{disc:limitations} and Appendix \ref{apx:acetylene_worlds}).

\section{Discussion}
\label{discussion}

\subsection{Comparison to mass-radius measurements}
\label{disc:mass_radius_diagrams}

\begin{figure}
    \centering
    \includegraphics[width=1\linewidth]{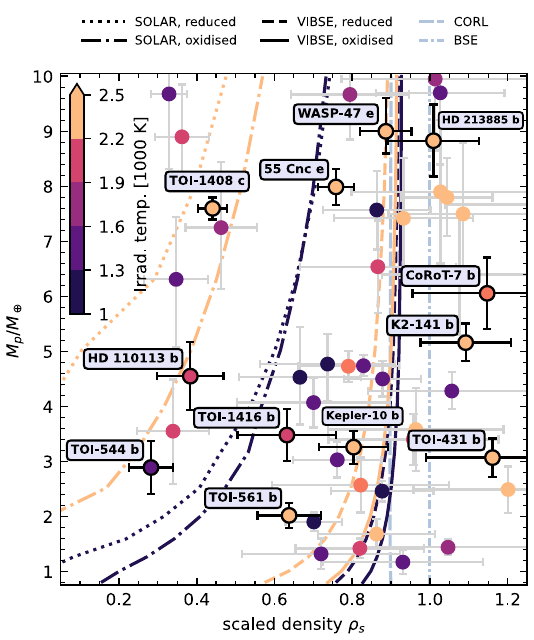}
    \caption{Mass versus the scaled density for observed (dots) and modelled planets (lines). Scaled density is defined as $\rho / \rho_\oplus$, where $\rho_\oplus (M)$ is the density of a planet with the same interior structure and composition as Earth at a given mass ($M$). The radii of the modelled planets are based on the radius of the photosphere, which was averaged over the Kepler bandpass. The curves for atmosphere-bearing planets are computed for endmember cases (\metal = VIBSE, SOLAR; \dIW{} = -3, +3; \tirr = 1250~K, 2500~K, at constant \logvmf = 0). Curves for two atmosphere-free cases (a coreless planet; CORL and a BSE-like planet, BSE) are also shown. The points denote a collection of measured masses and radii for planets $R \leq 4 R_\oplus$, $M \leq 10 M_\oplus$ and $T_{irr} \geq 1000$ K, obtained from the NASA Exoplanet Archive; we omit objects with an uncertainty $\sigma_{\rho_\text{scaled}}>0.25$ from the plot.}
    \label{fig:mr_diagram}
\end{figure}

\begin{figure}
    \centering
    \includegraphics[width=\linewidth]{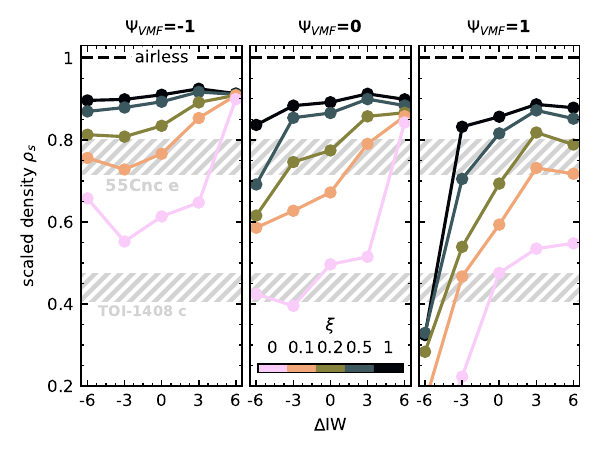}
    \caption{Average scaled density of the synthetic models from Fig. \ref{fig:spectra_transmission} in the Kepler/TESS bandpass (400-900 nm), shown as a function of \dIW{} and \logvmf (coloured lines and dots). The colour corresponds to \metal. The simulations are compared to the inferred scaled densities for 55 Cancri e and TOI-1408 c, which are shown as grey bands; their width denotes the $1\sigma$ limit. 
    %Hollow markers indicate solutions outside of the $1\sigma$ interval constrained by the mass-radius-observation. 
    The dashed black line indicates the scaled density of an atmosphere-free (airless) planet, i.e. $R_\text{p,MAI}$ (Sec. \ref{methods:pipeline:outgassing}).}
    \label{fig:radius_constraints}
\end{figure}

Since our model predicts the radius of a planet, it can be used to provide first-order estimates on the required atmospheric properties of observed HREs given their measured mass and radius.
We therefore selected planets with periods $<$4 days from the NASA Exoplanet Archive\footnote{\url{https://exoplanetarchive.ipac.caltech.edu/}} and show them in a mass-density diagram (Fig. \ref{fig:mr_diagram}), in which the scaled density $\rho_s$ (Eq. \ref{eq:scaled_density}) is plotted.
The radius, needed to compute $\rho_s$, was averaged over the Kepler bandpass (400-900 nm).
Overlain on the observational data are synthetic mass-density curves for a SOLAR and VIBSE composition, evaluated at reducing (\dIW{-3}) and oxidising (\dIW{+3}) conditions and for two temperatures, \tirr=1250 K and \tirr=2500 K to mirror the range of HREs.
Note that in this context, \tirr is equal to the average dayside temperature, corresponding to a dilution factor of $\mathfrak{f}=2/3$ \citep{hansen2008absorption}, motivated by the findings of \citet{hammond2017linking} and \citet{zhang2017effects} that the hot-spots of high-MMW atmospheres remain confined to the dayside.
\logvmf= 0 was held constant, but its effect on $\rho_s$ can be inferred from Fig. \ref{fig:spectra_transmission}.

We identify a population of under-dense planets with $\rho_s < 0.5$, all of which lie to the left of the $M-\rho_s$ curves defined by the SOLAR-composition lines (dotted and dash dotted lines).
Prominent examples are TOI-1408 c, TOI-544 b and HD 110131 b.
This marks these objects as likely metal-poor, as greater atmospheric extent is found in low-\metal atmospheres (Sec. \ref{results:spectra:transmission}); high-\metal atmospheres reach such low $\rho_s$ only when they are massive (\logvmf= 1) and highly reducing (\dIW{}$\sim -6$; see Fig. \ref{fig:spectra_transmission}), such that most sulfur and nitrogen has dissolved and carbon has been sequestered from the gas phase via graphite formation (cf. Fig. \ref{fig:outgassing}).
These planets might be likened to gas dwarfs if reducing \citep[e.g.][]{lous2024accretion} or waterworlds if oxidising \citep[e.g.][]{dorn2021hidden}.
TOI-1408 c stands out due to its \tirr of $\sim$2500~K, $M_p\sim 7.6 M_\oplus$ and Sun-like host star, which makes the planet similar to 55 Cnc e and thus comparable to the simulations from Sec. \ref{results:spectra}.
In Fig. \ref{fig:radius_constraints}, we compare the modelled to the observed $\rho_s$ for TOI-1408 c and 55 Cnc e.
For TOI-1408 c, only solutions with \logvmf $\geq$ 0 fit the observed $\rho_s$, with the degree of \metal enrichment being tied to \fOtwo and favouring C-N-S-poor, reducing solutions. Intermediate to oxidised scenarios (\dIW{}$\geq 0$) akin to a waterworld scenario remain possible within 2 $\sigma$.
The higher \tirr and the accompanying atmospheric size increase implies that TOI-1408 c requires less volatile material compared to typical sub-Neptunes to explain its low density. Thus, it may bear strong signatures of mineral gases, particularly SiO at 9 \micron, if it sustains a magma ocean below its atmosphere (Sec. \ref{results:spectra:transmission}).
However, our assessment hinges on the assumption of $\mathfrak{f}= 2/3$. Heat redistribution in low-MMW atmospheres may cause the dayside temperature to approach the equilibrium temperature instead \citep{hammond2017linking}, which would imply we may overestimate atmospheric extent (Eq. \ref{eq:scale_height}) and thus underestimate the volatile content.
It also stands to mention that planets in this category are highly inflated and H-He rich, which could leave them vulnerable to (rapid) atmospheric loss \citep[e.g.][]{salz2016energy, bourrier2018_55cnce}. If primarily H-He are lost, \metal and \fOtwo may increase \citep[cf.][]{cherubim2025oxidation}.

A population of intermediate density HREs plots in between the M-$\rho_s$ trends set by the SOLAR and VIBSE endmembers (0.6 $<$ $\rho / \rho_\oplus$ $<$ 0.8). 
This indicates a substantial atmosphere, whose nature is not defined from mass-density considerations alone.
Examples of HREs that might adhere to this category are 55 Cnc e\footnote{For 55 Cancri e, we use the mass-radius data from \citep{bourrier2018_55cnce}, which utilises data from HST/STIS. The wavelength coverage of this instrument is 115-1000 nm, larger than the assumed Kepler bandpass of 400-900 nm over which we average the radius. However, an observation with TESS (600-1000 nm) gave similar radii \citep{zhao2023measured}, implying the data remain compatible.}, TOI-561 b and TOI-1416 b.
To first order, the reduced, SOLAR-like case (dotted lines) and the oxidised VIBSE-like case (solid lines) are poor fits for planets in this category, as they would be too extended or compressed, respectively, even under varying \logvmf (see Fig. \ref{fig:spectra_transmission}).
Their compositions are thus more likely to lie intermediate to SOLAR and VIBSE (\metal $\sim 0.2-0.5$), and to be degenerate with \fOtwo such that higher-\metal solutions only fit when they are reduced (H$_2$-atmospheres enriched in either CO+SiO when \logvmf=-1 or CO+hydrocarbons when \logvmf$\geq0$; see Appendix \ref{apx:acetylene_worlds}), while oxidised cases are only consistent with their radii when \metal is low (H$_2$O-rich).
This is demonstrated in detail for 55 Cnc e in Fig. \ref{fig:radius_constraints}. As $\rho_s$ scales predictably with mass when $M_p > 2 M_\oplus$ (Fig. \ref{fig:mr_diagram}), similar conclusions hold for all planets with $\rho_s \sim 0.6-0.8$.

Planets of even higher scaled density (0.8 $<$ $\rho / \rho_\oplus$ $<$ 0.9) are best approximated by the VIBSE cases, either reduced (CO-dominated) or oxidised (CO$_2$-SO$_2$ dominated). 
Technically, their densities are sufficiently high that an atmosphere is not strictly required (within uncertainty), but the interior would be required to be coreless (CORL; dashed light-blue line in Fig. \ref{fig:mr_diagram}) \citep{elkins2008coreless}.
Falling into this category are WASP-47 e (according to updates in its mass-radius measurements from transit-timing variations, \citealt{nascimbeni2023new}) and Kepler-10 b.

Planets that fall closer to the $\rho_s=1$ line are consistent with a rocky interior similar to Earth's. However, the average uncertainties are typically too large to firmly distinguish such a planet from atmosphere-bearing cases, as exemplified by HD 213885 b.
Some planets, notably K2-141 b, TOI-431 b and CoRoT-7 b, plot to even higher scaled densities ($\rho / \rho_\oplus$ $>$ 1.0), which implies an interior structure with a higher core-mantle ratio than Earth \citep[e.g.][]{adibekyan2021compositional}.
However, this does not preclude the existence of an atmosphere, as the density decrease induced by an atmosphere could be offset by an even larger core.
Even if no volatile-rich atmosphere is present, the planets might still possess a pure mineral gas atmosphere that is highly compressed and confined to the dayside, therefore having no resolvable impact on the radius \citep{seidler2024impact}.

\subsection{Spectral constraints on the atmosphere of 55 Cancri e}
\label{disc:spectra}

\begin{figure}[!t]
    \centering
    \includegraphics[width=\linewidth]{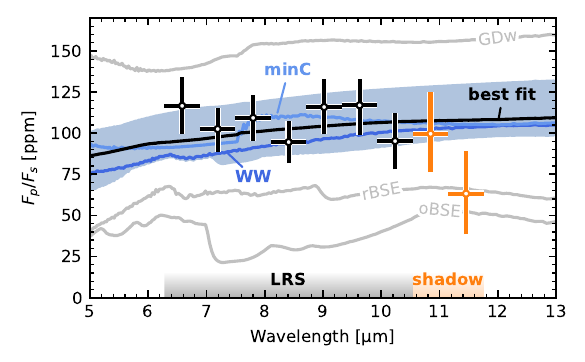}
    \caption{Observations from MIRI of 55 Cancri e (black markers, \citealt{hu2024secondary}), including the MIRI shadow region (orange markers), compared to a selection of synthetic emission spectra (coloured and grey lines). The blue shaded area indicates the range of spectra that matches the observation, based on a $\chi^2$-fit (excluding the shadow region). Three endmembers within the set of matching models are highlighted: the best fit (black, \logvmf=0, \dIW{-6} and \metal=0.5), a mineral-rich He-H$_2$-CO-SiO-atmosphere (minC, bright blue, \logvmf=-1, \dIW{-6} and \metal=0.1), and a waterworld (WW, dark blue, \logvmf=1, \dIW{+6} and \metal=0). Relevant non-fitting models are shown in grey and comprise the gas-dwarf scenario (GDw, \logvmf=1, \dIW{-6} and \metal=0) as well as a reduced (rBSE, \logvmf=0, \dIW{-3} and \metal=1) and an oxidised VIBSE scenario (oBSE, \logvmf=0, \dIW{+3} and \metal=1).}
    \label{fig:MIRI_best_fits}
\end{figure}

Recently, JWST observations of 55 Cnc e were published \citep{hu2024secondary, patel2024jwst}. The MIRI data presented by \citet{hu2024secondary} (black points; Fig. \ref{fig:MIRI_best_fits}) record a constant $F_p/F_s$ near $\sim$100~ppm, cited as evidence of an atmosphere owing to the suppression of the flux contrast relative to bare rock. This atmosphere might also be reducing, given that  the flat spectrum rules out an intense SO$_2$ feature (cf. Sec. \ref{results:spectra:emission}). 
In our study, we further support this conclusion.
The intensity predicted in this segment of the emission spectrum for models with visible SO$_2$ features (VIBSE, \dIW{+3}) is $\sim 2 \sigma$ lower than the observed $F_p/F_s$, even ruling out the scenario of an Earth-like, reduced (i.e. CO-bearing) atmosphere found by \citet{hu2024secondary}.
Instead, the MIRI observation can be fit by a degenerate set of solutions that include highly reduced (\dIW{-6}) CO-rich compositions at low to moderate \logvmf (-1 and 0) and moderate- to high-\metal (0.1 to 1), to comparatively oxidised (\dIW{} $\geq 0$), \logvmf= 0 H$_2$O-rich planets.
The best fit scenario from our discrete model grid, based on a $\chi^2$ estimate, is an atmosphere with \logvmf=0, \metal= 0.5 and \dIW{-6} (black curve in Fig. \ref{fig:MIRI_best_fits}) which is also consistent with the radius constraints (Sec. \ref{disc:mass_radius_diagrams}).
In addition to the VIBSE cases, massive (high \logvmf), reduced SOLAR-like cases (gas dwarfs) can be excluded, supporting the notion that this planet is neither a more massive Earth analogue, nor is it a planet that captured and maintained a primordial atmosphere.
A similar degeneracy, with pure CO or H$_2$O atmospheres fitting the MIRI observations, was also found by \citep{zilinskas2025characterising}. 

In terms of the NIRCam observations by \citet{hu2024secondary} and \citet{patel2024jwst}, no clear trend can be deduced.
Any attempt at fitting the observations, either individually or combined, was stifled by the variability of the target \citep{patel2024jwst} and the strong auto-correlation within the derived white light curves that does not allow for the deduction of the absolute flux value \citep[cf.][]{hu2024secondary, patel2024jwst, zilinskas2025characterising}.
The currently available data are thus unlikely to yield robust constraints, and further investigation is required to alleviate these shortcomings.

Besides JWST, 55 Cnc e has been the target of multiple missions. Of interest to our study are the observations by \citet{esteves2017search} and \citet{jindal2020characterization}, which limits the combinations of $\bar{m}$ and H$_2$O-VMR, potentially ruling out cases of \metal $\lesssim$ 0.1 and dIW{}>0. \citet{deibert2021near} rule out the most reduced "acetylene world" scenarios (Appendix \ref{apx:acetylene_worlds}) based on the non-detection of C$_2$H$_2$ and HCN at low $\bar{m}$; slightly more oxidising scenarios (\dIW{-3}) remain possible. \citet{tsiaras2016detection} however found hints for a HCN-bearing, H$_2$-rich atmosphere. \citet{ehrenreich2012hint} and \citet{zhang2021no} on the other hand fail to detect escaping H and He, respectively, an observation possibly at odds with the apparent \metal < 1 of our proposed cases for 55 Cnc e. However, the hot-spot shift seen with Spitzer \citep{demory2016map, angelo2017case} in conjunction with the atmospheric circulation models from \citet{hammond2017linking, zhang2017effects} again lend support to an H-bearing, low-$\bar{m}$ atmosphere consistent with the best-fit cases in Fig. \ref{fig:MIRI_best_fits}. Finally, high-resolution spectroscopy has not recovered Fe(g) in emission \citep{rasmussen2023nondetection}, and found no evidence of an extended Fe–Mg-bearing atmosphere in transmission \citep{keles2022pepsi}. This could potentially exclude clear, low-density, low-\metal models from Fig. \ref{fig:spectra_transmission} that show such lines at $\sim 0.5$ \micron; however, the direct comparability of our models to the template spectra used in their studies remains uncertain, as they were volatile free and probed a limited selection of silicate vapour compositions. 

In summary, current spectral observations provide only indirect and contradictory evidence of the presence and composition of an atmosphere on 55 Cnc e. Whether the apparent inconsistencies between observations and models arise primarily from observational challenges or from simplifying assumptions in atmospheric modelling remains unclear. Some major limitations of our approach are discussed in Sec.~\ref{disc:limitations}.

\subsection{Future observations}
\begin{figure}[!t]
    \centering
    \includegraphics[width=\linewidth]{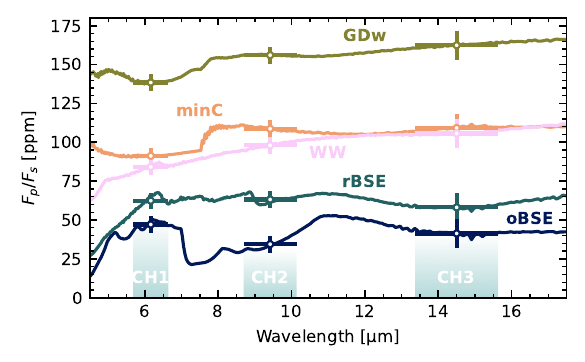}
    \caption{Prediction for MIRI MRS MEDIUM channel observations of 55 Cancri e for selected synthetic atmospheric models (identical to Fig. \ref{fig:MIRI_best_fits}). Uncertainties were simulated via the \texttt{pandeia.engine} package (v4.0) \citep[Exposure Time Calculator, ETC,][]{pandeia} with JWST reference data \citep{refdata}, assuming 7 groups and 19.4 seconds per exposure. Only channels 1, 2 and 3 are shown (CH1-CH3); channel 4 at 25 \micron is unlikely to yield useful constraints owing to its low through-put.}
    \label{fig:MRS_prediction}
\end{figure}

In Fig. \ref{fig:MRS_prediction}, we show five distinct scenarios for the atmosphere of 55 Cancri e, together with predicted measurements using the MIRI medium resolution spectroscopy (MRS) mode.
In contrast to the low-resolution (LRS) mode used by \citet{hu2024secondary}, the detector does not saturate, in spite of the star's brightness, and can thus be used to retrieve robust results.
The MRS mode offers four different channels with three subbands (SHORT, MEDIUM, and LONG) each.
To capture important features (CO$_2$ at 15 \micron, SO$_2$ and SiO at 9 \micron), the medium channel is required.
It becomes apparent that the predicted observations - when the spectrum in each channel is binned over the whole bandpass - is sufficient to distinguish between the endmember scenarios of gas dwarfs with primordial atmospheres (\logvmf=1, \dIW{-6}, \metal=0), steam planets (\logvmf=1, \dIW{0}, \metal=0), VIBSE-like planets and their redox states(\logvmf=0, \dIW{-3} (reduced) /\dIW{+3} (oxidised), \metal=1) and mineral gas enriched atmospheres (\logvmf=-1, \dIW{-6}, \metal=0).
A future observation of 55 Cancri e was approved in Cycle 4 GO, proposal 7875, \citealt{zhang2025only}.
However, it must be noted that a significant degree of information hinges on the baseline of the secondary eclipse flux, and any inference will be degenerate with the size of the condensed interior (Sec. \ref{disc:mass_radius_diagrams}) and the heat redistribution of the atmosphere \citep{zilinskas2025characterising}.
This information can be constrained by obtaining phase curves (which constrain the heat redistribution) and the primary transit, which determines the size of the atmosphere (Sec. \ref{results:spectra:transmission}).

\subsection{Limitations and improvements}
\label{disc:limitations}

Due to limitations in theory, observation, and experimental data, a number of simplifying assumptions had to be made. Below we list the ones we consider the most impactful on our results.

Incomplete or missing line lists represent a major limitation of this study. Some species occur in appreciable abundances in HRE atmospheres, yet lack short-wave opacities, for example acetylene (C$_2$H$_2$) and methane (CH$_4$). Both gases occur exclusively under highly reducing conditions (\dIW{}$\lesssim$ -3), with the latter emerging only in cooler atmospheres (\tirr $\lesssim$ 2000 K). Acetylene may, however, become a dominant absorber in massive atmospheres (\logvmf$\geq$1), but opacity data below 1 \micron are missing (Fig. \ref{fig:opacities}), meaning its inclusion would result in a spurious greenhouse effect (see Appendix \ref{apx:acetylene_worlds}). Complete opacities of important species in massive, reduced atmospheres is required. To eliminate these biases, the available line lists have to be extended via experimental and theoretical means, and evaluated over the whole pressure-temperature range of HRE atmospheres.

The solubility laws used in this work are described according to Henry's or Sieverts' law approximations, which may break down when the concentration of the solute increases, and pressures, temperatures and \fOtwo diverge from those at which the laws were calibrated (see Table \ref{table:target_species} or \citealt{bower2025diversity}, their Table 1). This is particularly relevant for the hotter HREs, for which surface temperatures can approach 3000~K. By contrast, experimental constraints are typically half that temperature; only H$_2$O  was measured to applicable temperatures ($\sim$2173~K, \citealt{sossi2023solubility}). Further, the Henry's law constants for all species except H$_2$O \citep{sossi2020redox} were obtained for basaltic or andesitic compositions of the melt, which, though reasonable approximations, are unlikely to hold for magma oceans on exoplanets. This may pose problems for species with composition-dependent solubilities, like S$_2$ (which is a complex function of melt FeO content, \citealt{o2002sulfide, BW23}) or H$_2$, N$_2$ and the noble gases that dissolve by physical incorporation \citep{carroll1994noble}. Furthermore, all solubility laws are derived for either pure gases or simple mixtures thereof, and thus the true solubilities may deviate from those reported in this study. More experimental measurements have to be obtained.

All the radiative transfer codes used here (\texttt{HELIOS} and \texttt{petitRADTRANS}) as well as \texttt{FastChem} assume an ideal gas equation of state, which is violated by heavy atmospheres ($\gtrsim$ 1000 bar). This issue is fundamental and goes beyond the scope of this study (cf. Sec. \ref{methods:pipeline:radtrans}), but only affects the layers below the photosphere ($P_{photo} \lesssim$ 1 bar in all cases).
Likely implications are \textit{i.)} miscalculation of atmospheric elemental ratios (e.g. C/O), \textit{ii.)} underestimation of dissolved content (since fugacity coefficients for major volatiles tend to exceed 1 at high pressures), \textit{iii.)} differences in speciation in the atmospheric column of the lower atmosphere (as computed by \texttt{FastChem}), and \textit{iv.)} an underestimation of the planets radius, as non-ideal species of the major volatiles are less compressible than ideal gases \citep{holland1991compensated,shi1992thermodynamic,bower2025diversity}. Issues \textit{i} and \textit{ii} can in principle be corrected for by passing a non-ideal equation of state to \atmodeller. Issues \textit{iii} and \textit{iv} cannot be mended without major adaptions to the other underlying codes.

While we included the effect of condensation on the removal of a gas phase, we did not include the formation of clouds or hazes. 
In a realistic HRE atmosphere with eddy diffusion and strong day-nightside thermal gradients \citep{hammond2017linking, zhang2017effects}, condensation may occur as gas is advected to cooler upper layers or the nightside \citep[cf.][]{nguyen2024clouds}. If present, clouds may substantially alter transmission spectra, potentially impeding observability of key spectral features \citep{janssen2026hot}. However, more detailed studies on the influence of condensation on P-T-structure and spectra in both 1D and 2D/3D models are required.

The heat redistribution factor $\mathfrak{f}$ might deviate in the presence of strong day-to-nightside advection from the herein assumed value of 2/3. It exerts a strong influence on the secondary eclipse spectra, with smaller $\mathfrak{f}$ lowering the simulated emission flux of 55 Cnc e by $\sim$20-50 ppm when changing $\mathfrak{f}$ from 2/3 (\tirr=2500 K, average dayside temperature) to 1/4 (\tirr=2000 K, equilibrium temperature).
This difference is crucial to consider when fitting the observed spectra, as they result in divergent best-fitting models \citep{zilinskas2025characterising}.
To constrain $\mathfrak{f}$, multi-wavelength full phase curve observations are required.

All preceding analysis was made under the assumption of a fully molten Earth-like interior structure and a restricted set of bulk volatile compositions. Deviations therefrom (e.g. changes in CMF, mantle and core composition, mantle melt fraction, or volatile compositions deviating from the mixing line defined by \metal) may result in significant changes to $R_p$ and transmission and emission spectra. However, the necessary changes in the interior structure would indicate that the outliers found in Sec. \ref{disc:mass_radius_diagrams}, including 55 Cnc e, TOI-1408c and TOI-561 b, would have to deviate from the upscaled Earth-like scenario regardless. The diversity in HRE composition and interior structure thus appears robust, but its details remain to be studied in more extensive work.

\section{Summary and conclusions}
\label{summary}

Despite varying bulk chemistry, volatile-rich mineral gas atmospheres are generally dominated by He, CO, CO$_2$, H$_2$, and H$_2$O. The \fOtwo controls the ratios of CO/CO$_2$ and H$_2$/H$_2$O, where CO and H$_2$ are favoured in reducing atmospheres and CO$_2$ and H$_2$O in oxidising atmospheres. The $p$SO$_2$ reaches a maximum near \dIW{+3}, owing to its dissolution as S$^{2-}$ at lower \fOtwo and SO$_4^{2-}$ at higher \fOtwo, rendering it the dominant gas in oxidising high-\metal atmospheres. H$_2$O as the prevailing gas only occurs in heavy low-\metal atmospheres due to its high solubility, while He, when present, resides almost exclusively in the atmosphere, as do CO and CO$_2$. Nitrogen is also insoluble (as N$_2$) except at low \fOtwo, at which point N$^{3-}$ is stable in the melt. Mineral gases, such as SiO, Mg, MgO, and Fe evaporate preferentially in highly reduced (\dIW{}$\lesssim -3$) and hot atmospheres (\tirr $\gtrsim$ 2000 K). Cooler, higher \logvmf and reducing atmospheres ($T<2000$ K) are more prone to form methane instead of CO, with the occurrence of CH$_4$, C$_2$H$_2$, and HCN. In reduced atmospheres at the highest temperatures and \logvmf, SiH$_4$ becomes abundant, but the ideal gas approximation is no longer valid in such cases. 

The atmospheric structures of HREs are typified by strong greenhouse effects in the lower atmosphere, induced by CO, CO$_2$, H$_2$O, and SO$_2$, which is nearly universally present but exacerbated in oxidised atmospheres (\dIW{}$\gtrsim 0$). In these cases, the surface of the planet is up to 800~K hotter than its irradiation temperature, which allows for magma oceans on planets with irradiation temperatures as low as 1000 K. The same gases cause cooling in the upper atmosphere, decreasing the upper atmospheric temperature by up to 1500 K compared to the irradiation temperature.
On the other hand, an upper atmosphere thermal inversion appears in atmospheres whose opacity structure is dominated by mineral gases, mainly SiO, which is the case in reduced and hot systems of low \logvmf.

Secondary eclipse depths are mostly sensitive to temperature and oxygen fugacity (\fOtwo) and to a lower degree to metallicity (\metal) and the volatile content (\logvmf). Key molecules with strong absorption features include SO$_2$, CO$_2$, and H$_2$O. When the atmosphere is inverted, CO and SiO appear as emission features. SO$_2$ is the most robust tracer of \fOtwo in the range $\Delta$IW +3$\pm$3, except at \metal$\sim 0$, where the feature appears weak due to low total sulfur abundances. CO$_2$ and H$_2$O are present across a wide range of conditions, making them poor stand-alone redox indicators, but they allow for constraints on metallicity (\metal) and volatile amounts (\logvmf). SiO emission at 9 \micron identifies hot, highly reducing low-metallicity atmospheres. Cooler (\tirr $\lesssim 2000$ K) and highly reduced atmospheres might show CH$_4$ features instead, which may be replaced by C$_2$H$_2$ features in a more realistic setting. The 4.5 \micron region (JWST NIRCam) is spectrally degenerate in CO$_2$, CO, SO$_2$, and SiO, while the JWST MIRI instrument can distinguish between redox regimes and temperatures by targeting distinct CO$_2$, SO$_2$, and SiO features in the 7–20 \micron band.

The transit depth is mainly governed by \metal: low-\metal atmospheres are significantly more extended than their compressed high-\metal counterparts. However, the identity and shape of spectral features in transmission is still dictated by \fOtwo, and they show \metal-sensitive H$_2$O and redox-sensitive SO$_2$ features in the mid-infrared to the same effect as emission spectra. H$_2$O bands between 1-2 \micron also become tracers of atmospheric mass (\logvmf) for oxidised low-\metal atmospheres, while planets with \metal$\sim 0$ show strong mid-IR SiO and CO features. The UVIS to NIR (300–1000 nm) shows a minimum in transit depth but hosts lines from SO, OH, Fe, Mg, MgO, SiO, and H$_2$O that vary with \fOtwo, offering a promising window for ground-based detection of magma oceans and retrievals of the \fOtwo.

In the cooler upper layers, condensation of silicates, oxides, and carbides might occur.
In \metal= 0 atmospheres, condensation is suppressed by the thermal inversion; in oxidised high-\metal atmospheres, condensation of silicates (and oxides at the highest \fOtwo) occurs in the upper atmospheric layers despite irradiation temperatures of 2500 K due to the cooling effect of the greenhouse gases. Reduced high-\metal planets condense SiC in their lower to middle atmosphere, and graphite at nearly all altitudes. \tirr < 2000 K progressively leads to condensation at greater atmospheric depths, and condensation in lower-\metal atmospheres.

While the majority of mass-radius relationships of the hottest HREs are consistent with atmospheres with a limited scale height (favoured at high \metal), some, such as 55 Cancri e and TOI-1408 c, have densities that are too low to be explicable by a bulk silicate Earth-like volatile budget (i.e. \metal = 1 and \logvmf =0). Hence, these HREs are consistent with intermediate \metal $\sim 0.1-0.5$ atmospheres in the case of 55 Cancri e or, in the case of TOI-1408 c, favour SOLAR-like (\metal =0), massive (\logvmf =1), and highly reducing (\dIW{}$\lesssim$-3) scenarios to explain their extended atmospheres. We find that mass-radius relations are degenerate with respect to \metal and \logvmf, which can only be resolved by spectroscopic means.

The MIRI spectra of 55 Cancri e obtained by \citet{hu2024secondary} preclude VIBSE-scenarios at any \logvmf and any \dIW{}$\geq -3$, as well as highly reduced (\dIW{-6}) SOLAR cases at \logvmf$\geq 1$, indicating that the planet is neither a heavier Earth-twin nor that it captured an extensive nebular atmosphere. Intermediate scenarios remain possible. This conclusion aligns with the constraints from the mass-radius data at a shorter wavelength. The NIRCam observations by \citet{hu2024secondary} and \citet{patel2024jwst} do not paint a consistent picture and remain inconclusive. Ground-based observations might disfavour H-rich (low-\metal) scenarios. Future observations with MIRI's MRS mode can place constraints on the possible scenarios but will remain degenerate with the interior structure and the temperature distribution of the planet.
\\

\noindent \textit{Code and Data availability.} The \texttt{MAGMA} code can be obtained from Bruce Fegley Jr. upon reasonable request. The other codes used in this study are distributed under permissive software licences and can be obtained from the respective sources (see below). The scripts used to produce this study can be found on github\footnote{\url{https://github.com/ExPlanetology/volbear-scripts}}, and the data products can be found on zenodo\footnote{\url{https://zenodo.org/records/19571079}}.
This work made use of the following codes: \atmodeller v1.0.0 \citep{bower2025diversity}, \texttt{phaethon} \citep{seidler2024impact}, \texttt{HELIOS} \citep{Malik:2017, Malik:2019}, \texttt{FastChem} \citep{kitzmann2024fastchem}, \texttt{MAGMA} \citep{fegley1987vaporization, schaefer2004thermodynamic}, \texttt{numpy} \citep{numpy_harris2020array}, \texttt{scipy} \citep{scipy_2020SciPy-NMeth}, \texttt{pandas} \citep{pandas_mckinney-proc-scipy-2010}, \texttt{matplotlib} \citep{matplotlib_Hunter:2007}, \texttt{seaborn} \citep{seaborn_Waskom2021},  \texttt{astropy} \citep{astropy:2013, astropy:2018, astropy:2022}, \texttt{bayes\_opt} \citep{bayes_opt}.
\\

\noindent \textit{Acknowledgements.}
We thank Merlin Zgraggen and Lukas Felix for their collaboration on an unaccepted JWST proposal; from their contribution, we deduced the uncertainties in the expected MRS observation shown in Fig. \ref{fig:MRS_prediction}.
This work was supported by the Swiss National Science Foundation (SNSF) through an Eccellenza Professorship (203668) and the Swiss State Secretariat for Education, Research and Innovation (SERI) under contract No. MB22.00033, a SERI-funded ERC Starting grant "2ATMO" to P.A.S. Parts of this work have been carried out within the framework of the National Centre of Competence in Research (NCCR) PlanetS supported by the SNSF under grant 51NF40\_205606. This publication makes use of The Data \& Analysis Center for Exoplanets (DACE), which is a facility based at the University of Geneva (CH) dedicated to extrasolar planets data visualisation, exchange and analysis. DACE is a platform of the Swiss NCCR PlanetS, federating  Swiss expertise in Exoplanet research. The DACE platform is available at \url{https://dace.unige.ch}.

\bibliographystyle{aa.bst}
\bibliography{bibliography}

@ARTICLE{astropy:2022,
       author = {{Astropy Collaboration} and {Price-Whelan}, Adrian M. and {Lim}, Pey
       Lian and {Earl}, Nicholas and {Starkman}, Nathaniel and {Bradley}, Larry and
       {Shupe}, David L. and {Patil}, Aarya A. and {Corrales}, Lia and {Brasseur}, C.~E.
       and {N{\"o}the}, Maximilian and {Donath}, Axel and {Tollerud}, Erik and {Morris},
       Brett M. and {Ginsburg}, Adam and {Vaher}, Eero and {Weaver}, Benjamin A. and
       {Tocknell}, James and {Jamieson}, William and {van Kerkwijk}, Marten H. and
       {Robitaille}, Thomas P. and {Merry}, Bruce and {Bachetti}, Matteo and
       {G{\"u}nther}, H. Moritz and {Aldcroft}, Thomas L. and {Alvarado-Montes}, Jaime
       A. and {Archibald}, Anne M. and {B{\'o}di}, Attila and {Bapat}, Shreyas and
       {Barentsen}, Geert and {Baz{\'a}n}, Juanjo and {Biswas}, Manish and {Boquien},
       M{\'e}d{\'e}ric and {Burke}, D.~J. and {Cara}, Daria and {Cara}, Mihai and
       {Conroy}, Kyle E. and {Conseil}, Simon and {Craig}, Matthew W. and {Cross},
       Robert M. and {Cruz}, Kelle L. and {D'Eugenio}, Francesco and {Dencheva}, Nadia
       and {Devillepoix}, Hadrien A.~R. and {Dietrich}, J{\"o}rg P. and {Eigenbrot},
       Arthur Davis and {Erben}, Thomas and {Ferreira}, Leonardo and {Foreman-Mackey},
       Daniel and {Fox}, Ryan and {Freij}, Nabil and {Garg}, Suyog and {Geda}, Robel and
       {Glattly}, Lauren and {Gondhalekar}, Yash and {Gordon}, Karl D. and {Grant},
       David and {Greenfield}, Perry and {Groener}, Austen M. and {Guest}, Steve and
       {Gurovich}, Sebastian and {Handberg}, Rasmus and {Hart}, Akeem and
       {Hatfield-Dodds}, Zac and {Homeier}, Derek and {Hosseinzadeh}, Griffin and
       {Jenness}, Tim and {Jones}, Craig K. and {Joseph}, Prajwel and {Kalmbach}, J.
       Bryce and {Karamehmetoglu}, Emir and {Ka{\l}uszy{\'n}ski}, Miko{\l}aj and
       {Kelley}, Michael S.~P. and {Kern}, Nicholas and {Kerzendorf}, Wolfgang E. and
       {Koch}, Eric W. and {Kulumani}, Shankar and {Lee}, Antony and {Ly}, Chun and
       {Ma}, Zhiyuan and {MacBride}, Conor and {Maljaars}, Jakob M. and {Muna}, Demitri
       and {Murphy}, N.~A. and {Norman}, Henrik and {O'Steen}, Richard and {Oman}, Kyle
       A. and {Pacifici}, Camilla and {Pascual}, Sergio and {Pascual-Granado}, J. and
       {Patil}, Rohit R. and {Perren}, Gabriel I. and {Pickering}, Timothy E. and
       {Rastogi}, Tanuj and {Roulston}, Benjamin R. and {Ryan}, Daniel F. and {Rykoff},
       Eli S. and {Sabater}, Jose and {Sakurikar}, Parikshit and {Salgado}, Jes{\'u}s
       and {Sanghi}, Aniket and {Saunders}, Nicholas and {Savchenko}, Volodymyr and
       {Schwardt}, Ludwig and {Seifert-Eckert}, Michael and {Shih}, Albert Y. and
       {Jain}, Anany Shrey and {Shukla}, Gyanendra and {Sick}, Jonathan and {Simpson},
       Chris and {Singanamalla}, Sudheesh and {Singer}, Leo P. and {Singhal}, Jaladh and
       {Sinha}, Manodeep and {Sip{\H{o}}cz}, Brigitta M. and {Spitler}, Lee R. and
       {Stansby}, David and {Streicher}, Ole and {{\v{S}}umak}, Jani and {Swinbank},
       John D. and {Taranu}, Dan S. and {Tewary}, Nikita and {Tremblay}, Grant R. and
       {Val-Borro}, Miguel de and {Van Kooten}, Samuel J. and {Vasovi{\'c}}, Zlatan and
       {Verma}, Shresth and {de Miranda Cardoso}, Jos{\'e} Vin{\'\i}cius and {Williams},
       Peter K.~G. and {Wilson}, Tom J. and {Winkel}, Benjamin and {Wood-Vasey}, W.~M.
       and {Xue}, Rui and {Yoachim}, Peter and {Zhang}, Chen and {Zonca}, Andrea and
       {Astropy Project Contributors}}, title = "{The Astropy Project: Sustaining and
       Growing a Community-oriented Open-source Project and the Latest Major Release
       (v5.0) of the Core Package}",
      journal = {\apj},
         year = 2022,
        month = aug,
       volume = {935},
       number = {2},
          eid = {167},
        pages = {167},
          doi = {10.3847/1538-4357/ac7c74},
archivePrefix = {arXiv},
       eprint = {2206.14220},
 primaryClass = {astro-ph.IM},
       adsurl = {https://ui.adsabs.harvard.edu/abs/2022ApJ...935..167A}
}

@ARTICLE{astropy:2018,
   author = {{Astropy Collaboration} and {Price-Whelan}, A.~M. and {Sip{\H o}cz}, B.~M. and
	{G{\"u}nther}, H.~M. and {Lim}, P.~L. and {Crawford}, S.~M. and
	{Conseil}, S. and {Shupe}, D.~L. and {Craig}, M.~W. and {Dencheva}, N. and
	{Ginsburg}, A. and {VanderPlas}, J.~T. and {Bradley}, L.~D. and
	{P{\'e}rez-Su{\'a}rez}, D. and {de Val-Borro}, M. and {Paper Contributors}, (. and
	{Aldcroft}, T.~L. and {Cruz}, K.~L. and {Robitaille}, T.~P. and
	{Tollerud}, E.~J. and {Coordination Committee}, (. and {Ardelean}, C. and
	{Babej}, T. and {Bach}, Y.~P. and {Bachetti}, M. and {Bakanov}, A.~V. and
	{Bamford}, S.~P. and {Barentsen}, G. and {Barmby}, P. and {Baumbach}, A. and
	{Berry}, K.~L. and {Biscani}, F. and {Boquien}, M. and {Bostroem}, K.~A. and
	{Bouma}, L.~G. and {Brammer}, G.~B. and {Bray}, E.~M. and {Breytenbach}, H. and
	{Buddelmeijer}, H. and {Burke}, D.~J. and {Calderone}, G. and
	{Cano Rodr{\'{\i}}guez}, J.~L. and {Cara}, M. and {Cardoso}, J.~V.~M. and
	{Cheedella}, S. and {Copin}, Y. and {Corrales}, L. and {Crichton}, D. and
	{D{\'A}vella}, D. and {Deil}, C. and {Depagne}, {\'E}. and
	{Dietrich}, J.~P. and {Donath}, A. and {Droettboom}, M. and
	{Earl}, N. and {Erben}, T. and {Fabbro}, S. and {Ferreira}, L.~A. and
	{Finethy}, T. and {Fox}, R.~T. and {Garrison}, L.~H. and {Gibbons}, S.~L.~J. and
	{Goldstein}, D.~A. and {Gommers}, R. and {Greco}, J.~P. and
	{Greenfield}, P. and {Groener}, A.~M. and {Grollier}, F. and
	{Hagen}, A. and {Hirst}, P. and {Homeier}, D. and {Horton}, A.~J. and
	{Hosseinzadeh}, G. and {Hu}, L. and {Hunkeler}, J.~S. and {Ivezi{\'c}}, {\v Z}. and
	{Jain}, A. and {Jenness}, T. and {Kanarek}, G. and {Kendrew}, S. and
	{Kern}, N.~S. and {Kerzendorf}, W.~E. and {Khvalko}, A. and
	{King}, J. and {Kirkby}, D. and {Kulkarni}, A.~M. and {Kumar}, A. and
	{Lee}, A. and {Lenz}, D. and {Littlefair}, S.~P. and {Ma}, Z. and
	{Macleod}, D.~M. and {Mastropietro}, M. and {McCully}, C. and
	{Montagnac}, S. and {Morris}, B.~M. and {Mueller}, M. and {Mumford}, S.~J. and
	{Muna}, D. and {Murphy}, N.~A. and {Nelson}, S. and {Nguyen}, G.~H. and
	{Ninan}, J.~P. and {N{\"o}the}, M. and {Ogaz}, S. and {Oh}, S. and
	{Parejko}, J.~K. and {Parley}, N. and {Pascual}, S. and {Patil}, R. and
	{Patil}, A.~A. and {Plunkett}, A.~L. and {Prochaska}, J.~X. and
	{Rastogi}, T. and {Reddy Janga}, V. and {Sabater}, J. and {Sakurikar}, P. and
	{Seifert}, M. and {Sherbert}, L.~E. and {Sherwood-Taylor}, H. and
	{Shih}, A.~Y. and {Sick}, J. and {Silbiger}, M.~T. and {Singanamalla}, S. and
	{Singer}, L.~P. and {Sladen}, P.~H. and {Sooley}, K.~A. and
	{Sornarajah}, S. and {Streicher}, O. and {Teuben}, P. and {Thomas}, S.~W. and
	{Tremblay}, G.~R. and {Turner}, J.~E.~H. and {Terr{\'o}n}, V. and
	{van Kerkwijk}, M.~H. and {de la Vega}, A. and {Watkins}, L.~L. and
	{Weaver}, B.~A. and {Whitmore}, J.~B. and {Woillez}, J. and
	{Zabalza}, V. and {Contributors}, (.},
    title = "{The Astropy Project: Building an Open-science Project and Status of the v2.0 Core Package}",
  journal = {\aj},
archivePrefix = "arXiv",
   eprint = {1801.02634},
 primaryClass = "astro-ph.IM",
     year = 2018,
    month = sep,
   volume = 156,
      eid = {123},
    pages = {123},
      doi = {10.3847/1538-3881/aabc4f},
   adsurl = {https://ui.adsabs.harvard.edu/abs/2018AJ....156..123T}
}

@ARTICLE{astropy:2013,
   author = {{Astropy Collaboration} and {Robitaille}, T.~P. and {Tollerud}, E.~J. and
    {Greenfield}, P. and {Droettboom}, M. and {Bray}, E. and {Aldcroft}, T. and
    {Davis}, M. and {Ginsburg}, A. and {Price-Whelan}, A.~M. and
    {Kerzendorf}, W.~E. and {Conley}, A. and {Crighton}, N. and
    {Barbary}, K. and {Muna}, D. and {Ferguson}, H. and {Grollier}, F. and
    {Parikh}, M.~M. and {Nair}, P.~H. and {Unther}, H.~M. and {Deil}, C. and
    {Woillez}, J. and {Conseil}, S. and {Kramer}, R. and {Turner}, J.~E.~H. and
    {Singer}, L. and {Fox}, R. and {Weaver}, B.~A. and {Zabalza}, V. and
    {Edwards}, Z.~I. and {Azalee Bostroem}, K. and {Burke}, D.~J. and
    {Casey}, A.~R. and {Crawford}, S.~M. and {Dencheva}, N. and
    {Ely}, J. and {Jenness}, T. and {Labrie}, K. and {Lian Lim}, P. and
    {Pierfederici}, F. and {Pontzen}, A. and {Ptak}, A. and {Refsdal}, B. and
    {Servillat}, M. and {Streicher}, O.},
    title = "{Astropy: A community Python package for astronomy}",
  journal = {\aap},
     year = 2013,
    month = oct,
   volume = 558,
      eid = {A33},
    pages = {A33},
      doi = {10.1051/0004-6361/201322068},
   adsurl = {https://ui.adsabs.harvard.edu/abs/2013A%26A...558A..33A}
}

@Misc{bayes_opt,
    author = {Fernando Nogueira},
    title = {{Bayesian Optimization}: Open source constrained global optimization tool for {Python}},
    year = {2014--},
    url = " https://github.com/bayesian-optimization/BayesianOptimization"
}

@article{Malik:2017,
doi = {10.3847/1538-3881/153/2/56},
url = {https://dx.doi.org/10.3847/1538-3881/153/2/56},
year = {2017},
month = {jan},
publisher = {The American Astronomical Society},
volume = {153},
number = {2},
pages = {56},
author = {Matej Malik and Luc Grosheintz and João M. Mendonça and Simon L. Grimm and Baptiste Lavie and Daniel Kitzmann and Shang-Min Tsai and Adam Burrows and Laura Kreidberg and Megan Bedell and Jacob L. Bean and Kevin B. Stevenson and Kevin Heng},
title = {HELIOS: AN OPEN-SOURCE, GPU-ACCELERATED RADIATIVE TRANSFER CODE FOR SELF-CONSISTENT EXOPLANETARY ATMOSPHERES},
journal = {AJ},
}

@article{Malik:2019,
doi = {10.3847/1538-3881/ab1084},
url = {https://dx.doi.org/10.3847/1538-3881/ab1084},
year = {2019},
month = {apr},
publisher = {The American Astronomical Society},
volume = {157},
number = {5},
pages = {170},
author = {Matej Malik and Daniel Kitzmann and João M. Mendonça and Simon L. Grimm and Gabriel-Dominique Marleau and Esther F. Linder and Shang-Min Tsai and Kevin Heng},
title = {Self-luminous and Irradiated Exoplanetary Atmospheres Explored with HELIOS},
journal = {AJ},
}

@article{angelo2017case,
  title={A case for an atmosphere on super-Earth 55 Cancri e},
  author={Angelo, Isabel and Hu, Renyu},
  journal={AJ},
  volume={154},
  number={6},
  pages={232},
  year={2017},
  publisher={IOP Publishing}
}

@article{abel2011collision,
  title={Collision-induced absorption by H2 pairs: From hundreds to thousands of kelvin},
  author={Abel, Martin and Frommhold, Lothar and Li, Xiaoping and Hunt, Katharine LC},
  journal={J. Phys. Chem. A},
  volume={115},
  number={25},
  pages={6805--6812},
  year={2011},
  publisher={ACS Publications}
}

@article{abel2012infrared,
  title={Infrared absorption by collisional H2--He complexes at temperatures up to 9000 K and frequencies from 0 to 20 000 cm- 1},
  author={Abel, Martin and Frommhold, Lothar and Li, Xiaoping and Hunt, Katharine LC},
  journal={J. Chem. Phys.},
  volume={136},
  number={4},
  year={2012},
  publisher={AIP Publishing}
}

@article{gustafsson2003h2,
  title={The H2--H infrared absorption bands at temperatures from 1000 K to 2500 K},
  author={Gustafsson, Magnus and Frommhold, Lothar},
  journal={A\&A},
  volume={400},
  number={3},
  pages={1161--1162},
  year={2003},
  publisher={EDP Sciences}
}

@article{gustafsson2001infrared,
  title={Infrared Absorption Spectra of Collisionally InteractingHe and H Atoms},
  author={Gustafsson, Magnus and Frommhold, Lothar},
  journal={ApJ},
  volume={546},
  number={2},
  pages={1168},
  year={2001},
  publisher={IOP Publishing}
}

@article{seidler2024impact,
  title={Impact of oxygen fugacity on the atmospheric structure and emission spectra of ultra-hot rocky exoplanets},
  author={Seidler, Fabian L and Sossi, Paolo A and Grimm, Simon L},
  journal={A\&A},
  volume={691},
  pages={A159},
  year={2024},
  publisher={EDP Sciences}
}

@article{bower2022,
  title={Retention of water in terrestrial magma oceans and carbon-rich early atmospheres},
  author={Bower, Dan J and Hakim, Kaustubh and Sossi, Paolo A and Sanan, Patrick},
  journal={Planet. Sci. J.},
  volume={3},
  number={4},
  pages={93},
  year={2022},
  publisher={IOP Publishing}
}

@article{bower2025diversity,
  title={Diversity of Low-mass Planet Atmospheres in the C--H--O--N--S--Cl System with Interior Dissolution, Nonideality, and Condensation: Application to TRAPPIST-1e and Sub-Neptunes},
  author={Bower, Dan J and Thompson, Maggie A and Hakim, Kaustubh and Tian, Meng and Sossi, Paolo A},
  journal={ApJ},
  volume={995},
  number={1},
  pages={59},
  year={2025},
  publisher={The American Astronomical Society}
}

@article{chaudhari2025hydrogen,
  title={The solubility of molecular hydrogen in silicate melts and the origin of hydrogen in the interiors of terrestrial planets},
  author={Chaudhari, Alok and Masotta, Matteo and Shcheka, Svyatoslav and Keppler, Hans},
  journal={Contrib. Mineral. Petrol.},
  volume={180},
  number={11},
  pages={1--20},
  year={2025},
  publisher={Springer}
}

@article{sossi2019evaporation,
  title={Evaporation of moderately volatile elements from silicate melts: experiments and theory},
  author={Sossi, Paolo A and Klemme, Stephan and O'Neill, Hugh St C and Berndt, Jasper and Moynier, Fr{\'e}d{\'e}ric},
  journal={Geochim.~Cosmochim.~Acta},
  volume={260},
  pages={204--231},
  year={2019},
  publisher={Elsevier}
}

@article{sossi2023solubility,
  title={Solubility of water in peridotite liquids and the prevalence of steam atmospheres on rocky planets},
  author={Sossi, Paolo A and Tollan, Peter ME and Badro, James and Bower, Dan J},
  journal={Earth Planet. Sci. Lett.},
  volume={601},
  pages={117894},
  year={2023},
  publisher={Elsevier}
}

@article{thompson2025water,
  title={Water solubility in silicate melts: The effects of melt composition under reducing conditions and implications for nebular ingassing on rocky planets},
  author={Thompson, Maggie A and Sossi, Paolo A and Bower, Dan J and Shahar, Anat and Liebske, Christian and Allaz, Julien},
  journal={Chem. Geol.},
  pages={123048},
  year={2025},
  publisher={Elsevier}
}

@article{monaghan2025low,
  title={Low 4.5 $\mu$m Dayside Emission Disfavors a Dark Bare-rock Scenario for the Hot Super-Earth TOI-431 b},
  author={Monaghan, Christopher and Roy, Pierre-Alexis and Benneke, Bj{\"o}rn and Crossfield, Ian JM and Coulombe, Louis-Philippe and Piaulet-Ghorayeb, Caroline and Kreidberg, Laura and Dressing, Courtney D and Kane, Stephen R and Dragomir, Diana and others},
  journal={AJ},
  volume={169},
  number={5},
  pages={239},
  year={2025},
  publisher={IOP Publishing}
}

@article{o2002sulfide,
  title={The sulfide capacity and the sulfur content at sulfide saturation of silicate melts at 1400 C and 1 bar},
  author={O’Neill, Hugh S.T.C. and Mavrogenes, John A.},
  journal={J. Petrol.},
  volume={43},
  number={6},
  pages={1049--1087},
  year={2002},
  publisher={Oxford University Press}
}

@article{liggins2022growth,
  title={Growth and evolution of secondary volcanic atmospheres: I. Identifying the geological character of hot rocky planets},
  author={Liggins, Philippa and Jordan, Sean and Rimmer, Paul B and Shorttle, Oliver},
  journal={J. Geophys. Res.: Planets},
  volume={127},
  number={7},
  pages={e2021JE007123},
  year={2022},
  publisher={Wiley Online Library}
}

@Article{dixon1995,
  author  = {Dixon, Jacqueline Eaby and Stolper, Edward M. and Holloway, John R.},
  journal = {J. Petrol.},
  title   = {{An Experimental Study of Water and Carbon Dioxide Solubilities in Mid-Ocean Ridge Basaltic Liquids. Part I: Calibration and Solubility Models}},
  year    = {1995},
  issn    = {0022-3530},
  month   = {12},
  number  = {6},
  pages   = {1607--1631},
  volume  = {36},
  doi     = {https://doi.org/10.1093/oxfordjournals.petrology.a037267},
  eprint  = {https://academic.oup.com/petrology/article-pdf/36/6/1607/4319247/36-6-1607.pdf},
  url     = {https://doi.org/10.1093/oxfordjournals.petrology.a037267},
}

@Article{yoshioka2019,
  author  = {Takahiro Yoshioka and Daisuke Nakashima and Tomoki Nakamura and Svyatoslav Shcheka and Hans Keppler},
  journal = {Geochim.~Cosmochim.~Acta},
  title   = {Carbon solubility in silicate melts in equilibrium with a {CO}-{CO$_2$} gas phase and graphite},
  year    = {2019},
  issn    = {0016-7037},
  pages   = {129--143},
  volume  = {259},
  doi     = {https://doi.org/10.1016/j.gca.2019.06.007},
}

@Article{ardia2013,
  author  = {P. Ardia and M.M. Hirschmann and A.C. Withers and B.D. Stanley},
  journal = {Geochim.~Cosmochim.~Acta},
  title   = {Solubility of {CH4} in a synthetic basaltic melt, with applications to atmosphere–magma ocean-core partitioning of volatiles and to the evolution of the {M}artian atmosphere},
  year    = {2013},
  issn    = {0016-7037},
  pages   = {52--71},
  volume  = {114},
  doi     = {https://doi.org/10.1016/j.gca.2013.03.028},
}

@article{lodders2021relative,
  title={Relative atomic solar system abundances, mass fractions, and atomic masses of the elements and their isotopes, composition of the solar photosphere, and compositions of the major chondritic meteorite groups},
  author={Lodders, Katharina},
  journal={Space Sci. Rev.},
  volume={217},
  number={3},
  pages={44},
  year={2021},
  publisher={Springer}
}

@article{bernadou2021nitrogen,
  title={Nitrogen solubility in basaltic silicate melt-Implications for degassing processes},
  author={Bernadou, Fabien and Gaillard, Fabrice and F{\"u}ri, Evelyn and Marrocchi, Yves and Slodczyk, Aneta},
  journal={Chem. Geol.},
  volume={573},
  pages={120192},
  year={2021},
  publisher={Elsevier}
}

@article{BW22,
	author = {Julien Boulliung and Bernard J. Wood},
	doi = {https://doi.org/10.1016/j.gca.2022.08.032},
	journal = {Geochim.~Cosmochim.~Acta},
	pages = {150--164},
	title = {SO$_{2}$ solubility and degassing behavior in silicate melts},
	volume = {336},
	year = {2022},}

@article{BW23,
	author = {Julien Boulliung and Bernard J. Wood},
	doi = {https://doi.org/10.1007/s00410-023-02033-9},
	journal = {Contrib. Mineral. Petrol. },
	number = {56},
	pages = {15},
	title = {Sulfur oxidation state and solubility in silicate melts },
	volume = {178},
	year = {2023},}

@Article{         numpy_harris2020array,
 title         = {Array programming with {NumPy}},
 author        = {Charles R. Harris and K. Jarrod Millman and St{\'{e}}fan J.
                 van der Walt and Ralf Gommers and Pauli Virtanen and David
                 Cournapeau and Eric Wieser and Julian Taylor and Sebastian
                 Berg and Nathaniel J. Smith and Robert Kern and Matti Picus
                 and Stephan Hoyer and Marten H. van Kerkwijk and Matthew
                 Brett and Allan Haldane and Jaime Fern{\'{a}}ndez del
                 R{\'{i}}o and Mark Wiebe and Pearu Peterson and Pierre
                 G{\'{e}}rard-Marchant and Kevin Sheppard and Tyler Reddy and
                 Warren Weckesser and Hameer Abbasi and Christoph Gohlke and
                 Travis E. Oliphant},
 year          = {2020},
 month         = sep,
 journal       = {Nature},
 volume        = {585},
 number        = {7825},
 pages         = {357--362},
 doi           = {10.1038/s41586-020-2649-2},
 publisher     = {Springer Science and Business Media {LLC}},
 url           = {https://doi.org/10.1038/s41586-020-2649-2}
}

@ARTICLE{scipy_2020SciPy-NMeth,
  author  = {Virtanen, Pauli and Gommers, Ralf and Oliphant, Travis E. and
            Haberland, Matt and Reddy, Tyler and Cournapeau, David and
            Burovski, Evgeni and Peterson, Pearu and Weckesser, Warren and
            Bright, Jonathan and {van der Walt}, St{\'e}fan J. and
            Brett, Matthew and Wilson, Joshua and Millman, K. Jarrod and
            Mayorov, Nikolay and Nelson, Andrew R. J. and Jones, Eric and
            Kern, Robert and Larson, Eric and Carey, C J and
            Polat, {\.I}lhan and Feng, Yu and Moore, Eric W. and
            {VanderPlas}, Jake and Laxalde, Denis and Perktold, Josef and
            Cimrman, Robert and Henriksen, Ian and Quintero, E. A. and
            Harris, Charles R. and Archibald, Anne M. and
            Ribeiro, Ant{\^o}nio H. and Pedregosa, Fabian and
            {van Mulbregt}, Paul and {SciPy 1.0 Contributors}},
  title   = {{{SciPy} 1.0: Fundamental Algorithms for Scientific
            Computing in Python}},
  journal = {Nat. Methods},
  year    = {2020},
  volume  = {17},
  pages   = {261--272},
  adsurl  = {https://rdcu.be/b08Wh},
  doi     = {10.1038/s41592-019-0686-2},
}

@Article{matplotlib_Hunter:2007,
  Author    = {Hunter, J. D.},
  Title     = {Matplotlib: A 2D graphics environment},
  Journal   = {Computing in Science \& Engineering},
  Volume    = {9},
  Number    = {3},
  Pages     = {90--95},
  publisher = {IEEE COMPUTER SOC},
  doi       = {10.1109/MCSE.2007.55},
  year      = 2007
}

@InProceedings{pandas_mckinney-proc-scipy-2010,
  author    = { {W}es {M}c{K}inney },
  title     = { {D}ata {S}tructures for {S}tatistical {C}omputing in {P}ython },
  booktitle = { {P}roceedings of the 9th {P}ython in {S}cience {C}onference },
  pages     = { 56 - 61 },
  year      = { 2010 },
  editor    = { {S}t\'efan van der {W}alt and {J}arrod {M}illman },
  doi       = { 10.25080/Majora-92bf1922-00a }
}

@article{seaborn_Waskom2021,
    doi = {10.21105/joss.03021},
    url = {https://doi.org/10.21105/joss.03021},
    year = {2021},
    publisher = {The Open Journal},
    volume = {6},
    number = {60},
    pages = {3021},
    author = {Michael L. Waskom},
    title = {seaborn: statistical data visualization},
    journal = {J. Open Source Software}
 }

@article{hu2024secondary,
  title={A secondary atmosphere on the rocky Exoplanet 55 Cancri e},
  author={Hu, Renyu and Bello-Arufe, Aaron and Zhang, Michael and Paragas, Kimberly and Zilinskas, Mantas and van Buchem, Christiaan and Bess, Michael and Patel, Jayshil and Ito, Yuichi and Damiano, Mario and others},
  journal={Nature},
  pages={1--2},
  year={2024},
  publisher={Nature Publishing Group UK London}
}

@article{kitzmann2024fastchem,
  title={fastchem cond: equilibrium chemistry with condensation and rainout for cool planetary and stellar environments},
  author={Kitzmann, Daniel and Stock, Joachim W and Patzer, A Beate C},
  journal={MNRAS},
  volume={527},
  number={3},
  pages={7263--7283},
  year={2024},
  publisher={Oxford University Press}
}

@article{kite2020atmosphere,
  title={Atmosphere origins for exoplanet sub-neptunes},
  author={Kite, Edwin S and Fegley Jr, Bruce and Schaefer, Laura and Ford, Eric B},
  journal={ApJ},
  volume={891},
  number={2},
  pages={111},
  year={2020},
  publisher={IOP Publishing}
}

@article{bourrier2018_55cnce,
  title={The 55 Cancri system reassessed},
  author={Bourrier, Vincent and Dumusque, Xavier and Dorn, Caroline and Henry, Gregory W and Astudillo-Defru, Nicola and Rey, Javiera and Benneke, Bj{\"o}rn and H{\'e}brard, Guillaume and Lovis, Christophe and Demory, Brice-Olivier and others},
  journal={A\&A},
  volume={619},
  pages={A1},
  year={2018},
  publisher={EDP Sciences}
}

@article{jindal2020characterization,
  title={Characterization of the Atmosphere of Super-Earth 55 Cancri e Using High-resolution Ground-based Spectroscopy},
  author={Jindal, Abhinav and de Mooij, Ernst JW and Jayawardhana, Ray and Deibert, Emily K and Brogi, Matteo and Rustamkulov, Zafar and Fortney, Jonathan J and Hood, Callie E and Morley, Caroline V},
  journal={AJ},
  volume={160},
  number={3},
  pages={101},
  year={2020},
  publisher={IOP Publishing}
}

@article{tsiaras2016detection,
  title={Detection of an atmosphere around the super-Earth 55 Cancri e},
  author={Tsiaras, A and Rocchetto, M and Waldmann, IP and Venot, O and Varley, R and Morello, G and Damiano, M and Tinetti, G and Barton, EJ and Yurchenko, SN and others},
  journal={ApJ},
  volume={820},
  number={2},
  pages={99},
  year={2016},
  publisher={IOP Publishing}
}

@article{bower2019linking,
  title={Linking the evolution of terrestrial interiors and an early outgassed atmosphere to astrophysical observations},
  author={Bower, Dan J and Kitzmann, Daniel and Wolf, Aaron S and Sanan, Patrick and Dorn, Caroline and Oza, Apurva V},
  journal={A\&A},
  volume={631},
  pages={A103},
  year={2019},
  publisher={EDP Sciences}
}

@article{charnoz2023effect,
  title={The effect of a small amount of hydrogen in the atmosphere of ultrahot magma-ocean planets: Atmospheric composition and escape},
  author={Charnoz, S{\'e}bastien and Falco, Aur{\'e}lien and Tremblin, Pascal and Sossi, Paolo and Caracas, Razvan and Lagage, Pierre-Olivier},
  journal={A\&A},
  volume={674},
  pages={A224},
  year={2023},
  publisher={EDP Sciences}
}

@article{wolf2023vaporock,
  title={VapoRock: thermodynamics of vaporized silicate melts for modeling volcanic outgassing and magma ocean atmospheres},
  author={Wolf, Aaron S and J{\"a}ggi, Noah and Sossi, Paolo A and Bower, Dan J},
  journal={ApJ},
  volume={947},
  number={2},
  pages={64},
  year={2023},
  publisher={IOP Publishing}
}

@article{elkins2008coreless,
  title={Coreless terrestrial exoplanets},
  author={Elkins-Tanton, Linda T and Seager, Sara},
  journal={ApJ},
  volume={688},
  number={1},
  pages={628},
  year={2008},
  publisher={IOP Publishing}
}

@article{peng2024puffy,
  title={Puffy Venuses: The Mass--Radius Impact of Carbon-rich Atmospheres on Lava Worlds},
  author={Peng, Bo and Valencia, Diana},
  journal={ApJ},
  volume={976},
  number={2},
  pages={202},
  year={2024},
  publisher={IOP Publishing}
}

@article{sossi2020redox,
  title={Redox state of Earth’s magma ocean and its Venus-like early atmosphere},
  author={Sossi, Paolo A and Burnham, Antony D and Badro, James and Lanzirotti, Antonio and Newville, Matt and O’neill, Hugh St C},
  journal={Sci. Adv.},
  volume={6},
  number={48},
  pages={eabd1387},
  year={2020},
  publisher={American Association for the Advancement of Science}
}

@article{seo2024role,
  title={Role of magma oceans in controlling carbon and oxygen of sub-Neptune atmospheres},
  author={Seo, Chanoul and Ito, Yuichi and Fujii, Yuka},
  journal={ApJ},
  volume={975},
  number={1},
  pages={14},
  year={2024},
  publisher={The American Astronomical Society}
}

@article{zilinskas2022observability,
  title={Observability of evaporating lava worlds},
  author={Zilinskas, Mantas and Van Buchem, CPA and Miguel, Yamila and Louca, Amy and Lupu, Roxana and Zieba, Sebastian and van Westrenen, Wim},
  journal={A\&A},
  volume={661},
  pages={A126},
  year={2022},
  publisher={EDP Sciences}
}

@article{arrhenius1889dissociationswarme,
  title={{\"U}ber die Dissociationsw{\"a}rme und den Einfluss der Temperatur auf den Dissociationsgrad der Elektrolyte},
  author={Arrhenius, Svante},
  journal={Z. Phys. Chem.},
  volume={4},
  number={1},
  pages={96--116},
  year={1889},
  publisher={De Gruyter Oldenbourg}
}

@article{patel2024jwst,
  title={JWST reveals the rapid and strong day-side variability of 55 Cancri e},
  author={Patel, JA and Brandeker, A and Kitzmann, D and dit de la Roche, DJM Petit and Bello-Arufe, A and Heng, K and Vald{\'e}s, E Meier and Persson, CM and Zhang, M and Demory, B-O and others},
  journal={A\&A},
  volume={690},
  pages={A159},
  year={2024},
  publisher={EDP Sciences}
}

@article{gaillard2022redox,
  title={Redox controls during magma ocean degassing},
  author={Gaillard, Fabrice and Bernadou, Fabien and Roskosz, Mathieu and Bouhifd, Mohamed Ali and Marrocchi, Yves and Iacono-Marziano, Giada and Moreira, Manuel and Scaillet, Bruno and Rogerie, Gregory},
  journal={Earth Planet. Sci. Lett.},
  volume={577},
  pages={117255},
  year={2022},
  publisher={Elsevier}
}

@article{zeng2016mass,
  title={Mass--radius relation for rocky planets based on PREM},
  author={Zeng, Li and Sasselov, Dimitar D and Jacobsen, Stein B},
  journal={ApJ},
  volume={819},
  number={2},
  pages={127},
  year={2016},
  publisher={IOP Publishing}
}

@article{zilinskas2023observability,
  title={Observability of silicates in volatile atmospheres of super-Earths and sub-Neptunes-Exploring the edge of the evaporation desert},
  author={Zilinskas, Mantas and Miguel, Yamila and van Buchem, CPA and Snellen, Ignas AG},
  journal={A\&A},
  volume={671},
  pages={A138},
  year={2023},
  publisher={EDP Sciences}
}

@article{piette2023rocky,
  title={Rocky planet or water world? Observability of low-density lava world atmospheres},
  author={Piette, Anjali AA and Gao, Peter and Brugman, Kara and Shahar, Anat and Lichtenberg, Tim and Miozzi, Francesca and Driscoll, Peter},
  journal={ApJ},
  volume={954},
  number={1},
  pages={29},
  year={2023},
  publisher={IOP Publishing}
}

@article{van2025lavatmos,
  title={LavAtmos 2.0-Incorporating volatile species in vaporisation models},
  author={Van Buchem, CPA and Zilinskas, Mantas and Miguel, Yamila and van Westrenen, Wim},
  journal={A\&A},
  volume={695},
  pages={A154},
  year={2025},
  publisher={EDP Sciences}
}

@article{dorn2021hidden,
  title={Hidden water in magma ocean exoplanets},
  author={Dorn, Caroline and Lichtenberg, Tim},
  journal={ApJL},
  volume={922},
  number={1},
  pages={L4},
  year={2021},
  publisher={IOP Publishing}
}

@incollection{palme2013cosmochemical,
  title={Cosmochemical estimates of mantle composition},
  author={Palme, H and O'Neill, HSTC},
  booktitle={The mantle and core},
  pages={1--39},
  year={2013},
  publisher={Elsevier}
}

@article{schaefer2004thermodynamic,
  title={A thermodynamic model of high temperature lava vaporization on Io},
  author={Schaefer, Laura and Fegley Jr, Bruce},
  journal={Icarus},
  volume={169},
  number={1},
  pages={216--241},
  year={2004},
  publisher={Elsevier}
}

@article{fegley1987vaporization,
  title={A vaporization model for iron/silicate fractionation in the Mercury protoplanet},
  author={Fegley Jr, Bruce and Cameron, AGW},
  journal={Earth Planet. Sci. Lett.},
  volume={82},
  number={3-4},
  pages={207--222},
  year={1987},
  publisher={Elsevier}
}

@article{molliere2019petitradtrans,
  title={petitRADTRANS-A Python radiative transfer package for exoplanet characterization and retrieval},
  author={Molli{\`e}re, P and Wardenier, JP and Van Boekel, R and Henning, Th and Molaverdikhani, K and Snellen, IAG},
  journal={A\&A},
  volume={627},
  pages={A67},
  year={2019},
  publisher={EDP Sciences}
}

@article{cherubim2025oxidation,
  title={An Oxidation Gradient Straddling the Small Planet Radius Valley},
  author={Cherubim, Collin and Wordsworth, Robin and Bower, Dan J and Sossi, Paolo A and Adams, Danica and Hu, Renyu},
  journal={ApJ},
  volume={983},
  number={2},
  pages={97},
  year={2025},
  publisher={IOP Publishing}
}

@article{nicholls2024magma,
  title={Magma ocean evolution at arbitrary redox state},
  author={Nicholls, Harrison and Lichtenberg, Tim and Bower, Dan J and Pierrehumbert, Raymond},
  journal={J. Geophys. Res.: Planets},
  volume={129},
  number={12},
  pages={e2024JE008576},
  year={2024},
  publisher={Wiley Online Library}
}

@article{nicholls2025convective,
  title={Convective shutdown in the atmospheres of lava worlds},
  author={Nicholls, Harrison and Pierrehumbert, Raymond T and Lichtenberg, Tim and Soucasse, Laurent and Smeets, Stef},
  journal={MNRAS},
  volume={536},
  number={3},
  pages={2957--2971},
  year={2025},
  publisher={Oxford University Press}
}

@article{Gueymard:2003,
title = {The sun’s total and spectral irradiance for Sol. Energy applications and solar radiation models},
journal = {Sol. Energy},
volume = {76},
number = {4},
pages = {423-453},
year = {2004},
issn = {0038-092X},
doi = {https://doi.org/10.1016/j.solener.2003.08.039},
url = {https://www.sciencedirect.com/science/article/pii/S0038092X03003967},
author = {Christian A. Gueymard},
}

@article{ledley1999climate,
  title={Climate change and greenhouse gases},
  author={Ledley, Tamara S and Sundquist, Eric T and Schwartz, Stephen E and Hall, Dorothy K and Fellows, Jack D and Killeen, Timothy L},
  journal={Eos Trans. AGU},
  volume={80},
  number={39},
  pages={453--458},
  year={1999},
  publisher={Wiley Online Library}
}

@article{bougher1999comparative,
  title={Comparative terrestrial planet thermospheres: 2. Solar cycle variation of global structure and winds at equinox},
  author={Bougher, SW and Engel, S and Roble, RG and Foster, B},
  journal={J. Geophys. Res.: Planets},
  volume={104},
  number={E7},
  pages={16591--16611},
  year={1999},
  publisher={Wiley Online Library}
}

@article{schlichting2022chemical,
  title={Chemical equilibrium between cores, mantles, and atmospheres of super-earths and sub-neptunes and implications for their compositions, interiors, and evolution},
  author={Schlichting, Hilke E and Young, Edward D},
  journal={Planet. Sci. J.},
  volume={3},
  number={5},
  pages={127},
  year={2022},
  publisher={IOP Publishing}
}

@article{zilinskas2025characterising,
  title={Characterising the atmosphere of 55 Cancri e-1D forward model grid for current and future JWST observations},
  author={Zilinskas, Mantas and van Buchem, CPA and Zieba, Sebastian and Miguel, Yamila and Sandford, Emily and Hu, Renyu and Patel, JA and Bello-Arufe, Aaron and Janssen, LJ and Tsai, S-M and others},
  journal={A\&A},
  volume={697},
  pages={A34},
  year={2025},
  publisher={EDP Sciences}
}

@article{hammond2017linking,
  title={Linking the climate and thermal phase curve of 55 Cancri e},
  author={Hammond, Mark and Pierrehumbert, Raymond T},
  journal={ApJ},
  volume={849},
  number={2},
  pages={152},
  year={2017},
  publisher={IOP Publishing}
}

@article{zhang2017effects,
  title={Effects of bulk composition on the atmospheric dynamics on close-in exoplanets},
  author={Zhang, Xi and Showman, Adam P},
  journal={ApJ},
  volume={836},
  number={1},
  pages={73},
  year={2017},
  publisher={IOP Publishing}
}

@article{demory2016variability,
  title={Variability in the super-Earth 55 Cnc e},
  author={Demory, Brice-Olivier and Gillon, Michael and Madhusudhan, Nikku and Queloz, Didier},
  journal={MNRAS},
  volume={455},
  number={2},
  pages={2018--2027},
  year={2016},
  publisher={Oxford University Press}
}

@article{demory2016map,
  title={A map of the large day--night temperature gradient of a super-Earth exoplanet},
  author={Demory, Brice-Olivier and Gillon, Michael and De Wit, Julien and Madhusudhan, Nikku and Bolmont, Emeline and Heng, Kevin and Kataria, Tiffany and Lewis, Nikole and Hu, Renyu and Krick, Jessica and others},
  journal={Nature},
  volume={532},
  number={7598},
  pages={207--209},
  year={2016},
  publisher={Nature Publishing Group UK London}
}

@article{hansen2008absorption,
  title={On the absorption and redistribution of energy in irradiated planets},
  author={Hansen, Brad MS},
  journal={ApJS},
  volume={179},
  number={2},
  pages={484},
  year={2008},
  publisher={IOP Publishing}
}

@article{salz2016energy,
  title={Energy-limited escape revised-The transition from strong planetary winds to stable thermospheres},
  author={Salz, M and Schneider, PC and Czesla, S and Schmitt, JHMM},
  journal={A\&A},
  volume={585},
  pages={L2},
  year={2016},
  publisher={EDP Sciences}
}

@article{grimm2015helios,
  title={HELIOS-K: an ultrafast, open-source opacity calculator for radiative transfer},
  author={Grimm, Simon L and Heng, Kevin},
  journal={ApJ},
  volume={808},
  number={2},
  pages={182},
  year={2015},
  publisher={IOP Publishing}
}

@article{grimm2021helios,
  title={HELIOS-K 2.0 opacity calculator and open-source opacity database for exoplanetary atmospheres},
  author={Grimm, Simon L and Malik, Matej and Kitzmann, Daniel and Guzm{\'a}n-Mesa, Andrea and Hoeijmakers, H Jens and Fisher, Chloe and Mendon{\c{c}}a, Jo{\~a}o M and Yurchenko, Sergey N and Tennyson, Jonathan and Alesina, Fabien and others},
  journal={ApJS},
  volume={253},
  number={1},
  pages={30},
  year={2021},
  publisher={IOP Publishing}
}

@article{nguyen2024clouds,
  title={Clouds in Partial Atmospheres of Lava Planets and Where to Find Them},
  author={Nguyen, T Giang and Cowan, Nicolas B and Dang, Lisa},
  journal={AJ},
  volume={168},
  number={6},
  pages={287},
  year={2024},
  publisher={IOP Publishing}
}

@article{nascimbeni2023new,
  title={A new dynamical modeling of the WASP-47 system with CHEOPS observations},
  author={Nascimbeni, Valerio and Borsato, Luca and Zingales, Tiziano and Piotto, Giampaolo and Pagano, Isabella and Beck, Mathias and Broeg, Christopher and Ehrenreich, David and Hoyer, Sergio and Majidi, Fatemeh Z and others},
  journal={A\&A},
  volume={673},
  pages={A42},
  year={2023},
  publisher={EDP Sciences}
}

@article{adibekyan2021compositional,
  title={A compositional link between rocky exoplanets and their host stars},
  author={Adibekyan, Vardan and Dorn, Caroline and Sousa, S{\'e}rgio G and Santos, Nuno C and Bitsch, Bertram and Israelian, Garik and Mordasini, Christoph and Barros, Susana CC and Delgado Mena, Elisa and Demangeon, Olivier DS and others},
  journal={Science},
  volume={374},
  number={6565},
  pages={330--332},
  year={2021},
  publisher={American Association for the Advancement of Science}
}

@article{katz2003new,
  title={A new parameterization of hydrous mantle melting},
  author={Katz, Richard F and Spiegelman, Marc and Langmuir, Charles H},
  journal={Geochem. Geophys. Geosyst.},
  volume={4},
  number={9},
  year={2003},
  publisher={Wiley Online Library}
}

@article{deibert2021near,
  title={A near-infrared chemical inventory of the atmosphere of 55 Cancri e},
  author={Deibert, Emily K and De Mooij, Ernst JW and Jayawardhana, Ray and Ridden-Harper, Andrew and Sivanandam, Suresh and Karjalainen, Raine and Karjalainen, Marie},
  journal={AJ},
  volume={161},
  number={5},
  pages={209},
  year={2021},
  publisher={IOP Publishing}
}

@article{wang2019enhanced,
  title={Enhanced constraints on the interior composition and structure of terrestrial exoplanets},
  author={Wang, Haiyang S and Liu, Fan and Ireland, Trevor R and Brasser, Ramon and Yong, David and Lineweaver, Charles H},
  journal={MNRAS},
  volume={482},
  number={2},
  pages={2222--2233},
  year={2019},
  publisher={Oxford University Press}
}

@article{wang2022detailed,
  title={Detailed chemical compositions of planet-hosting stars--II. Exploration of the interiors of terrestrial-type exoplanets},
  author={Wang, Haiyang S and Quanz, Sascha P and Yong, David and Liu, Fan and Seidler, Fabian and Acu{\~n}a, Lorena and Mojzsis, Stephen J},
  journal={MNRAS},
  volume={513},
  number={4},
  pages={5829--5846},
  year={2022},
  publisher={Oxford University Press}
}

@article{spaargaren2023plausible,
  title={Plausible constraints on the range of bulk terrestrial exoplanet compositions in the solar neighborhood},
  author={Spaargaren, Rob J and Wang, Haiyang S and Mojzsis, Stephen J and Ballmer, Maxim D and Tackley, Paul J},
  journal={ApJ},
  volume={948},
  number={1},
  pages={53},
  year={2023},
  publisher={IOP Publishing}
}

@article{leger2011extreme,
  title={The extreme physical properties of the CoRoT-7b super-Earth},
  author={L{\'e}ger, Alain and Grasset, O and Fegley, B and Codron, F and Albarede, AF and Barge, P and Barnes, R and Cance, P and Carpy, Sabrina and Catalano, F and others},
  journal={Icarus},
  volume={213},
  number={1},
  pages={1--11},
  year={2011},
  publisher={Elsevier}
}

@article{falco2024hydrogenated,
  title={Hydrogenated atmospheres of lava planets: Atmospheric structure and emission spectra},
  author={Falco, Aur{\'e}lien and Tremblin, Pascal and Charnoz, S{\'e}bastien and Ridgway, Robert J and Lagage, Pierre-Olivier},
  journal={A\&A},
  volume={683},
  pages={A194},
  year={2024},
  publisher={EDP Sciences}
}

@article{shorttle2024distinguishing,
  title={Distinguishing oceans of water from magma on mini-Neptune K2-18b},
  author={Shorttle, Oliver and Jordan, Sean and Nicholls, Harrison and Lichtenberg, Tim and Bower, Dan J},
  journal={ApJL},
  volume={962},
  number={1},
  pages={L8},
  year={2024},
  publisher={IOP Publishing}
}

@article{libourel2003nitrogen,
  title={Nitrogen solubility in basaltic melt. Part I. Effect of oxygen fugacity},
  author={Libourel, G and Marty, B and Humbert, F},
  journal={Geochim.~Cosmochim.~Acta},
  volume={67},
  number={21},
  pages={4123--4135},
  year={2003},
  publisher={Elsevier}
}

@article{dasgupta2022fate,
  title={The fate of nitrogen during parent body partial melting and accretion of the inner solar system bodies at reducing conditions},
  author={Dasgupta, Rajdeep and Falksen, Emily and Pal, Aindrila and Sun, Chenguang},
  journal={Geochim.~Cosmochim.~Acta},
  volume={336},
  pages={291--307},
  year={2022},
  publisher={Elsevier}
}

@article{maurice2024volatile,
  title={Volatile atmospheres of lava worlds},
  author={Maurice, Maxime and Dasgupta, Rajdeep and Hassanzadeh, Pedram},
  journal={A\&A},
  volume={688},
  pages={A47},
  year={2024},
  publisher={edp Sciences}
}

@article{dash2025detectability,
  title={Detectability of oxygen fugacity regimes in the magma ocean world 55 Cancri e at high spectral resolution},
  author={Dash, Spandan and Brogi, Matteo and Seidler, Fabian Lukas and Sossi, Paolo A and Gandhi, Siddharth and Panwar, Vatsal and Lafarga, Marina and Wheatley, Peter J},
  journal={MNRAS},
  volume={538},
  number={4},
  pages={3042--3066},
  year={2025},
  publisher={Oxford University Press}
}

@article{zhang2025only,
  title={The only known atmosphere on a rocky exoplanet?},
  author={Zhang, Michael and Savel, Arjun Baliga and Steinrueck, Maria E and Bean, Jacob L and Coy, Brandon Park and Fu, Guangwei and Hu, Renyu and Ih, Jegug and Kempton, Eliza M-R and Kite, Edwin S and others},
  journal={JWST Proposal. Cycle 4},
  pages={7875},
  year={2025}
}

@article{hirschmann2021iron,
  title={Iron-w{\"u}stite revisited: A revised calibration accounting for variable stoichiometry and the effects of pressure},
  author={Hirschmann, MM},
  journal={Geochim.~Cosmochim.~Acta},
  volume={313},
  pages={74--84},
  year={2021},
  publisher={Elsevier}
}

@article{amundsen2017treatment,
  title={Treatment of overlapping gaseous absorption with the correlated-k method in hot Jupiter and brown dwarf atmosphere models},
  author={Amundsen, David S and Tremblin, Pascal and Manners, James and Baraffe, Isabelle and Mayne, Nathan J},
  journal={A\&A},
  volume={598},
  pages={A97},
  year={2017},
  publisher={EDP Sciences}
}

@article{holland1991compensated,
  title={A Compensated-Redlich-Kwong (CORK) equation for volumes and fugacities of CO2 and H2O in the range 1 bar to 50 kbar and 100--1600 C},
  author={Holland, Tim and Powell, Roger},
  journal={Contrib. Mineral. Petrol.},
  volume={109},
  number={2},
  pages={265--273},
  year={1991},
  publisher={Springer}
}

@article{shi1992thermodynamic,
  title={Thermodynamic modeing of the CHOS fluid system},
  author={Shi, Pingfang and Saxena, SK},
  journal={Am. Mineral.},
  volume={77},
  number={9-10},
  pages={1038--1049},
  year={1992},
  publisher={Mineralogical Society of America}
}

@article{gillmann2024interior,
  title={Interior controls on the habitability of rocky planets},
  author={Gillmann, Cedric and Hakim, Kaustubh and Louren{\c{c}}o, Diogo and Quanz, Sascha P and Sossi, Paolo A},
  journal={Space Sci. Technol.},
  volume={4},
  pages={0075},
  year={2024},
  publisher={AAAS}
}

@article{luque2022density,
  title={Density, not radius, separates rocky and water-rich small planets orbiting M dwarf stars},
  author={Luque, Rafael and Pall{\'e}, Enric},
  journal={Science},
  volume={377},
  number={6611},
  pages={1211--1214},
  year={2022},
  publisher={American Association for the Advancement of Science}
}

@article{carroll1994noble,
  title={Noble gases as trace elements in magmatic processes},
  author={Carroll, Michael R and Draper, David S},
  journal={Chem. Geol.},
  volume={117},
  number={1-4},
  pages={37--56},
  year={1994},
  publisher={Elsevier}
}

@article{foustoukos2025molecular,
  title={Molecular H2 in silicate melts},
  author={Foustoukos, Dionysis I},
  journal={Geochim.~Cosmochim.~Acta},
  volume={389},
  pages={125--135},
  year={2025},
  publisher={Elsevier}
}

@article{hirschmann2012solubility,
  title={Solubility of molecular hydrogen in silicate melts and consequences for volatile evolution of terrestrial planets},
  author={Hirschmann, Marc M and Withers, AC and Ardia, P and Foley, NT},
  journal={Earth Planet. Sci. Lett.},
  volume={345},
  pages={38--48},
  year={2012},
  publisher={Elsevier}
}

@article{hughes2023sulfur,
  title={The sulfur solubility minimum and maximum in silicate melt},
  author={Hughes, Ery C and Saper, Lee M and Liggins, Philippa and O'Neill, Hugh St C and Stolper, Edward M},
  journal={J. Geol. Soc.},
  volume={180},
  number={3},
  pages={jgs2021--125},
  year={2023},
  publisher={The Geological Society of London}
}

@article{misener2023atmospheres,
  title={Atmospheres as windows into sub-Neptune interiors: coupled chemistry and structure of hydrogen--silane--water envelopes},
  author={Misener, William and Schlichting, Hilke E and Young, Edward D},
  journal={MNRAS},
  volume={524},
  number={1},
  pages={981--992},
  year={2023},
  publisher={Oxford University Press}
}

@article{bello2025evidence,
  title={Evidence for a Volcanic Atmosphere on the Sub-Earth L 98-59 b},
  author={Bello-Arufe, Aaron and Damiano, Mario and Bennett, Katherine A and Hu, Renyu and Welbanks, Luis and MacDonald, Ryan J and Seligman, Darryl Z and Sing, David K and Tokadjian, Armen and Oza, Apurva V and others},
  journal={ApJL},
  volume={980},
  number={2},
  pages={L26},
  year={2025},
  publisher={IOP Publishing}
}

@article{jambon1986solubility,
  title={Solubility of He, Ne, Ar, Kr and Xe in a basalt melt in the range 1250--1600 C. Geochemical implications},
  author={Jambon, Albert and Weber, Hartwig and Braun, Otto},
  journal={Geochim.~Cosmochim.~Acta},
  volume={50},
  number={3},
  pages={401--408},
  year={1986},
  publisher={Elsevier}
}

@article{polyansky2018exomol,
  title={ExoMol molecular line lists XXX: a complete high-accuracy line list for water},
  author={Polyansky, Oleg L and Kyuberis, Aleksandra A and Zobov, Nikolai F and Tennyson, Jonathan and Yurchenko, Sergei N and Lodi, Lorenzo},
  journal={MNRAS},
  volume={480},
  number={2},
  pages={2597--2608},
  year={2018},
  publisher={Oxford University Press}
}

@article{yurchenko2022exomol,
  title={ExoMol line lists--XLIV. Infrared and ultraviolet line list for silicon monoxide (28Si16O)},
  author={Yurchenko, Sergei N and Tennyson, Jonathan and Syme, Anna-Maree and Adam, Ahmad Y and Clark, Victoria HJ and Cooper, Bridgette and Dobney, C Pria and Donnelly, Shaun TE and Gorman, Maire N and Lynas-Gray, Anthony E and others},
  journal={MNRAS},
  volume={510},
  number={1},
  pages={903--919},
  year={2022},
  publisher={Oxford University Press}
}

@article{li2019exomol,
  title={ExoMol line lists--XXXII. The rovibronic spectrum of MgO},
  author={Li, Heng Ying and Tennyson, Jonathan and Yurchenko, Sergei N},
  journal={MNRAS},
  volume={486},
  number={2},
  pages={2351--2365},
  year={2019},
  publisher={Oxford University Press}
}

@article{chubb2020exomol,
  title={ExoMol molecular line lists--XXXVII. Spectra of acetylene},
  author={Chubb, Katy L and Tennyson, Jonathan and Yurchenko, Sergei N},
  journal={MNRAS},
  volume={493},
  number={2},
  pages={1531--1545},
  year={2020},
  publisher={Oxford University Press}
}

@article{brady2024exomol,
  title={ExoMol line lists--LVI. The SO line list, MARVEL analysis of experimental transition data and refinement of the spectroscopic model},
  author={Brady, Ryan P and Yurchenko, Sergei N and Tennyson, Jonathan and Kim, Gap-Sue},
  journal={MNRAS},
  volume={527},
  number={3},
  pages={6675--6690},
  year={2024},
  publisher={Oxford University Press}
}

@article{yurchenko2014exomol,
  title={ExoMol line lists--IV. The rotation--vibration spectrum of methane up to 1500 K},
  author={Yurchenko, Sergei N and Tennyson, Jonathan},
  journal={MNRAS},
  volume={440},
  number={2},
  pages={1649--1661},
  year={2014},
  publisher={The Royal Astronomical Society}
}

@article{li2015rovibrational,
  title={Rovibrational line lists for nine isotopologues of the CO molecule in the X1$\Sigma$+ ground electronic state},
  author={Li, Gang and Gordon, Iouli E and Rothman, Laurence S and Tan, Yan and Hu, Shui-Ming and Kassi, Samir and Campargue, Alain and Medvedev, Emile S},
  journal={ApJS},
  volume={216},
  number={1},
  pages={15},
  year={2015},
  publisher={IOP Publishing}
}

@article{yurchenko2020exomol,
  title={ExoMol line lists--XXXIX. Ro-vibrational molecular line list for CO2},
  author={Yurchenko, SN and Mellor, Thomas M and Freedman, Richard S and Tennyson, J},
  journal={MNRAS},
  volume={496},
  number={4},
  pages={5282--5291},
  year={2020},
  publisher={Oxford University Press}
}

@article{qu2021exomol,
  title={Exomol molecular line lists--XLII. Rovibronic molecular line list for the low-lying states of NO},
  author={Qu, Qianwei and Yurchenko, Sergei N and Tennyson, Jonathan},
  journal={MNRAS},
  volume={504},
  number={4},
  pages={5768--5777},
  year={2021},
  publisher={Oxford University Press}
}

@article{roueff2019full,
  title={The full infrared spectrum of molecular hydrogen},
  author={Roueff, E and Abgrall, H and Czachorowski, P and Pachucki, K and Puchalski, Mariusz and Komasa, Jacek},
  journal={A\&A},
  volume={630},
  pages={A58},
  year={2019},
  publisher={EDP Sciences}
}

@article{kurucz2017including,
  title={Including all the lines: data releases for spectra and opacities},
  author={Kurucz, Robert L},
  journal={Can. J. Phys.},
  volume={95},
  number={9},
  pages={825--827},
  year={2017},
  publisher={NRC Research Press}
}

@article{mitev2025exomol,
  title={ExoMol line lists--LXI. A trihybrid line list for rovibronic transitions of the hydroxyl radical (OH)},
  author={Mitev, Georgi B and Bowesman, Charles A and Zhang, Jingxin and Yurchenko, Sergei N and Tennyson, Jonathan},
  journal={MNRAS},
  volume={536},
  number={4},
  pages={3401--3420},
  year={2025},
  publisher={Oxford University Press}
}

@article{western2018spectrum,
  title={The spectrum of N2 from 4,500 to 15,700 cm- 1 revisited with PGOPHER},
  author={Western, Colin M and Carter-Blatchford, Luke and Crozet, Patrick and Ross, Amanda J and Morville, J{\'e}r{\^o}me and Tokaryk, Dennis W},
  journal={J. Quant. Spectrosc. Radiat. Transfer},
  volume={219},
  pages={127--141},
  year={2018},
  publisher={Elsevier}
}

@article{underwood2016exomol,
  title={ExoMol molecular line lists--XIV. The rotation--vibration spectrum of hot SO2},
  author={Underwood, Daniel S and Tennyson, Jonathan and Yurchenko, Sergei N and Huang, Xinchuan and Schwenke, David W and Lee, Timothy J and Clausen, S{\o}nnik and Fateev, Alexander},
  journal={MNRAS},
  volume={459},
  number={4},
  pages={3890--3899},
  year={2016},
  publisher={Oxford University Press}
}

@article{azzam2016exomol,
  title={ExoMol molecular line lists--XVI. The rotation--vibration spectrum of hot H2S},
  author={Azzam, Ala'a AA and Tennyson, Jonathan and Yurchenko, Sergei N and Naumenko, Olga V},
  journal={MNRAS},
  volume={460},
  number={4},
  pages={4063--4074},
  year={2016},
  publisher={Oxford University Press}
}

@article{gorman2019exomol,
  title={ExoMol molecular line lists XXXVI: X 2$\Pi$--X 2$\Pi$ and A 2$\Sigma$+--X 2$\Pi$ transitions of SH},
  author={Gorman, Maire N and Yurchenko, Sergei N and Tennyson, Jonathan},
  journal={MNRAS},
  volume={490},
  number={2},
  pages={1652--1665},
  year={2019},
  publisher={Oxford University Press}
}

@article{gordon2017hitran2016,
  title={The HITRAN2016 molecular spectroscopic database},
  author={Gordon, Iouli E and Rothman, Laurence S and Hill, Christian and Kochanov, Roman V and Tan, Y and Bernath, Peter F and Birk, Manfred and Boudon, V and Campargue, Alain and Chance, KV and others},
  journal={J. Quant. Spectrosc. Radiat. Transfer},
  volume={203},
  pages={3--69},
  year={2017},
  publisher={Elsevier}
}

@inproceedings{pandeia,
author = {Klaus M. Pontoppidan and Timothy E. Pickering and Victoria G. Laidler and Karoline Gilbert and Christopher D. Sontag and Christine Slocum and Mark J. Sienkiewicz Jr. and Christopher Hanley and Nicholas M. Earl and Laurent Pueyo and Swara Ravindranath and Diane M. Karakla and Massimo Robberto and Alberto Noriega-Crespo and Elizabeth A. Barker},
title = {{Pandeia: a multi-mission exposure time calculator for JWST and WFIRST}},
volume = {9910},
booktitle = {Observatory Operations: Strategies, Processes, and Systems VI},
editor = {Alison B. Peck and Robert L. Seaman and Chris R. Benn},
organization = {International Society for Optics and Photonics},
publisher = {SPIE},
pages = {991016},
year = {2016},
doi = {10.1117/12.2231768},
URL = {https://doi.org/10.1117/12.2231768}
}

@online{refdata,
  author = {Space Telescope Science Institute {STScI}},
  title = {pandeia reference data v4.0},
  year = {2024},
  url = {https://stsci.app.box.com/v/pandeia-refdata-v4p0-jwst},
  urldate = {2024-10-13}
}

@article{zhao2023measured,
  title={Measured spin--orbit alignment of ultra-short-period super-Earth 55 Cancri e},
  author={Zhao, Lily L and Kunovac, Vedad and Brewer, John M and Llama, Joe and Millholland, Sarah C and Hedges, Christina and Szymkowiak, Andrew E and Roettenbacher, Rachael M and Cabot, Samuel HC and Weiss, Sam A and others},
  journal={Nat. Astron},
  volume={7},
  number={2},
  pages={198--205},
  year={2023},
  publisher={Nature Publishing Group UK London}
}

@article{aguichine2021mass,
  title={Mass--radius relationships for irradiated ocean planets},
  author={Aguichine, Artyom and Mousis, Olivier and Deleuil, Magali and Marcq, Emmanuel},
  journal={ApJ},
  volume={914},
  number={2},
  pages={84},
  year={2021},
  publisher={IOP Publishing}
}

@article{nicholls2025self,
  title={Self-limited tidal heating and prolonged magma oceans in the L 98-59 system},
  author={Nicholls, Harrison and Guimond, Claire Marie and Hay, Hamish CFC and Chatterjee, Richard D and Lichtenberg, Tim and Pierrehumbert, Raymond T},
  journal={MNRAS},
  volume={541},
  number={3},
  pages={2566--2584},
  year={2025},
  publisher={Oxford University Press}
}

@article{matsukage2005density,
  title={Density of hydrous silicate melt at the conditions of Earth's deep upper mantle},
  author={Matsukage, Kyoko N and Jing, Zhicheng and Karato, Shun-ichiro},
  journal={Nature},
  volume={438},
  number={7067},
  pages={488--491},
  year={2005},
  publisher={Nature Publishing Group UK London}
}

@article{boley2023fizzy,
  title={Fizzy super-Earths: impacts of magma composition on the bulk density and structure of lava worlds},
  author={Boley, Kiersten M and Panero, Wendy R and Unterborn, Cayman T and Schulze, Joseph G and Mart{\'\i}nez, Romy Rodr{\'\i}guez and Wang, Ji},
  journal={ApJ},
  volume={954},
  number={2},
  pages={202},
  year={2023},
  publisher={IOP Publishing}
}

@article{lous2024accretion,
  title={Accretion of primordial H--He atmospheres in mini-Neptunes: The importance of envelope enrichment},
  author={Mol Lous, Marit and Mordasini, Christoph and Helled, Ravit},
  journal={A\&A},
  volume={685},
  pages={A22},
  year={2024},
  publisher={EDP Sciences}
}

@article{zhang2021no,
  title={No escaping helium from 55 Cnc e},
  author={Zhang, Michael and Knutson, Heather A and Wang, Lile and Dai, Fei and Oklopcic, Antonija and Hu, Renyu},
  journal={AJ},
  volume={161},
  number={4},
  pages={181},
  year={2021},
  publisher={IOP Publishing}
}

@article{ehrenreich2012hint,
  title={Hint of a transiting extended atmosphere on 55 Cancri b},
  author={Ehrenreich, David and Bourrier, Vincent and Bonfils, Xavier and Des Etangs, A Lecavelier and H{\'e}brard, Guillaume and Sing, David K and Wheatley, Peter J and Vidal-Madjar, Alfred and Delfosse, Xavier and Udry, St{\'e}phane and others},
  journal={A\&A},
  volume={547},
  pages={A18},
  year={2012},
  publisher={EDP Sciences}
}

@article{rasmussen2023nondetection,
  title={A nondetection of iron in the first high-resolution emission study of the lava planet 55 Cnc e},
  author={Rasmussen, Kaitlin C and Currie, Miles H and Hagee, Celeste and van Buchem, Christiaan and Malik, Matej and Savel, Arjun B and Brogi, Matteo and Rauscher, Emily and Meadows, Victoria and Mansfield, Megan and others},
  journal={AJ},
  volume={166},
  number={4},
  pages={155},
  year={2023},
  publisher={IOP Publishing}
}

@article{esteves2017search,
  title={A search for water in a super-earth atmosphere: high-resolution optical spectroscopy of 55Cancri e},
  author={Esteves, Lisa J and De Mooij, Ernst JW and Jayawardhana, Ray and Watson, Chris and De Kok, Remco},
  journal={AJ},
  volume={153},
  number={6},
  pages={268},
  year={2017},
  publisher={IOP Publishing}
}

@article{keles2022pepsi,
  title={The PEPSI exoplanet transit survey (PETS) I: investigating the presence of a silicate atmosphere on the super-earth 55 Cnc e},
  author={Keles, Engin and Mallonn, Matthias and Kitzmann, Daniel and Poppenhaeger, Katja and Hoeijmakers, H Jens and Ilyin, Ilya and Alexoudi, Xanthippi and Carroll, Thorsten A and Alvarado-Gomez, Julian and Ketzer, Laura and others},
  journal={MNRAS},
  volume={513},
  number={1},
  pages={1544--1556},
  year={2022},
  publisher={Oxford University Press}
}

@article{janssen2026hot,
  title={Hot and cloudy: High temperature clouds in super-Earths and sub-Neptunes},
  author={Janssen, Leoni J. and Miguel, Yamila and Min, Michiel and Huang, Helong and Zilinskas, Mantas and van Buchem, Christiaan},
  journal={MNRAS},
  year={2026},
  note={accepted (arXiv:2601.15927)},
  eprint={2601.15927},
  archivePrefix={arXiv},
  primaryClass={astro-ph.EP}
}

@article{teske2025thick,
  title={A Thick Volatile Atmosphere on the Ultrahot Super-Earth TOI-561 b},
  author={Teske, Johanna K and Wallack, Nicole L and Piette, Anjali AA and Dang, Lisa and Lichtenberg, Tim and Plotnykov, Mykhaylo and Pierrehumbert, Raymond and Postolec, Emma and Boucher, Samuel and McGinty, Alex and others},
  journal={ApJL},
  volume={995},
  number={2},
  pages={L39},
  year={2025},
  publisher={IOP Publishing}
}

@article{guillot2010radiative,
  title={On the radiative equilibrium of irradiated planetary atmospheres},
  author={Guillot, Tristan},
  journal={A\&A},
  volume={520},
  pages={A27},
  year={2010},
  publisher={EDP Sciences}
}

@ARTICLE{hakim2026silane,
       author = {{Hakim}, Kaustubh and {Bower}, Dan J. and {Seidler}, Fabian L. and {Sossi}, Paolo A.},
        title = "{Silane─methane competition in sub-Neptune atmospheres as a diagnostic of metallicity and magma oceans}",
      journal = {\mnras},
         year = 2026,
        month = feb,
       volume = {546},
       number = {2},
          eid = {stag133},
        pages = {stag133},
          doi = {10.1093/mnras/stag133},
archivePrefix = {arXiv},
       eprint = {2508.19235},
 primaryClass = {astro-ph.EP},
       adsurl = {https://ui.adsabs.harvard.edu/abs/2026MNRAS.546ag133H}
}

\begin{appendix}

\onecolumn

\section{Chemical networks}
\label{apx:atmosphere_interior_exchange}

\begin{table*}[!h]
\centering
\caption{Target species with solubility laws for the \atmodeller chemical network. Table adopted from \citet{bower2025diversity}.}
\begin{tabular}{lccccc}
\toprule
Species & Solubility law & Composition & Pressure & Temperature & fO$_2$ rel. IW \\
& & & (kbar) & (K) & (log10 units) \\
\midrule
He & \citet{jambon1986solubility} & Basalt (tholeiitic) & 0.001 & 1523-1873 & 7.3 to 10.7 \\
H$_2$ & \citet{hirschmann2012solubility} & Basalt, Andesite & 7-30 & 1673-1773 & -1 to 3.8 \\
H$_2$O & \citet{sossi2023solubility} & Peridotite & 0.001 & 2173 & -1.9 to 6.0 \\
CO & \citealt{yoshioka2019} & Basalt, Rhyolite & 2-30 & 1473-1773 & 1 to 3.8\\
CO$_2$ & \citet{dixon1995} & Basalt & 0.21-0.98 & 1473 & 4.2 to 5.5 \\
CH$_4$ & \citet{ardia2013} & Basalt (Fe-free) & 7-30 & 1673-1723 & -9.50 to -1.36 \\
C$_2$H$_2$ & no sol. & - & - & - & - \\
HCN & no sol. & - & - & - & - \\
N$_2$ & \citet{dasgupta2022fate}  & Basalt & 0.001-82 & 1323-2600 & -8.3 to 8.7 \\
NH$_3$ & no sol. & - & - & - & - \\
{S$_2$}$^\ddagger$ & \citet{BW22, BW23} & Basalt, Andesite & 0.001 & 1473-1773 & -0.14 to 10.9 \\
SO & no sol. & - & - & - & - \\
SO$_2$ & no sol. & - & - & - & - \\
HS & no sol. & - & - & - & - \\
H$_2$S & no sol. & - & - & - & - \\
O$_2$ & no sol. & - & - & - & - \\
SiO & no sol. & - & - & - & - \\
SiH$_4$ & no sol. & - & - & - & - \\
MgO & no sol. & - & - & - & - \\
Mg & no sol. & - & - & - & - \\
Fe & no sol. & - & - & - & - \\
FeO & no sol. & - & - & - & - \\
graphite (cr) & - & - & - & - & - \\
\bottomrule
\end{tabular}
\label{table:target_species}
\end{table*}

\begin{table}[!h]
\centering
\caption{Reactions used in \texttt{FastChem} (see Sec. \ref{methods:pipeline:radtrans}).}
\label{tab:fastchem_reactions}

\begin{tabular}{lll}
\hline
Elem & Gas species & Condensates \\
\hline
H    & \begin{tabular}[c]{@{}l@{}}H, H$_{2}$, H$_{2}$+, H$_{2}$-, H$_{2}$MgO$_{2}$, H$_{2}$N, H$_{2}$N$_{2}$, H$_{2}$O, H$_{2}$O$_{2}$, H$_{2}$S,\\ H$_{2}$SO$_{4}$, H$_{2}$Si, H$_{3}$O+, H$_{3}$N, H$_{3}$Si, H$_{4}$N$_{2}$, H$_{4}$Si, H+, HO+, H-,\\ HO-, HMg, HMgO+, HMgO, HN, HNO, HNO$_{2}$(trans), HNO$_{3}$, HNO(cis), HO, \\ HO$_{2}$, HS, HS-, HSi, HSi+\end{tabular}                                                                                                                                                                                                                                                                                   & \begin{tabular}[c]{@{}l@{}}H$_{10}$SO$_{8}$(s,l), H$_{2}$Mg(s), H$_{2}$MgO$_{2}$(s),\\ H$_{2}$O(s,l), H$_{2}$SO$_{4}$(s,l), H$_{3}$N(s,l),\\ H$_{4}$N$_{2}$(l), H$_{4}$SO$_{5}$(s,l), H$_{6}$SO$_{6}$(s,l),\\ H$_{8}$SO$_{7}$(s,l)\end{tabular} \\
\hline
He   & He, He+                                                                                                                                                                                                                                                                                                                                                                                                                                                                                                                                                                                                                                                      &                                                                                                                                                                                                                                                 \\
\hline
C    & \begin{tabular}[c]{@{}l@{}}C, C$_{2}$, C$_{2}$-, C$_{2}$H, C$_{2}$H$_{2}$, C$_{2}$H$_{2}$O$_{2}$, C$_{2}$H$_{2}$O$_{4}$, C$_{2}$H$_{4}$, C$_{2}$H$_{4}$O,\\ C$_{2}$H$_{4}$O$_{3}$, C$_{2}$H$_{6}$O$_{2}$, C$_{2}$N, C$_{2}$N$_{2}$, C$_{2}$NO, C$_{2}$O, C$_{2}$Si, C$_{2}$Si$_{2}$, C$_{3}$,\\ C$_{3}$H, C$_{3}$N$_{2}$O, C$_{3}$O$_{2}$, C$_{4}$, C$_{4}$H$_{6}$O$_{4}$, C$_{4}$N$_{2}$, C$_{5}$, C$_{5}$FeO$_{5}$, \\ C+, C-, CO$_{2}$-, CH, CH$_{2}$, CH$_{2}$O, CH$_{3}$, CH$_{4}$, CH$_{4}$O$_{2}$, CH+, CHO+, CH-, \\ CHN, CHNO, CHO, CN, CN$_{2}$(cnn), CN$_{2}$(ncn), CN+, CN-, CNO, CO, CO$_{2}$,\\ CS, CS$_{2}$, CSO, CSi, CSi$_{2}$\end{tabular} & \begin{tabular}[c]{@{}l@{}}C$_{2}$Mg(s), C$_{3}$Mg$_{2}$(s), C$_{5}$FeO$_{5}$(l),\\ C(s), CH$_{4}$(s,l), CMgO$_{3}$(s),\\ CO$_{2}$(s,l), CO(l), CSi(s)\end{tabular}                                                                             \\
\hline
N    & \begin{tabular}[c]{@{}l@{}}N, N$_{2}$, N$_{2}$+, N$_{2}$O+, N$_{2}$-, N$_{2}$O, N$_{2}$O$_{3}$, N$_{2}$O$_{4}$, N$_{2}$O$_{5}$, N$_{3}$,\\ N+, NO+, N-, NO$_{2}$-, NO, NO$_{2}$, NO$_{3}$, NS, NSi, NSi$_{2}$\end{tabular}                                                                                                                                                                                                                                                                                                                                                                                                                                   & N$_{2}$(s,l), N$_{2}$O$_{4}$(s,l), N$_{4}$Si$_{3}$(s)                                                                                                                                                                                           \\
\hline
O    & O, O$_{2}$, O$_{3}$, O+, O$_{2}$+, O-, O$_{2}$-                                                                                                                                                                                                                                                                                                                                                                                                                                                                                                                                                                                                              &                                                                                                                                                                                                                                                 \\
\hline
Mg   & Mg, Mg$_{2}$, Mg+, MgN, MgO, MgS                                                                                                                                                                                                                                                                                                                                                                                                                                                                                                                                                                                                                             & \begin{tabular}[c]{@{}l@{}}Mg$_{2}$Si(s,l), Mg$_{2}$SiO$_{4}$(s,l), Mg$_{3}$N$_{2}$(s), \\ Mg(s,l), MgO(s,l), MgS(s),\\  MgSO$_{4}$(s,l), MgSiO$_{3}$(s,l)\end{tabular}                                                                          \\
\hline
Si   & Si, Si$_{2}$, Si$_{3}$, Si+, Si-, SiO, SiO$_{2}$, SiS                                                                                                                                                                                                                                                                                                                                                                                                                                                                                                                                                                                                        & Si(s,l), SiO$_{2}$(s,l)                                                                                                                                                                                                                         \\
\hline
S    & S, S$_{2}$, S$_{2}$O, S$_{3}$, S$_{4}$, S$_{5}$, S$_{6}$, S$_{7}$, S$_{8}$, S+, S-, SO, SO$_{2}$, SO$_{3}$                                                                                                                                                                                                                                                                                                                                                                                                                                                                                                         & S$_{2}$Si(s,l), S(s,l)                                                                                                                                                                                                                          \\
\hline
Fe   & Fe, Fe+, Fe-, FeH, FeH$_{2}$O$_{2}$, FeO, FeS                                                                                                                                                                                                                                                                                                                                                                                                                                                                                                                                                                                                                & \begin{tabular}[c]{@{}l@{}}Fe$_{2}$O$_{3}$(s), Fe$_{2}$S$_{3}$O$_{12}$(s), Fe$_{2}$SiO$_{4}$(s),\\ Fe$_{3}$O$_{4}$(s), Fe(s,l), FeH$_{2}$O$_{2}$(s),\\ FeH$_{3}$O$_{3}$(s), FeO(s,l), FeS$_{2}$(s),\\ FeS(s,l), FeSO$_{4}$(s)
\end{tabular} \\
\hline
\end{tabular}

\end{table}

\twocolumn

\FloatBarrier
\section{More details on \texttt{phaethon}}

\subsection{root finding}
\label{apx:rootfinding}

As seen in Sec. \ref{methods:pipeline}, the melt temperature (equivalent to $T_\text{MAI}$) sets the atmospheric composition, which in turn affects the P–T profile and thus $T_\text{BOA}$, which must equal $T_\text{MAI}$ in equilibrium. This circular dependency is resolved by finding the root of the residual $\Delta T = T_\text{BOA} - T_\text{MAI}$ via iterative P–T profile computations. Because radiative transfer calculations are computationally expensive, minimising the number of iterations is essential.

We find heuristically that $\Delta T$, to first order, behaves approximately linearly with $T_\text{MAI}$, which we exploit in the root-finding algorithm. During each iteration $n$, \texttt{phaethon} records the pair $(T_{\text{MAI},n}, \Delta T_n)$, where $\Delta T_n := T_{\mathrm{BOA},n} - T_{\text{MAI},n}$. The initial guess $T_{\text{MAI},1}$ is typically the planets average dayside temperature (or a user-defined value, if desired). The second estimate is set to $T_{\text{MAI},2} = T_{\mathrm{BOA},1}$.

In subsequent iterations, if a sign change in $\Delta T_n$ is detected (relative to any previous $\Delta T_k$, $k<n$), the next $T_{\text{MAI},n+1}$ is computed by finding the root of a linear interpolation between $(T_{\text{MAI},n}, \Delta T_n)$ and $(T_{\text{MAI},m^\ast}, \Delta T_{m^\ast})$, where $m^\ast$ is defined such that $T_{\text{MAI},m^\ast}$ is closest to $T_{\text{MAI},n}$, i.e. $m^\ast = \underset{m \neq n}{\arg\min} \left| T_{\text{MAI},n} - T_{\text{MAI},m} \right|$ and $\operatorname{sgn}(\Delta T_{n}) = -\operatorname{sgn}(\Delta T_{m^\ast})$. 
If no sign change occurs, the algorithm proceeds with a secant-like method by linearly extrapolation the last two residuals, $(T_{\text{MAI},n-1}, \Delta T_{n-1})$ and $(T_{\text{MAI},n}, \Delta T_n)$, and defining the root of this extrapolation as $T_{\text{MAI}, n+1}$.
If the algorithm detects cycling or fails to converge within a predefined number of iterations, it switches to Bayesian optimisation \citep{bayes_opt} to escape the local region, and resumes the linear search once a new viable estimate is found.
Convergence is reached when $|\Delta T_n| \leq \Delta T_\mathrm{tol}$, typically within 3–5 iterations. Final residuals are often below 10 K, and in many cases below 5 K.

\subsection{Wavelength dependent radius}
\label{apx:radius_wavelength_dependent}

The occultation depth is written as

\begin{equation}
    \label{eq:occultation_depth}
    \frac{F_p}{F_\star} = \left(\frac{R_p}{R_\star} \right)^2 \frac{\varepsilon_p(\lambda)}{\varepsilon_\star(\lambda)}
\end{equation}

where $F$ is the flux ($J/m^2$) of the respective body emitted at its surface, $R$ its radius and $\varepsilon$ the emissivity, in $J/m^{-3}$. Most often, the assumption of $R_p= \text{const}$ is made, which is valid under the assumption that the atmosphere is significantly smaller than the planet itself. However, in many models shown in this study, we find that the atmospheric extension presents itself as a significant portion of the planets size, especially in the hydrogen-rich and reduced cases (see Fig. \ref{fig:spectra_transmission}). Further, as we see in the opacity structure (Fig. \ref{fig:atmospheric_structure}), the photosphere changes its height with wavelength; for extended atmospheres, this effect can be drastic.
As a result, the planet will appear bigger in wavebands where the photosphere is higher up and smaller in cases were it lies at depth. This will have an effect on the spectral features based on Eq. \ref{eq:occultation_depth}. To account for this, we directly find $F_p(\lambda)$ by integrating the contribution function times the altitude in each wavelength:

\begin{equation}
    \label{eq:wavl_dependent_radius}
    F_p(\lambda) = 4 \pi \int_0^{z_{max}} c(z, \lambda) \cdot (R_c + x)^2 dz
\end{equation}

where $z$ is the altitude above the surface ($z=0$ corresponds to the radius of the condensed parts) $c(z, \lambda)$ is the contribution function in $J/m^2$ per wavelength (total units $J/m^3$), and $z_{max}$ is the maximal vertical extend of the atmosphere (here, the height of the top layer, which always corresponds to $P_{TOA}=10^{-8}$ bar).
Note that the altitude obtained with HELIOS depends on the local gravitational acceleration, which is held constant at the surface value and thus may lead to more compressed atmospheres, further underestimating its true radius. 
The effect of using a fixed planetary radius is shown in Fig. \ref{fig:spectra_const_radius}.
Generally, only atmospheres with low \metal are affected, as the expansion effect is strongly correlated to the MMW (see Sec. \ref{results:spectra:transmission}).

\begin{figure}[!h]
    \centering
    \includegraphics[width=\linewidth]{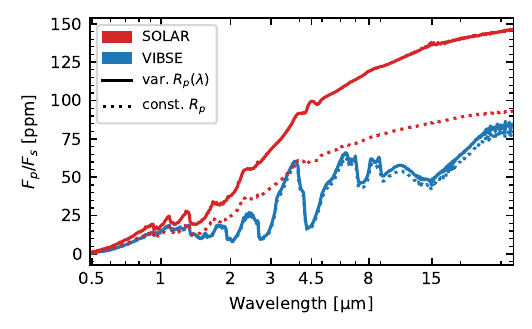}
    \caption{Comparison between emission flux computed with constant $R_p=1.74$ by HELIOS (dotted) and including the wavelength-dependent extension in radius $R_p(\lambda)$ by the atmosphere, obtained with Eq. \ref{eq:wavl_dependent_radius}. The synthetic spectra are for a SOLAR and a VIBSE case, respectively, both with \tirr=2500 K, $M_p=8M_\oplus$, \dIW{0} and \logvmf=0.}
    \label{fig:spectra_const_radius}
\end{figure}

\section{Average mean molecular weight}

\begin{figure}[!h]
    \centering
    \includegraphics[width=\linewidth]{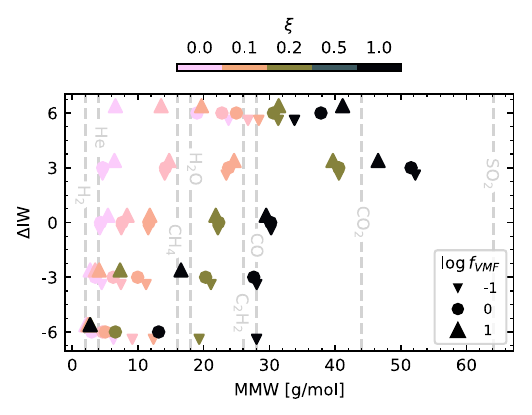}
    \caption{Average MMW of atmospheres from our grid at $T_{irr}=2500$ K. The MMW does not vary significantly with altitude. Note that the individual cases in \logvmf have been plotted with small offsets on the y-axis (\dIW{}) to enhance visual clarity .}
    \label{fig:mmw_2500}
\end{figure}

\FloatBarrier
\onecolumn
\section{Opacities}
\label{apx:opacity}

\begin{table*}[!h]
\centering
\caption{Opacity species, sources and validity ranges.}
\label{tab:opacities}
\begin{tabular}{|l|llllcc|}
\hline
Set & Species & Line list & Line list reference & Origin & T [K] & P [bar]\\
\hline
\multirow{19}{*}{all} & H & Kurucz & \citealt{kurucz2017including}& DACE & 2500-6100 & $10^{-8}$ \\
& H$_2$ & RACPPK & \citealt{roueff2019full} & DACE & 50-4500 & $10^{-8}-10^{3}$\\
& H$_2$O  & POKAZATEL & \citealt{polyansky2018exomol} & ESP & 50-8900 & $10^{-8}-10^{3}$\\
& OH & MoLLIST & \citealt{mitev2025exomol} & DACE & 50-4900 & $10^{-8}-10^{3}$\\
& He & Kurucz & \citealt{kurucz2017including} & DACE & 2500-6100 & $10^{-8}$\\
& CO$_2$ & UCL-4000 & \citealt{yurchenko2020exomol} & ESP & 50-8900 & $10^{-8}-10^{3}$\\
& CO & Li2015 & \citealt{li2015rovibrational} & DACE & 50-8900 & $10^{-8}-10^{3}$\\
& CH$_4$ & YT10to10 & \citealt{yurchenko2014exomol} & DACE & 50-2900 & $10^{-8}-10^{3}$\\
& N$_2$ & WCCRMT & \citealt{western2018spectrum} & DACE & 50-4500 & $10^{-8}-10^{3}$\\
& NO & XABC & \citealt{qu2021exomol} & DACE & 50-4500 & $10^{-8}-10^{3}$ \\
& O$_2$ & HITRAN & \citealt{gordon2017hitran2016} & ESP & 1500-4500 & $10^{-8}-10^{3}$\\
& SO & SOLIS & \citealt{brady2024exomol} & DACE & 50-4900 & $10^{-8}-10^{3}$\\
& SO$_2$ & ExoAmes & \citealt{underwood2016exomol} & DACE & 50-1900 & $10^{-8}-10^{3}$\\
& SH & GYT & \citealt{gorman2019exomol} & DACE & 50-4500 & $10^{-8}-10^{3}$\\
& H$_2$S & AYT2 & \citealt{azzam2016exomol} & DACE & 50-2900 & $10^{-8}-10^{3}$\\
& SiO  & SiOUVeNIR  & \citealt{yurchenko2022exomol} & ESP & 1500-4500 & $10^{-8}-10^{3}$\\
& Mg & Kurucz & \citealt{kurucz2017including} & DACE & 2500-6100 & $10^{-8}$\\
& MgO & LiTY & \citealt{li2019exomol} & ESP & 50-4500 & $10^{-8}-10^{3}$\\
& Fe & Kurucz & \citealt{kurucz2017including} & DACE & 2500-6100 & $10^{-8}$\\
\hline
\multirow{5}{*}{special} & C$_2$H$_2$ & aCeTY & \citealt{chubb2020exomol} & DACE & 50-2900 & $10^{-8}-10^{3}$ \\
& H$_2$-H$_2$ & HITRAN & \citealt{abel2011collision} & HELIOS-K & 50-5100 & $10^{-8}-10^{4}$ \\
& H$_2$-He & HITRAN & \citealt{abel2012infrared} & HELIOS-K & 50-5100 & $10^{-8}-10^{4}$\\
& H$_2$-H & HITRAN & \citealt{gustafsson2003h2} & HELIOS-K & 50-5100 & $10^{-8}-10^{4}$\\
& He-H & HITRAN & \citealt{gustafsson2001infrared} & HELIOS-K & 50-5100 & $10^{-8}-10^{4}$\\
\hline
\end{tabular}
% \tablefoot{
% Species in \textit{Set: all} appear in all atmospheres, whereas \textit{Set: special} species only in edge cases (see Appendix \ref{apx:acetylene_worlds} and \ref{apx:cia}). ESP stands the former Exoclimes simulation platform  (\url{chaldene.unibe.ch}), DACE for the DACE platform (\url{dace.unige.ch}; no longer online), and HELIOS-K for direct generation using \texttt{HELIOS-K} (see Sec. \ref{methods:opacities}).
% }
\tablefoot{
Species in set "all" are included in all atmospheric models, whereas species in set "special" are only considered in specific edge cases (see Appendix \ref{apx:acetylene_worlds} and \ref{apx:cia}). "ESP" refers to the former Exoclimes simulation platform (\url{chaldene.unibe.ch}; no longer available), "DACE" denotes the DACE platform (\url{dace.unige.ch}), and "HELIOS-K" indicates direct opacity generation using \texttt{HELIOS-K} (see Section \ref{methods:opacities}).
}
\end{table*}

\begin{figure*}[!h]
    \centering
    \includegraphics[width=\linewidth]{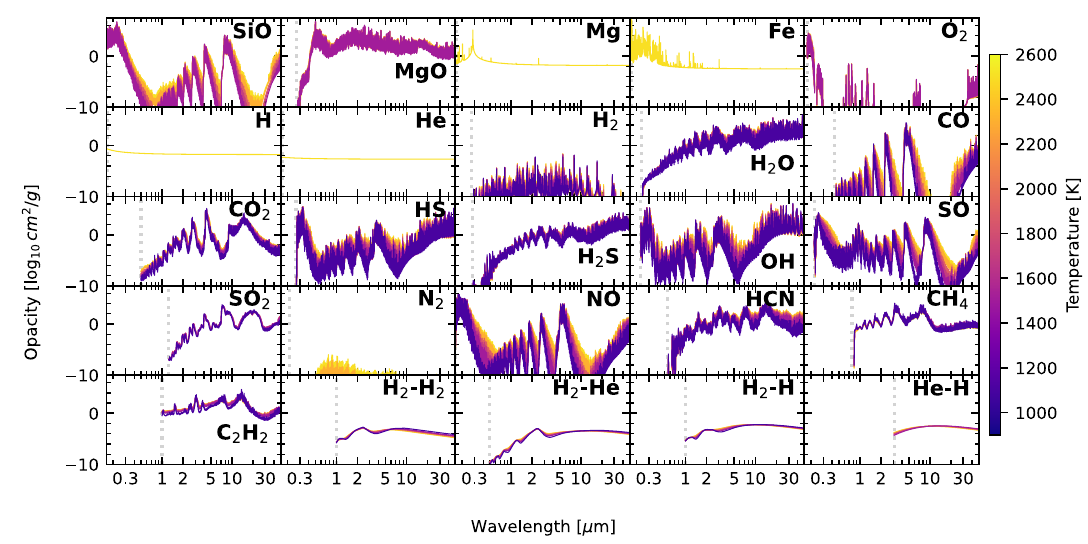}
    \caption{Unweighted opacities used in this study. Species are shown over a range of temperatures, if available. The pressure is 0.01 bar for molecules and CIAs, but for the atomic species, only $10^{-8}$ bar are available.}
    \label{fig:opacities}
\end{figure*}

\twocolumn

\section{Collision-induced absorption}
\label{apx:cia}

Here, we assess the effect of CIA relative to the opacities of dominant molecular species. The available CIA pairs relevant for the high-temperature conditions in HRE atmospheres are H$_2$–H$_2$, H$_2$–H, and H$_2$–He.
The collision line-lists of other dominant gases, e.g. CO$_2$-CO$_2$, have no experimental or theoretical support in the temperature range of interest.
CIA scales primarily with pressure, in contrast to the opacities of triatomic molecules such as H$_2$O, CO$_2$, and SO$_2$ that show stronger response to temperature.

As shown in Sec. \ref{results:atmo_struc:opac_and_thermal}, the photospheres are located at low pressure (typically in the $\sim 10^{-1}$–$10^{-6}$ bar range) for any combination of \metal, \logvmf, and \fOtwo. These altitudes are well above the pressure levels at which CIA opacities become comparable to molecular opacities ($\sim 1-100$ bar). At such high pressures, HRE atmospheres can reach temperatures of $\sim 3000$ K, for which even the high-temperature CIA line lists are currently unavailable. As a result, our model cannot self-consistently include CIA effects under these conditions. However, given the relatively grey nature of CIA opacity in the infrared, we expect their primary effect to be a general greying of the spectrum in the wavelength range where most energy transport occurs (i.e. at infrared wavelengths corresponding to $T \sim 3000$ K). Consequently, CIAs are unlikely to significantly affect the radiative transfer in atmospheres that are already optically thick due to strong absorbers like CO$_2$, SO$_2$, and especially H$_2$O.

To test this hypothesis, we performed simulations for a \logvmf = 0, \dIW{-6} atmosphere at both \metal=0 (SOLAR) and \metal=1 (VIBSE), including CIA contributions from H$_2$–H$_2$ and H$_2$–He. The results (Fig. \ref{fig:cia}) confirm that the effect of CIA is negligible at both \tirr = 2500 K and 1500 K. This specific choice of \logvmf=0 and \dIW{-6} corresponds to the part of the parameter space were we expect CIAs to be most pronounced, i.e. H$_2$-rich atmospheres.

\begin{figure}[!h]
    \centering
    \includegraphics[width=\linewidth]{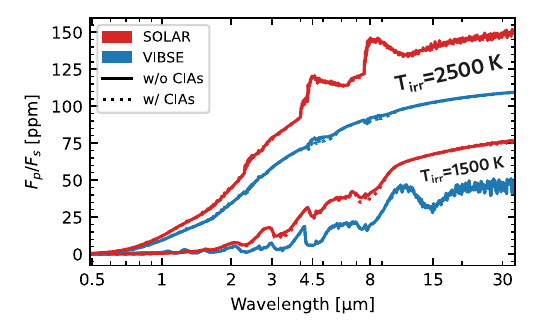}
    \caption{Negligible effect of CIAs on emission spectra.}
    \label{fig:cia}
\end{figure}

\section{Intrinsic temperature}
\label{apx:intrinsic_temp}

Earlier results on HRE atmospheres showed that elevated \tint can lead to heating of lower layers and onset of convection \citep{zilinskas2023observability, nicholls2025self}. We tested these findings by varying \tint from 0 to 300 K for SOLAR and VIBSE atmospheres at \dIW{-3} and \dIW{+3}, but recover little impact for SOLAR-like atmospheres for \tint < 200 K and no impact on VIBSE-like atmospheres at any \tint (Fig. \ref{fig:t_int}).
The oxidised SOLAR-case seems to be marginally more sensitive to the choice of \tint.

A leave-one-out-test on the opacities (not shown) confirms that mainly H$_2$O is responsible for the increase in temperature of the bottom layers seen in SOLAR-cases at \tint $\gtrsim$ 200 K in Fig. \ref{fig:t_int}.
As underlying reason we identify its strong and effectively grey opacity compared to other major gases like H$_2$, He, CO$_2$, CO and SO$_2$ (Fig. \ref{fig:opacities}).
In highly opaque atmospheres, the contribution of the intrinsic temperature is enhanced and becomes dominant above a certain value of \tint, based on the opacity structure \citep[cf.][their Eq. 29]{guillot2010radiative}; comparison of these contributions indicate that in SOLAR-like atmospheres, \tint starts to dominate  at $\sim$200-300 K, while VIBSE-like atmospheres should remain unaffected (i.e. dominated by \tirr), consistent with our findings.
Since H$_2$O is more abundant in SOLAR-like atmospheres (see Fig. \ref{fig:atmospheric_structure}), and there particularly in oxidised cases, these atmospheres tend to be most sensitive to \tint.
Once the intrinsic temperature modifies the P-T-profile, the atmosphere quickly enters a positive feedback loop where rising temperatures force greater \fOtwo for a given \dIW{}, in turn producing more H$_2$O.
Hence, no converging solution for \tint=300 K could be found for the oxidising SOLAR case in Fig. \ref{fig:t_int}.

Regarding convective stability, we found that the sharp increase in surface temperature pushes the atmosphere towards violation of the Schwarzschild-criterion (albeit not exceeding it), potentially inducing some small-scale convection in the lower layers of the planetary atmosphere, close to the magma ocean.
However, this analysis is only valid for the tested case, \tirr=2500 K.
Cooler atmospheres host lower radiation fluxes, and thus, a given value of \tint should have a stronger effect, potentially leading to significant modification of the atmospheric P-T-profile, including the onset of convection which we found to be suppressed in the canonical models studied here \citep[cf.][]{nicholls2025self}.

\begin{figure}
    \centering
    \includegraphics[width=\linewidth]{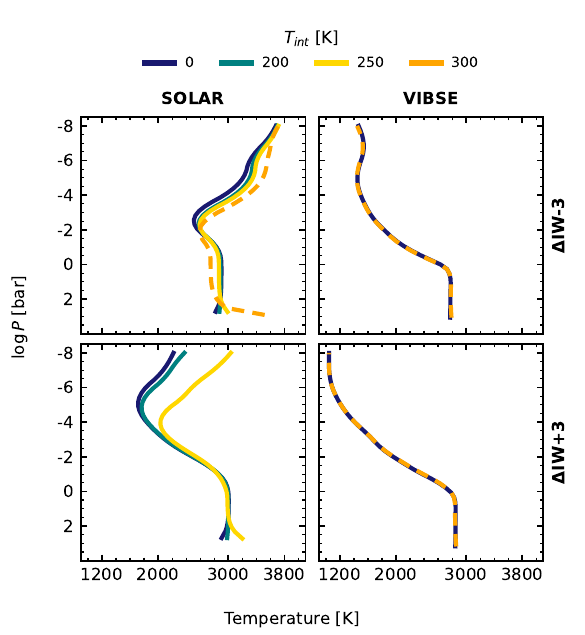}
    \caption{Effect of intrinsic temperature, \tint, on P-T-profiles.}
    \label{fig:t_int}
\end{figure}

\FloatBarrier
\onecolumn
\section{Extended vapour series}
\begin{figure*}[!h]
    \centering
    \includegraphics[width=\linewidth]{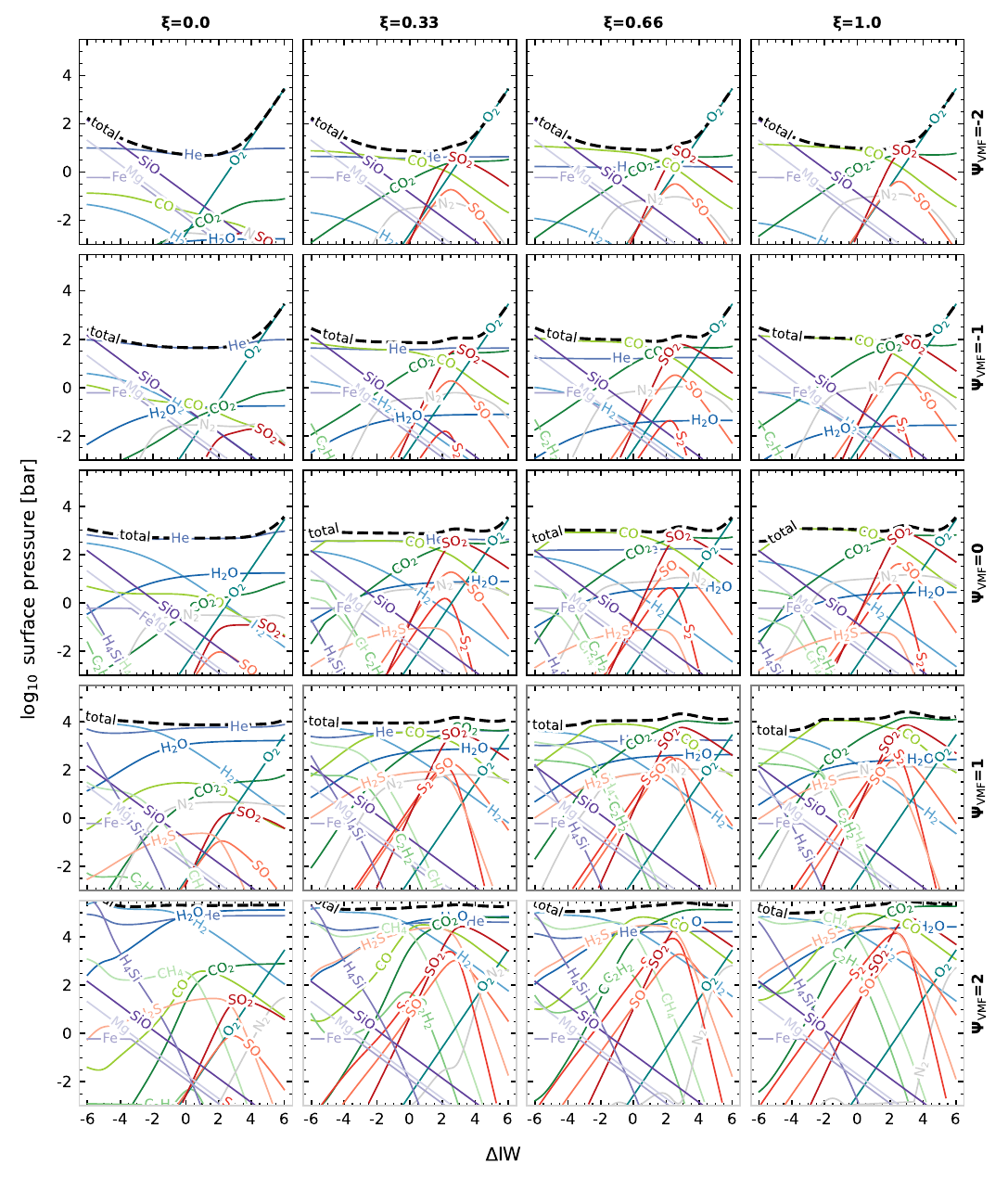}
    \caption{Partial pressures of exsolved vapours for a wide range of conditions. The temperature at the MAI is 3000 K.  Greyed-out axes highlight the onset of non-ideality, with greater deviation from the ideal gas law indicated by more desaturated grey.}
    \label{fig:vapser_batch_T3000}
\end{figure*}

\begin{figure*}[!h]
    \centering
    \includegraphics[width=\linewidth]{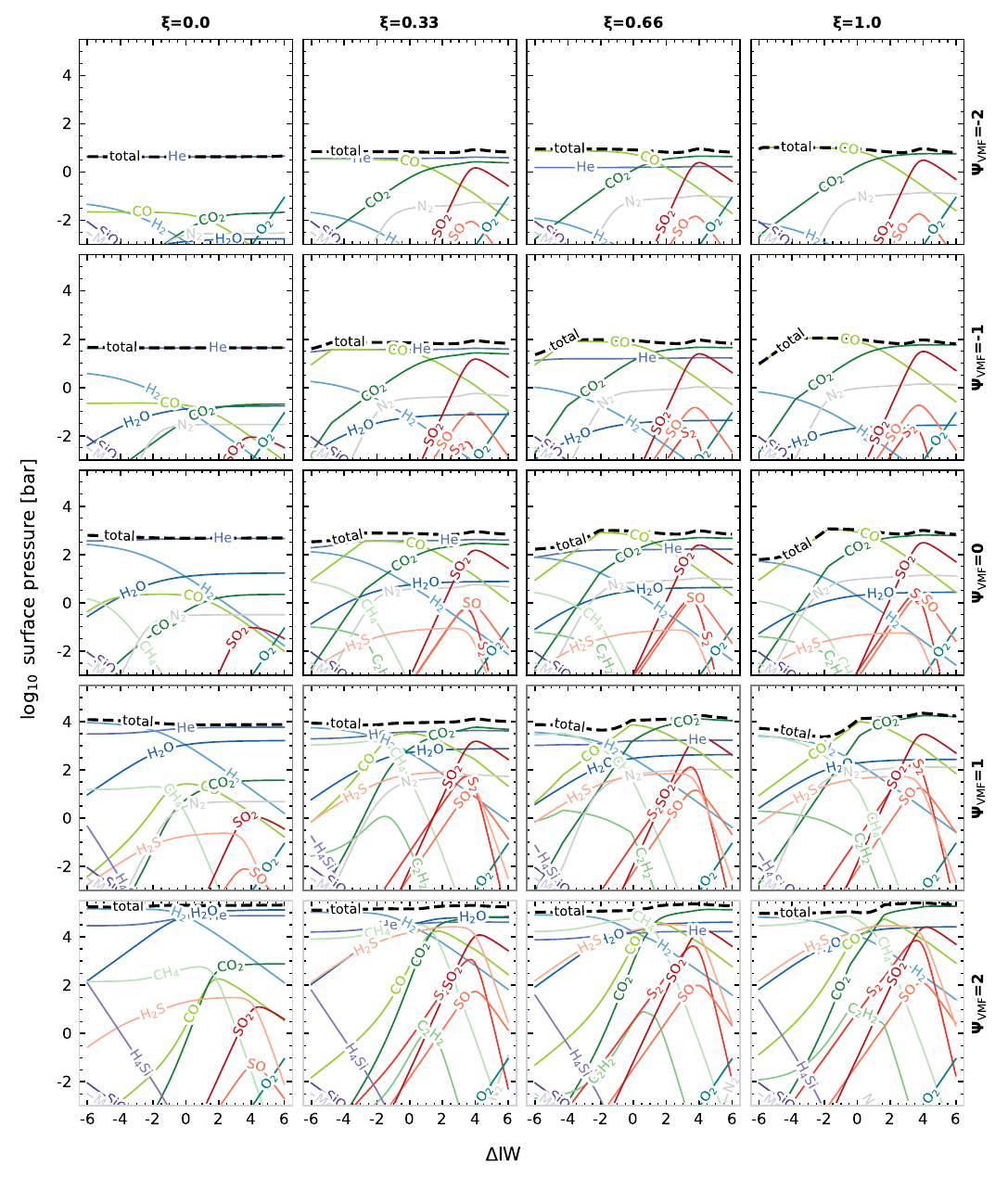}
    \caption{Similar to Fig. \ref{fig:vapser_batch_T3000}, but with T=2000 K.}
    \label{fig:vapser_batch_T2000}
\end{figure*}
\twocolumn

\FloatBarrier
\onecolumn
\section{Emission spectra at varying temperatures}
\label{apx:spectra_emission_cooler}

\begin{figure*}[!h]
    \centering
    \includegraphics[width=\linewidth]{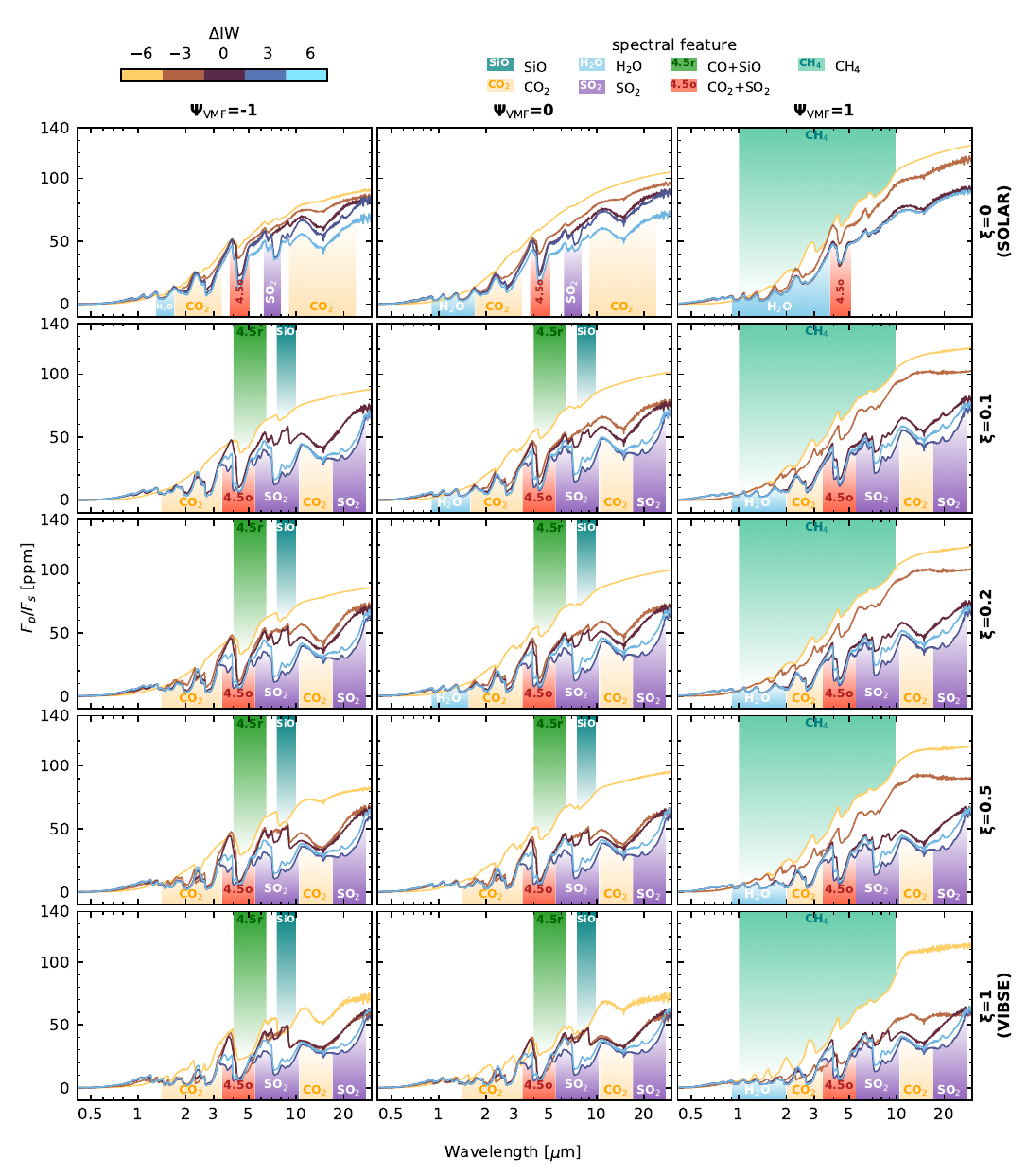}
    \caption{Similar to Fig. \ref{fig:spectra_emission}, but for \tirr=2000 K.}
    \label{fig:spectra_emission_2000K}
\end{figure*}

\begin{figure*}[!h]
    \centering
    \includegraphics[width=\linewidth]{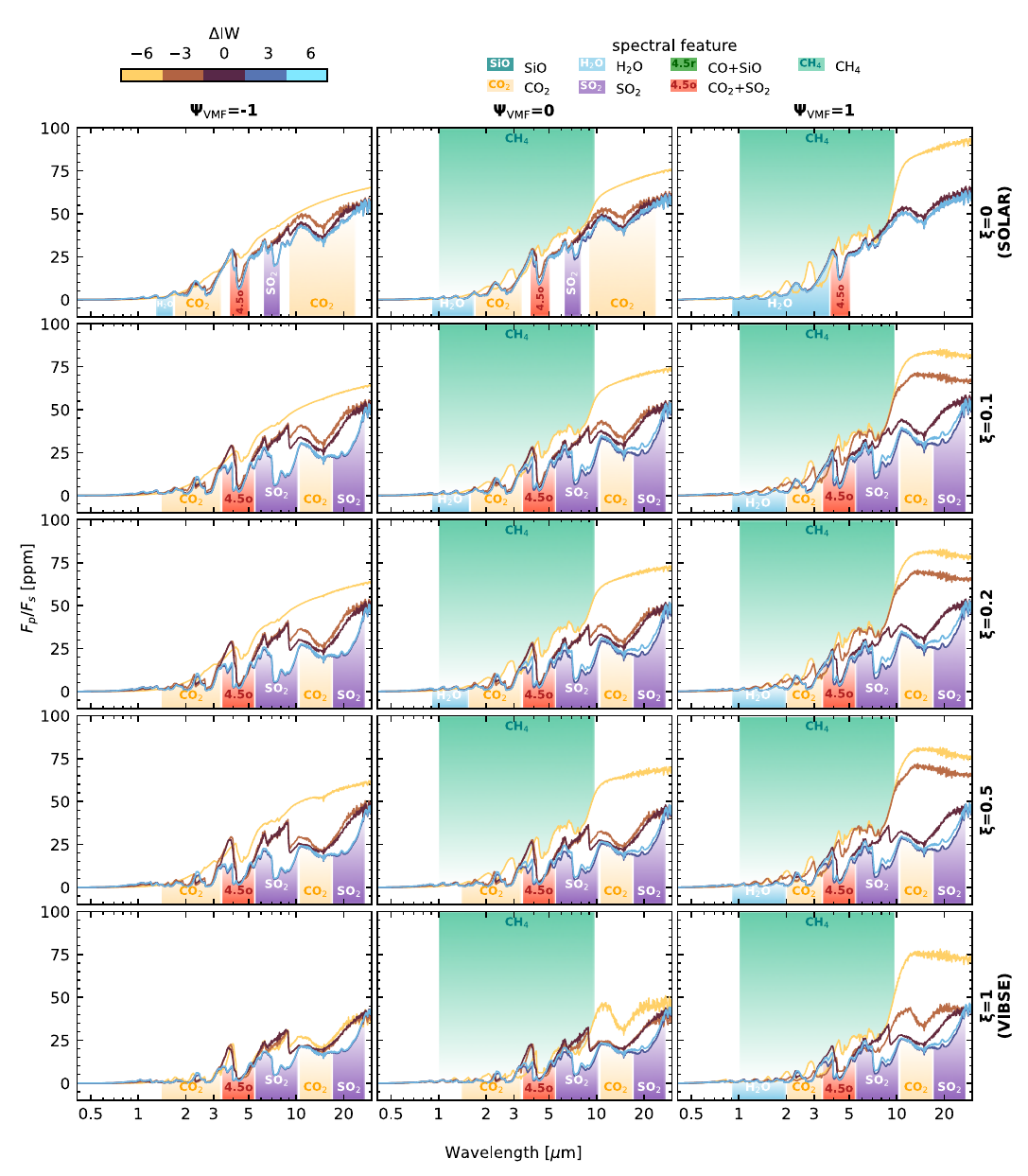}
    \caption{Similar to Fig. \ref{fig:spectra_emission}, but for \tirr=1500 K.}
    \label{fig:spectra_emission_1500K}
\end{figure*}
\twocolumn

\FloatBarrier
\twocolumn

\section{55 Cnc e, MIRI constraints}

Here we report the detailed results from the $\chi^2$-analysis presented in Sec. \ref{disc:spectra}.
We evaluated the $\chi^2$ for the MIRI observation \citep{hu2024secondary} for all emission spectra from Fig. \ref{fig:spectra_emission}, and show them in Fig. \ref{fig:chi2_posterior}.
The upper threshold value of $\chi^2$ was chosen such that the percentile point function is 0.95; a model that has a $\chi^2$ worse than this is considered non-fitting.
Our model has seven degrees of freedom (corresponding to the number of data points in the MIRI observation, excluding the shadow region; see Fig. \ref{fig:MIRI_best_fits}), which yields a threshold value of $\sim 14$. 

\begin{figure}[!h]
    \centering
    \includegraphics[width=\linewidth]{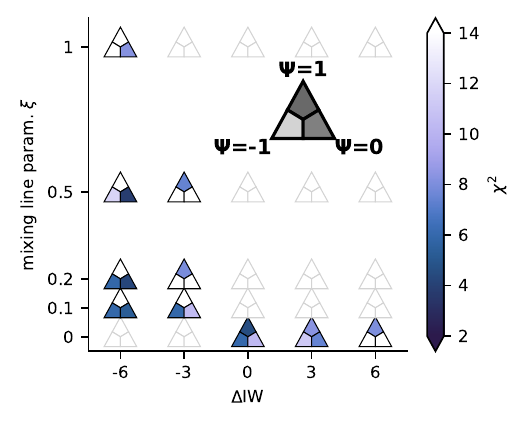}
    \caption{Models that fit the MIRI LRS observation, ranked by the $\chi^2$ metric. Each segment of the triangular glyphs reports the $\chi^2$ of all tested \logvmf (-1, 0, 1) at given \dIW{} (x-axis) and \metal (y-axis). Cases were no fitting solution was found are greyed out.}
    \label{fig:chi2_posterior}
\end{figure}

\section{Acetylene Worlds}
\label{apx:acetylene_worlds}

Acetylene, C$_2$H$_2$, emerges under C-rich but H- and O-depleted conditions (see Fig. \ref{fig:acetylene_chemistry}), typically found in highly reducing (\dIW{}  $\lesssim -3$), high \logvmf ($\gtrsim$ 0) and VIBSE-like ($z\sim1$) scenarios.
Compared to methane (CH$_4$), acetylene appears preferentially in hot atmospheres ($\gtrsim$ 2000 K, cf. Fig. \ref{fig:vapser_batch_T2000} and \ref{fig:vapser_batch_T3000}).
As major opacity source in the infra red (Fig. \ref{fig:opacities}), it may become the dominant absorber, overpowering HCN; in cooler atmospheres (\tirr $\lesssim 2000$ K), methane begins to replace C$_2$H$_2$ (Fig. \ref{fig:spectra_emission_1500K}).

However, the line lists of both CH$_4$ \citep{yurchenko2014exomol} and C$_2$H$_2$ \citep{chubb2020exomol} terminate at 0.8/1 \micron, respectively, and thus lack opacity in the vital UVIS where a significant fraction of the stars energy is input into the atmosphere.
The rapid cut-off induces an artificial and pronounced greenhouse-effect; the BOA temperatures are raised by $\sim 250$ K by a scenario where C$_2$H$_2$ is included in the radiative transfer simulation versus a scenario without (Fig. \ref{fig:acetylne_pt}, left column).
The abrupt change in opacity at 1 \micron becomes evident in the emission spectrum as artificial feature that strongly affects both tested oxidation states equally, \dIW{-3} and \dIW{-6} (Fig. \ref{fig:acetylne_pt}).
The effect of both acetylene and methane quickly dies off with increasing oxygen content, as both gases vanish; atmospheres with \dIW{}$\gtrsim$ -2 or \logvmf<0 are never affected.

Due to the limitation in the line list and the artificial changes to the pressure-temperature structure and the spectra of ultra-HREs, we omitted C$_2$H$_2$ from the opacity, but remark that the affected atmospheres in Fig. \ref{fig:spectra_emission} and \ref{fig:spectra_transmission} are \dIW{-6} for \logvmf=0 and \dIW{-6} and \dIW{-3} for \logvmf=1; in all other atmospheres, the effect of acetylene is not relevant.
In the absence of C$_2$H$_2$ as absorber, HCN and CO take precedence (see Fig. \ref{fig:spectra_transmission}).
However, these spectra are likely inaccurate, and subject to change once newer line lists can be obtained.

\begin{figure}[!h]
    \centering
    \includegraphics[width=\linewidth]{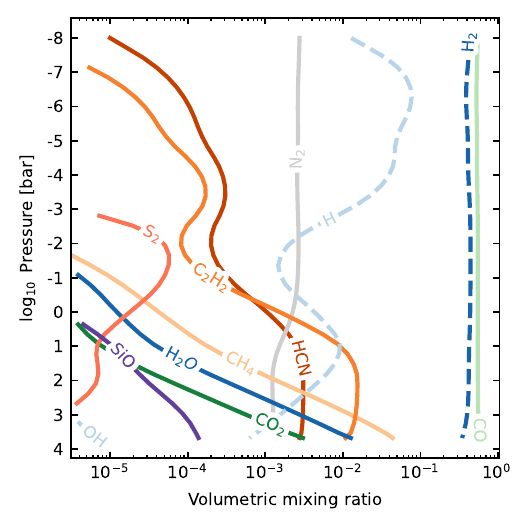}
    \caption{Chemistry as function of altitude in a \logvmf=1, \metal=1, \dIW{-3} planet at \tirr=2500 K.}
    \label{fig:acetylene_chemistry}
\end{figure}

\begin{figure}[!h]
    \centering
    \includegraphics[width=\linewidth]{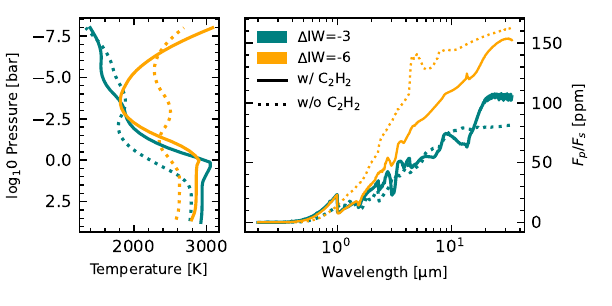}
    \caption{Effect of acetylene on the atmospheric structure and spectra of reducing and heavy atmospheres (\dIW{}$\leq$ -3, \logvmf=1, \metal=1).}
    \label{fig:acetylne_pt}
\end{figure}

\end{appendix}

\end{document}